\documentclass[twocolumn,preprintnumbers,superscriptaddress,amsmath,amssymb,longbibliography]{revtex4-2}

\usepackage{color}
\usepackage{graphicx}
\usepackage{dcolumn}
\usepackage{float} 
\usepackage{bbm}

\usepackage{setspace}
\usepackage{verbatim}

\usepackage[dvipsnames]{xcolor}

\newcommand \ket[1]{\left|#1 \right>}
\newcommand \ketbra[2]{\left|#1 \right>\left<#2 \right|}
\newcommand \braket[2]{\left<#1 \mid #2 \right>}

\newcommand{\textD}{\text{D} }

\begin{document}

\title{Photon-resolved Floquet theory approach to   spectroscopic quantum sensing }

\author{Georg Engelhardt}
\email{georg-engelhardt-research@outlook.com}

\affiliation{International Quantum Academy, Shenzhen 518048, China}
\affiliation{Shenzhen Institute for Quantum Science and Engineering, Southern University of Science and Technology, Shenzhen 518055, China}
\affiliation{Guangdong Provincial Key Laboratory of Quantum Science and Engineering, Southern University of Science and Technology, Shenzhen, 518055, China}

\author{Konstantin Dorfman}

\email{dorfmank@hainanu.edu.cn}

\affiliation{Center for Theoretical Physics and School of Physics and Optoelectronic Engineering, Hainan University, Haikou, China, 570228, China}
\affiliation{Himalayan Institute for Advanced Study, Unit of Gopinath Seva Foundation, MIG 38, Avas Vikas, Rishikesh, Uttarakhand 249201, India}

\author{Zhedong Zhang}
\email{zzhan26@cityu.edu.hk}

\affiliation{Department of Physics, City University of Hong Kong, Kowloon, Hong Kong SAR}

\affiliation{Shenzhen Research Institute, City University of Hong Kong, Shenzhen, Guangdong 518057, China}

\date{\today}

\pacs{
  }

\begin{abstract}
Spectroscopic methods play a vital role in quantum sensing, which uses the quantized nature of atoms or molecules to reach astonishing precision for sensing of, e.g., electric or magnetic fields. In the theoretical treatment, one typically invokes  semiclassical methods to describe the light-matter interaction between quantum emitters, e.g., atoms or  molecules, and a  strong coherent laser field. However, these semiclassical approaches struggle to predict the stochastic measurement fluctuations beyond the mean value, necessary to predict the sensitivity of spectroscopic quantum sensing protocols. Here, we develop  a theoretical framework based on the recently developed Photon-resolved Floquet theory (PRFT) which is capable to predict the measurement statistics describing higher order statistics of coherent quantum states of light. The PRFT constructs flow equations for the cumulants of the photonic measurement statistics utilizing only the  semiclassical dynamics of the matter system. We apply the PRFT to spectroscopic quantum sensing using dissipative two-level and four-level systems (describing electric field sensing with Rydberg atoms), and demonstrate how to calculate the Fisher information of the measurement statistics with respect to various system parameters. In doing so, we demonstrate that the PRFT is a flexible tool allowing to improve the sensitivity of spectroscopic quantum sensing devices by several orders of magnitudes.
\end{abstract}

\maketitle

\allowdisplaybreaks


\section{Introduction}

\begin{figure}
	\includegraphics[width=\linewidth]{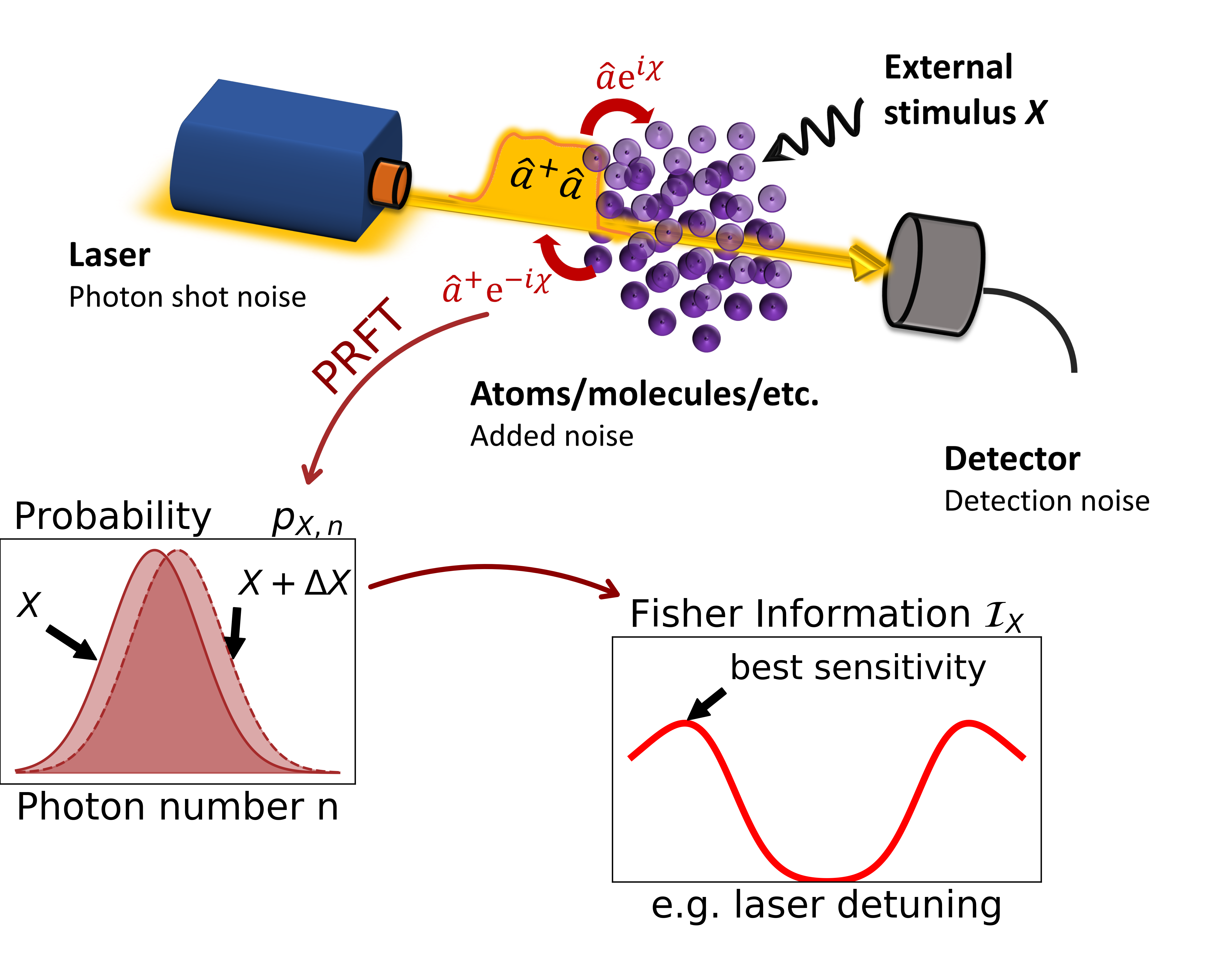}
	\caption{Sketch of the spectroscopic setup, in which laser light interacts with an atomic ensemble, before it is finally measured by a detector.   The PRFT includes counting fields $\chi$ into the semiclassical dynamics allowing to predict the measured photonic probability distribution. The presence of an external stimulus $X$ changes the measurement statistics from which on can estimate the strength of $X$. The Fisher information predicted by the PRFT assists in finding  experimental parameters (e.g., the laser detuning) which optimize the measurement sensitivity.}
	\label{figSpectroscopySetup}
\end{figure}

It is hard to overstate the importance of spectroscopy in science and technology, reaching from medicine, environmental monitoring, pharmaceutical research, material science to public safety, to mention but a few. The advent of the laser enabled experimentalists to generate light sources with outstanding properties, e.g.,  pulses featuring a  duration of few Femto seconds, an extreme narrow bandwidth below $1\,\text{Hz}$, or peak intensities exceeding $10^{20}\;\text{W}/\text{cm}^2$. These technological advances have allowed researchers to study  atomic and molecular processes, such as isomerization~\cite{Baraban2015}, photo dissociation~\cite{Colley2023}, conical intersections~\cite{Domcke2012},  polariton dynamics~\cite{Zhang2022a,Reitz2025}, as well as free-space superradiance~\cite{Ferioli2023,Agarwal2024,Goncalves2025} with astonishing precision.

 Moreover,  spectroscopic methods find vital applications in quantum sensing, which deploys either coherence, entanglement or the quantized structure of matter to improve the measurement precision beyond classical means~\cite{Degen2017,Pezze2018,Ye2024}. Experiments using the coherence of cold-atom matter-wave interferometers are currently being developed for precision measurement of the gravitational field, requiring spectroscopic methods for state preparation and readout~\cite{Graham2016,Hogan2016,Panda2024}. Squeezed light  assists in the detection of gravitation waves at LIGO~\cite{Jia2024}. Protocols deploying entangled photons have the potential to improve spectroscopic precision measurements~\cite{Li2022} as they can circumvent the time-frequency uncertainty relation~\cite{Dorfman2021,Zhang2022,Fan2024}. The quantized nature of atoms is the basis for the already ultra-precise but ever-improving definition of the time-standard using atomic and optical clocks~\cite{Ludlow2015,Robinson2024}.

 Likewise, optical magnetometry, which measures the rotation of linearly polarized light due to the Faraday effect induced by the interaction with the quantized-level structure of alkali atoms, is a highly-precise spectroscopic quantum sensing method for the measurement of magnetic fields~\cite{Budker2013,Wolfgramm2010,Ma2024,Xu2024,Ma2025,Wang2025}. In combination with noble atoms, optical magnetometry is presently employed in the search for the elusive dark matter~\cite{Budker2014,Aybas2021,Bloch2022}. Moreover, the huge polarizibility of highly-excited states of alkali atoms makes  these so-called Rydberg atoms extremely susceptible to  electromagnetic fields in the microwave  and radio wavelength regimes~\cite{Loew2012}. The resulting change in the optical absorption of an optical probe laser can then be measured using  electromagnetically-induced transparency protocols~\cite{Holloway2017,Cox2018,Meyer2020,Jing2020,Meyer2021,Holloway2022,Liu2022,Borowka2024,Kumar2023}. This approach finds also application in the search of dark matter~\cite{Ghosh2024}. Such spectroscopic quantum sensing protocols using coherent laser light have a very simple experimental setup, as sketched in Fig.~\ref{figSpectroscopySetup}. A laser emits coherent light which interacts with an ensemble of atoms or other quantum emitters before its intensity, phase or polarization rotation is being measured by a photon detector setup. The presence of an external stimulus (denoted by $X$ in Fig.~\ref{figSpectroscopySetup}) changes the optical properties of the atoms, resulting in a modified measurement statistics at the photon detector, from which one can statistically infer the strength of that stimulus.

To theoretically predict the mean values of  spectroscopic signals, a series of powerful  methods for nonlinear spectroscopy have been devised. For instance,  diagrammatic techniques have proven  useful to match each order of the measured intensity to a term in the perturbative expansion~\cite{Mukamel1995}. In the high-intensity regime, one usually resorts to semiclassical approaches like the quantum master equation~\cite{Scully1997} in combination with the Maxwell-Bloch theory~\cite{Hammerer2010}. Even though these established semiclassical methods  reliably predict the mean value of the measured spectroscopic signal, they  do not directly provide information about the   fluctuations of the experimental measurements.

 Typically, there are three distinct sources of noise in a spectroscopic setup as sketched in Fig.~\ref{figSpectroscopySetup}: (a) the photon-shot noise of the laser-light source; (b) the noise  occurring during the light-matter interaction, which might have a thermal origin; (c) the detection noise appearing due to an imperfect measurement process. While the estimation of the photon-shot noise and the detection noise is straightforward, the prediction of the noise added by the light-matter interaction is nontrivial, and bears fundamental limitations on the measurement sensitivity. To predict the measurement statistics beyond the mean value, more advanced methods such as phase-space methods~\cite{Scully1997}, quantum trajectories~\cite{Wiseman1993}, and the input-output theory~\cite{Gardiner1985} can be employed. While general, these methods become difficult to evaluate in practice: Phase space methods require to solve complex partial differential equations. Input-output theory   will lead to multi-time correlation functions of the polarization operator, which are  nontrivial to evaluate even within a semiclassical approximation~\cite{Gardiner2004}.

 In this paper, we develop an alternative method to predict the measurement statistic in spectroscopic experiments based on a full-counting statistics approach \cite{Dorfman2018}. The derivation of this methods takes advantage of the recently developed Photon-resolved Floquet theory (PRFT), which is a semiclassical method of light-matter interaction, not only capable of predicting the state of the driven matter system, but also the detailed photon statistics of the driving field~\cite{Engelhardt2024,Engelhardt2024b,Engelhardt2024c}. It thus constitutes a suitable tool to calculate the noise added by the light-matter interaction as sketched in Fig.~\ref{figSpectroscopySetup}. Using the PRFT, it is then possible to project the  measurement sensitivity $\left<\Delta X^2\right> $  of a spectroscopic quantum sensing protocol via the celebrated Cramer-Rao bound~\cite{Rao1945,Cramer1946}
\begin{equation}
	 \left<\Delta X^2\right>  \geq \frac{1}{\mathcal I_X}, 
	 \label{eq:cramerRaoBound}
\end{equation}
where  the Fisher information 
\begin{eqnarray}
\mathcal I_{X} &=& \sum_{\boldsymbol  n }  p_{ X, \boldsymbol  n } \left( \partial_{X} \ln p_{X, \boldsymbol  n } \right)  ^2 ,
\label{eq:def:FisherInformation}
\end{eqnarray}
is a function of the $X$-dependent photonic probabilities  predicted by the PRFT. Using the Fisher information, it is then possible to optimize the measurement precision by finding  experimental parameters, e.g., the laser detuning,  maximizing the Fisher information as illustrated in Fig.~\ref{figSpectroscopySetup}.

 The PRFT  takes advantage of the rich methodologies of full-counting statistics of electron transport through mesoscopic structures in electronic systems~\cite{Levitov1996,Platero2004,Schoenhammer2007,Brandes2008,Esposito2009,Hussein2014,Flindt2009,Ridley2019,Pollock2022,Kleinherbers2018,Benito2016,Schaller2018,Xu2023,Landi2024}, where a spatially confined quantum system, such as quantum dots or single-electron transistors, is coupled to two or  more electronic leads. A thermodynamic bias between the leads drives an electron current through the quantum system, whose statistics can be traced by implementing so-called counting fields into the equations of motion of the reduced density matrix of the confined quantum system. More recently, full-counting statistics has been also applied in the study of heat transport through quantum systems~\cite{Wang2017,Cerrillo2016,Restrepo2018,Soret2025}. 

Similarly, the PRFT includes counting fields into the semiclassical equations of motion of the quantum emitters, which trace the photon exchange between several coherent driving fields. In its original form, the PRFT has been  a generic theoretical framework which has not been applied to realistic physical systems so far~\cite{Engelhardt2024,Engelhardt2024c,Engelhardt2024b}. Moreover, the semiclassical approximation invoked in the establishment of the PRFT, does not take  into account the photon number dependence of the creation and annihilation operators.  These  two  major issues will be addressed in this work.

The theoretical framework developed here, which applies the PRFT to spectroscopic measurements,  relies  only on semiclassical methods, in which light and matter are, respectively,  described quantum mechanically and classically,  and is nevertheless capable to predict the measurement statistics of prominent spectroscopic observables, namely, the intensity, polarization direction and phase shifts of the probe light in a coherent quantum state. We concentrate thereby on the calculation of the corresponding mean values and the covariance matrix, as they typically dominate the measurement statistic for a large number of non-interacting  quantum emitters. We further  derive flow equations for  low-order cumulants, which are reminiscent of the Maxwell-Bloch equation for the mean value of the electric field.  These flow equations can be easily integrated numerically, and in some special cases can be solved analytically.

We apply the framework to an ensemble of two-level systems which are subject to a laser field, whose measurement statistics in spectroscopic quantum sensing is to be determined. In a full  quantum description, this system can be represented as a dissipative Tavis-Cummings model~\cite{TAVIS1969}, which cannot be solved exactly (neither analytically nor numerically) because of the large number of photons in the driving field. However, using the PRFT approach, it is possible to analytically study the stochastic properties of the system. Using these findings, we derive analytical expressions for the Fisher information of  the atomic density and the probe field detuning, which allows to determine an experimental configuration  optimizing the measurement sensitivity. To demonstrate the broad applicability of our method, we carry out a similar analysis for an atomic four-level system, which is presently employed for spectroscopic quantum sensing using Rydberg atoms~\cite{Jing2020,Meyer2020}. The PRFT thereby predicts that the measurement precision can be improved by several order of magnitudes by invoking a phase measurement instead of the commonly used intensity measurements.

The rest of this article is organized as follows: In Sec.~\ref{sec:framework}, we introduce the generic system under investigation and   the  theoretical framework. In Sec.~\ref{sec:spectroscopicObservables}, we explain how to utilize the framework to calculate the measurement statistic of important spectroscopic observables, namely, the intensity, the polarization direction and phase shift of a coherent light field, from which one can calculate the Fisher information of spectroscopic sensing protocols.  In Sec.~\ref{sec:tavisCummingsModel}, we apply the framework to analyze the spectroscopic properties of an ensemble of dissipative two-level systems. In Sec.~\ref{sec:fourLevelSystem}, we evaluate the measurement sensitivity of an ensemble of four-level atoms. In Sec.~\ref{sec:conclusions} we conclude and discuss our findings. Details and derivations are given in the Appendices.

\section{Framework}

\label{sec:framework}

In this section, we introduce the generic light-matter system for which we develop methods to predict the measurement statistics in spectroscopic measurements with coherent light. Thereby, we apply the PRFT by introducing counting-fields into the semiclassical equation of motions, which predict the major characteristics of the photonic probability distribution required to calculate the Fisher information.

\subsection{System}

 The spectroscopic setup investigated in this work   is  sketched in Fig.~\ref{figSpectroscopySetup}, in which   strong laser light propagates through an ensemble of quantum emitters (e.g., atoms or molecules), to which we will refer as the matter system in the following. For concreteness,  we specify the discussion to atoms but remark that the  framework has broader applicability. In a microscopic quantum optical description, the  Hamiltonian is given by
\begin{equation}
\hat H   = \sum_{m=1}^{N} \hat H_{m} + \sum_{k=1}^{D}  \omega_k  \hat a_k^\dagger  \hat a_k  +  \sum_{m,k} g_{m,k} \hat V_m \left(  \hat a_k^\dagger   + \hat a_k \right),
\label{eq:hamiltonian:quantum}
\end{equation}
where $\hat H_{m} $ describes the quantum emitter $m$ in the absence of the laser fields,  and $\hat V_m$ denotes the corresponding polarization operator. The photonic creation and annihilation operators $\hat a_k^\dagger,  \hat a_k$ quantize the  light emitted by  lasers with frequencies $\omega_k$. The light-matter interaction strength is parameterized by the couplings $g_{m,k}$. After transversing the atomic ensemble, the transmitted light is measured by a detector setup. 

More precisely, we  assume that the laser light consists  of a sequence of pulses with  duration $t_{\text{P}}$ within the total measurement time $t_{\text{M}}$. A specific pulse is quantized by a set of photonic operators  $\hat a_k^\dagger,  \hat a_k$ whose statistics is measured afterwards in the detection setup. These photonic operators might quantize, e.g., different frequencies  of the laser light, the polarization direction of linearly polarized light, or some other quantized photonic observables.  We envision that the laser pulses  interact  with a particular atom $m$, during which photons can be either absorbed by the atom, or redistributed between the photonic modes. This interaction process can be parameterized by the time-depended coupling function $g_{m,k}(t)$, whose time dependence is determined by the   amplitude and shape of the pulse. A  microscopic motivation of the Hamiltonian in Eq.~\eqref{eq:hamiltonian:quantum} is given in Appendix~\ref{sec:lightMatterHamiltonian}.

Prior to the  interaction of a pulse with the ensemble of atoms, the joint system is assumed to be in the product state
\begin{equation}
\rho_{\text{tot} }(t=0) = \rho_{\text{M}}(0) \otimes \left|\alpha \right>\left< \alpha\right|,
\label{eq:InitalState}
\end{equation}
in which the matter state 
\begin{equation}
\rho_{\text{M} }(t=0) =  \bigotimes_{j=1}^N \rho_{\text{M} ,j}(0) 
\label{eq:InitalStateMatter}
\end{equation}
is  a tensor product of the density matrices of the individual atoms.  The  initial state of the photonic modes is assumed to be a pure state, which is parameterized by
\begin{eqnarray}
\left|\alpha \right> &=& \sum_{ \boldsymbol n } a_{\mathbf n }   \left| \boldsymbol n \right>\nonumber,\\
a_{\mathbf n } &=& \frac{1}{(2\pi )^{\frac{D}{4} } \sqrt{\det \boldsymbol \Sigma} } e^{-\frac{1}{4}\left(\mathbf n- \overline { \mathbf n }  \right) \boldsymbol \Sigma^{-2}  \left(\mathbf n- \overline {\mathbf n } \right)^T   }  e^{i\overline {\boldsymbol \varphi} \cdot \boldsymbol n},
\label{eq:initialStateMatterSubsystem}
\end{eqnarray}
where we have introduced the vector notation $ \boldsymbol n =(n_1,\dots, n_{D} )$  for the photon numbers $n_k$, which label the Fock states $\left| \boldsymbol n \right> =\bigotimes_{k=1}^{D}\left|  n_k \right> $. 
The photonic state is assumed to have  well-defined mean photon numbers $\overline { \mathbf n }$ and  mean phases $\overline {\boldsymbol \varphi} =( \overline \varphi_1,\dots, \overline \varphi_{D} )$. The covariance matrix $\boldsymbol \Sigma^{2} $ parameterize the photon-shot noise  and mutual correlations  between the photonic modes.

In the following, we investigate the light-matter interaction in the semiclassical regime, which can be defined by a diverging covariance matrix $\boldsymbol \Sigma^{2} $. To make this limit precise, we assume that $\boldsymbol \Sigma^{2} = \sigma \tilde{ \boldsymbol \Sigma}^{2}  $, where the matrix $\tilde{ \boldsymbol \Sigma}^{2}$ shall be constant. The semiclassical limit can then be controlled by assuming the limit $\sigma \rightarrow \infty$. 

In the semiclassical limit, the photonic creation and annihilation in   Eq.~\eqref{eq:hamiltonian:quantum} can be approximated by their values at the mean-photon numbers $ \overline { \mathbf n }$. More precisely, after a transformation of the Hamiltonian in Eq.~\eqref{eq:hamiltonian:quantum} into a rotating frame defined by $U^{(\text{free})}= \exp\left( -i\sum_k \omega_k \hat a_k^\dagger \hat a_k t \right)$, we replace the photonic operators in the Hamiltonian with complex-valued numbers, i.e.,
\begin{eqnarray}
 \hat a_k &\rightarrow& \sqrt{\overline n_k} e^{i\varphi_k-i\omega_k t}. 
\end{eqnarray}
The corresponding Hamiltonian in the semiclassical limit is given by
\begin{eqnarray}
\hat {\mathcal H}_{\boldsymbol \varphi} (t) &=& \sum_{m=1}^{N} \hat {\mathcal H}_{m ,\boldsymbol \varphi} (t), \nonumber \\
\hat {\mathcal H}_{m,\boldsymbol \varphi} (t)  &=& \hat H_{m}    +  \sum_{k}\Omega_{m,k} (t)\hat V_m \cos \left(\omega_k t -\varphi_k \right) ,
\label{eq:hamiltonian:semiclassical}
\end{eqnarray}
which acts only on the matter system. Thereby, we have introduced the time-dependent Rabi frequencies  $\Omega_{m,k} (t) =2 g_{m,k} (t)\sqrt{\overline n_k}$.  As we assume noninteracting quantum emitters, the Hamiltonian can be expressed as a sum over the quantum emitter index $m$. For each semiclassical quantum-emitter  Hamiltonian $\hat {\mathcal H}_{m,\boldsymbol \varphi} (t) $, the time-evolution operator given by
\begin{equation}
\hat {\mathcal U}_{\overline { \boldsymbol  n},m,\boldsymbol \varphi} (t)  = \hat{\mathcal T} e^{-i \int_0^{t} \hat {\mathcal H}_{m,\boldsymbol \varphi} (t^\prime) dt^\prime}
\label{eq:timeEvolutionOperator:semiclassical}
\end{equation}
describes the time-evolution of the matter system in the semiclassical regime. As we  assume identical quantum emitters in this paper, we will neglect the index $m$ hereafter.  For later purpose, we have added the mean-photon number vector $\overline { \boldsymbol  n} $ as an index to the time-evolution operator. Despite the absence of the photonic operators in Eq.~\eqref{eq:hamiltonian:semiclassical}, the PRFT can  predict the detailed information about the photonic state solely based on the semiclassical time-evolution operator Eq.~\eqref{eq:timeEvolutionOperator:semiclassical} as we will show in the remaining part of this paper.

\subsection{Full-counting statistics}

 Before introducing the PRFT,  we first explain the concepts of full-counting statistics, on which the PRFT relies on.  The cumulant-generating function $K_{\boldsymbol \chi} (t)$ of the photonic probability distribution is defined by
\begin{eqnarray}
K_{\boldsymbol \chi} (t)  &=&  \log  \left< e^{-i \sum_k  \chi_k \hat a_k^\dagger \hat a_k   }\right>,
\label{eq:def:generatingFcts}
\end{eqnarray}
 where $\left< \bullet \right>$ denotes the expectation value with regard to an arbitrary quantum state. For instance, for the Gaussian state in Eq.~\eqref{eq:initialStateMatterSubsystem}, the cumulant-generating function is given by
 \begin{eqnarray}
 K_{\boldsymbol \chi} &=& e^{-i\overline {\boldsymbol n }\cdot \boldsymbol \chi  -  \boldsymbol \chi  \boldsymbol \Sigma^{2}  \boldsymbol \chi  }.
 \label{eq:initialCGF}
 \end{eqnarray}
 For brevity, we have represented the counting fields using a vector notation $ {\boldsymbol \chi} =( \overline \chi_1,\dots, \overline \chi_{D} )$. 

The  cumulants of the photonic probability distribution are defined as the derivatives of the  cumulant-generating function with respect to the counting fields
\begin{eqnarray}
\label{eq:def:cumulants}
\kappa_{l_1\dots l_D} &\equiv& i^{l_1+\dots l_D}  \frac{d^{l_1}}{d\chi_1 ^{l_1}}  \dots  \frac{d^{l_D}}{d\chi_D ^{l_D}}   K_{\boldsymbol \chi= \boldsymbol 0}  ,
\end{eqnarray}
where the $l_k$ are positive integer-valued numbers.

The cumulants allow for a geometric interpretation of the characteristic features of the probability distribution, and have convenient mathematical properties. For example, the first and second cumulants, which explicitly read
\begin{eqnarray}
\overline  n_k   &=& \kappa_{0\dots l_k =1 \dots 0}  \nonumber  \\
 &=&  \sum_{\boldsymbol n } n_k  p_{\boldsymbol n }, \nonumber  \\
\sigma_{k_1k_2}^2  &=& \kappa_{0\dots l_{k_1} =1 \dots l_{k_2}=1 \dots 0} \nonumber \\
&=&  \sum_{\boldsymbol n } \left[n_{k_1}  - \overline  n_{k_1} \right]  \left[n_{k_2}  - \overline  n_{k_2}(t) \right]p_{\boldsymbol n }  (t).
\label{eq:meanAndVariancePRFT}
\end{eqnarray}
are the means and the covariances of the probability distribution. 
For a notational reason, we combine the  $\sigma_{k_1k_2}^2 (t)$ into the covariance matrix
$
	 \left( \boldsymbol \Sigma^2 \right)_{k_1,k_2 } = \sigma_{k_1k_2}^2,
$
whose diagonal elements represents the variances of the probability distribution.

\subsection{Photon-resolved Floquet theory}

\label{eq:photonResovledFloquetTheory}

 Floquet theory is a theoretical method to analyze periodically-driven quantum systems, where the Hamiltonian
\begin{equation}
 \hat H(t) = \hat H(t+\tau)
\end{equation}
is time periodic with period $\tau$~\cite{Shirley1965}. Floquet theory finds important applications in the control of non-equilibrium quantum-phase transition~\cite{Bastidas2012,Naji2022,Esin2020,Luan2022}, in the classification  of topological phases of matter~\cite{Lindner2011,Katan2013,Benito2014,Roy2017}, and in the analysis of quantum-time crystals~\cite{Else2016,Khemani2016,Yao2017}, to mention but a few. Floquet theory can be regarded as a semiclassical method of light-matter interaction, which treats the matter as quantum and the coherent light field driving the dynamics as classical. Considering a mono-chromatic driving field with frequency $\omega=2\pi/\tau$ and defining the energy-flow operator as
\begin{equation}
	\hat {\mathcal J}(t) = \frac{d}{dt}  \hat H(t) ,
\end{equation}
we can define the operator counting the number of photons which the matter systems has interchanged with the coherent driving field of frequency $\omega $ as
\begin{equation}
 \Delta \hat n = \frac{1}{\omega} \int_{0}^{t} 	\hat{\mathcal J}(t^\prime)dt^\prime
\end{equation}
during the interaction time $t$. Using this operator, one can calculate the cumulants of the photonic probability distribution. The generalization to polychromatic driving is straightforward~\cite{Crowley2019,Long2021}. However, the evaluation of higher-order cumulants is numerically cumbersome, as it requires  high-dimensional integration  of multi-time correlation functions for $\hat{\mathcal J}(t)$.

To circumvent this obstacle, the PRFT has been developed recently~\cite{Engelhardt2024c}.  Like the Floquet theory, the PRFT  relies only on the semiclassical time evolution of the matter system given by the Hamiltonian in Eq.~\eqref{eq:hamiltonian:semiclassical}, but is still capable to predict the measurement statistics of the coherent photonic field by introducing counting-fields into the semiclassical Hamiltonian. This is possible as we consider  coherent quantum states of light, which feature relatively small fluctuations around  the mean-photon numbers and the mean phases. The aim of the PRFT is to express the  photonic measurement statistics  in terms of the cumulant-generating function in Eq.~\eqref{eq:def:generatingFcts}. For the interaction with a single-quantum emitter in Eq.~\eqref{eq:hamiltonian:semiclassical}, the PRFT shows that the cumulant-generating function in Eq.~\eqref{eq:def:generatingFcts} can be expressed as
\begin{equation}
	K_{\boldsymbol \chi} (t)   = \mathcal K_{\overline{\boldsymbol n}, \boldsymbol\chi} (t) K_{\boldsymbol \chi}(0),
\end{equation}
where $K_{\boldsymbol \chi}(0)$ denotes the initial cumulant-generating function given in Eq.~\eqref{eq:initialCGF}. The information about the light-matter interaction is contained in the dynamical-moment-generating function 
\begin{eqnarray}
\mathcal K_{\overline{\boldsymbol n}, \boldsymbol\chi} (t)  = \log \left<     \hat   {\mathcal U }  ^\dagger  _{\overline { \boldsymbol  n},\overline{\boldsymbol  \varphi} -\frac{\boldsymbol \chi}{2} }   (t)  \hat   {\mathcal U }   _{\overline{\boldsymbol  n},  \overline{\boldsymbol  \varphi}     +\frac{\boldsymbol \chi}{2} } (t)   \right>_{0},
\label{eq:def:auxiliaryGenerator}
\end{eqnarray}
 where the semiclassical time-evolution operator is given in Eq.~\eqref{eq:timeEvolutionOperator:semiclassical}. Thus, the PRFT introduces for each coherent photonic driving field $k$, a corresponding counting field $\chi_k$, which allows to obtain the photon statistics of the light-matter interaction dependent on the driving frequency $\omega_k$ or any other photonic quantum number encoded by $k$.
 
 Importantly,   the PRFT becomes exact in Sambe space in the semiclassical limit $\sigma\rightarrow \infty$~\cite{Engelhardt2024c}.  The  Sambe space is equivalent to replacing $\hat a_k^\dagger  \hat a_k \rightarrow \sum_{n_k} n_k \ketbra{n_k}{n_k}$ and $ \hat a_k \rightarrow \sum_{n_k} \sqrt{\overline n_k} \ketbra{n_k}{n_k+1}$~\cite{Sambe1973}.  Intriguingly, the PRFT predicts a macroscopic light-matter entanglement effect, in which both subsystems become entangled in the Floquet basis. 

\begin{figure*}
	\includegraphics[width=\linewidth]{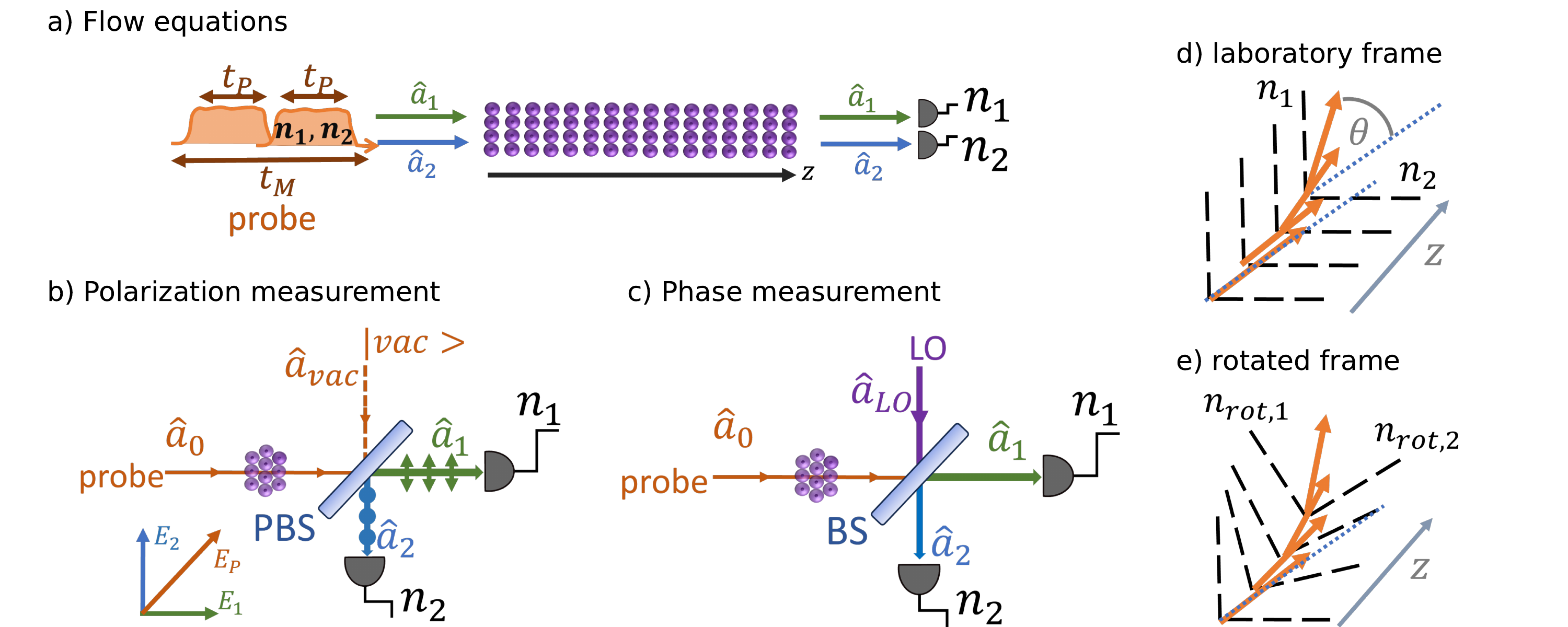}
	\caption{(a) Sketch of the general setup: A probe field, which can be quantized in terms of the operators $\hat a_k$ for $k=1,2$ propagates through a cloud of atoms (or any other kind of quantum emitters). This generic setup can, e.g.,  represent  a polarization direction measurement as sketched in (b), or a phase measurement as sketched in (c).	(d) Illustration of the fixed laboratory frame used to measure linearly polarized light. The polarization direction rotates by an angle of $\overline \theta$. (c) Rotated measurement basis,  in which the polarization direction always encloses an angel of $\pi/4$ with both measurement directions.
	}
	\label{figOverview}
\end{figure*}

 To apply the PRFT to spectroscopy, in which the light-field interacts with a macroscopic number of quantum emitters, we derive a flow equation that describes the cumulant-generating function   as a function of the propagation distance $z$ of the pulses through the quantum emitter ensemble. This basic concept is sketched in Fig.~\ref{figOverview}(a). As shown in Appendix~\ref{app:derivation:CumulantFlowEquation}, this flow equation for the cumulant-generating function is given by
\begin{eqnarray}
\frac{d}{dz}K_{\overline{\boldsymbol n},\boldsymbol\chi  } (z) = \rho_A \mathcal A   K_{\text{aux},\overline{\boldsymbol n} , \boldsymbol\chi}(z) ,
\label{eq:cumulantFlowEquation}
\end{eqnarray}
where $\rho_A$ denotes the density of the quantum emitters and $ \mathcal A$ is the (effective) cross section of the laser. The initial condition is given by the Gaussian cumulant-generating function in Eq.~\eqref{eq:initialCGF}. The function $ K_{\text{aux}, \overline{\boldsymbol n},\boldsymbol \chi } $ will be denoted as the auxiliary-generating function and is defined as
\begin{multline}
	K_{\text{aux}, \overline{\boldsymbol n},\boldsymbol \chi } (z)  =  \sum_{r=0}^{\infty}  \sum_{\iota_1+ \dots\iota_{D } =r }     \left\lbrace \frac{\partial_{\overline n_1}^{\iota_1 }\cdots  \partial_{\overline n_\textD }^{\iota_D }  }{\iota_1 !\cdots \iota_\textD! }  \mathcal K_{\overline{\boldsymbol n},\boldsymbol\chi}  (z) 
	\right\rbrace     \\
	\times \prod_{k=1}^{D} \left[ \partial_{-i\chi_k} K_{\overline{ \boldsymbol n},\boldsymbol\chi}(z) -\overline n_k\right] ^{\iota_k}.
	\label{eq:dynamicalCumulantGeneratingFunction}
\end{multline}
In the following, we  explain the elements in the auxiliary-generating function in more detail:

The central building block appearing in the auxiliary-generating function is the dynamical cumulant-generating function  in Eq.~\eqref{eq:def:auxiliaryGenerator} evaluated at  the measurement time $t = t_{\text{M} }$.  In Eq.~\eqref{eq:dynamicalCumulantGeneratingFunction}, we express the dynamical cumulant-generating function as a function of the propagation distance $z$ to highlight a possible position dependence.

The auxiliary-generating function in Eq.~\eqref{eq:dynamicalCumulantGeneratingFunction} contains a summation over $r$, which appears due to a Taylor expansion of the dynamical-cumulant generating function in Eq.~\eqref{eq:def:auxiliaryGenerator}. The subscript of the second summation ($\iota_1+ \dots+\iota_{D }=r$) means that we have to sum over all combinations $\iota_1, \dots,\iota_{D }$ whose index sum equals  $r$. In this context, the symbols $\partial{\overline n_k }^{\iota_k }  $ denote the $\iota_k $-th derivatives of the dynamical cumulant-generation function with respect to the mean photon number $\overline n_k$.

 The $r=0$ term just equals the auxiliary generator in Eq.~\eqref{eq:def:auxiliaryGenerator}. This term describes the photon statistics correctly if one would replace the photonic operators in the Hamiltonian in Eq.~\eqref{eq:hamiltonian:quantum} by their counterparts in Sambe space. The higher-order terms $r>0$ in Eq.~\eqref{eq:dynamicalCumulantGeneratingFunction} appear because of the photon-number dependence of the photonic creation and annihilation operators $\left< n_k- 1\right| \hat a_k \left| n_k\right>  =\left< n_k\right| \hat a_k^\dagger  \left| n_k-1 \right> =\sqrt{n_k} $. As these terms are related to the deformation of the photonic state in phase space, we will refer to them as phase-space terms in the following. Its interpretation will be discussed in more detail in Sec.~\ref{eq:lowOrderCumulants}.

Moreover, the symbol $\partial_{-i\chi_k}$ denotes the derivative with respect to the counting field $\chi_k$, which act on the cumulant-generating function at position $z$. Because of this term, the flow equation in Eq.~\eqref{eq:cumulantFlowEquation} becomes a non-linear partial differential equation, which cannot be solved, neither numerically nor analytically. The usage of Eq.~\eqref{eq:cumulantFlowEquation} will be discussed in the next section.

\subsection{Mean and variance}

\label{eq:lowOrderCumulants}

In general, the flow equation for the cumulant-generating function in Eq.~\eqref{eq:cumulantFlowEquation} cannot be integrated. However, it can be considered as an \textit{equation-generating equation}, as we will demonstrate here. More precisely, by deriving Eq.~\eqref{eq:cumulantFlowEquation} with respect to the counting fields, we can generate flow equations for  low-order cumulants. We will demonstrate this for the first two cumulants, and comment on the construction of higher-order flow equations.

 (i) Deriving  Eq.~\eqref{eq:cumulantFlowEquation} with respect to each counting field $\chi_k$, we obtain the flow equation for the mean photon numbers 
\begin{eqnarray}
\frac{d}{dz}\boldsymbol {\overline n}  
&=&     \boldsymbol  I_{\boldsymbol {\overline n}   } ,
\label{eq:tra4:momGenFunction} 
\end{eqnarray}
where the coefficient on the right-hand side is explicitly given by
\begin{equation}
\left[ \boldsymbol  I_{\boldsymbol {\overline n}    } \right]_k  =   \rho_A  \mathcal  A  \partial_{\chi_{k} }   \mathcal K_{\boldsymbol {\overline n}, \boldsymbol  \chi =0 } (z) .
\label{eq:fluxEvaluation}
\end{equation}
Importantly, this flow equation is a nonlinear differential equation, as $ \boldsymbol  I_{\boldsymbol {\overline n}} $ depends on the mean photon numbers in a nonlinear fashion. Thus, we have to continuosly update $\overline  { \boldsymbol n} =\overline  { \boldsymbol n}(z)  $ when (numerically) solving Eq.~\eqref{eq:tra4:momGenFunction}.

 (ii) Deriving  Eq.~\eqref{eq:cumulantFlowEquation} two times with respect to the counting fields $\chi_k$, we obtain the flow equation for  the covariance matrix
\begin{equation}
\frac{d}{dz}  \boldsymbol \Sigma^2 
=\boldsymbol  D_{\boldsymbol {\overline n} }  +   \boldsymbol   C_{\boldsymbol {\overline n} }  \boldsymbol   \Sigma^2   +\boldsymbol  \Sigma^2\boldsymbol   C_{\boldsymbol {\overline n}}  ,
\label{eq:varianceFlow} 
\end{equation}
where
\begin{eqnarray}
\left[	 \boldsymbol  D_{\boldsymbol {\overline n} }   \right]_{k_1,k_2}   &=&   \rho_A  \mathcal  A \;  \partial_{ \chi_{k_1} }\partial_{\chi_{k_2} }      \mathcal K_{\boldsymbol {\overline n},\boldsymbol \chi =0  } (z) \nonumber ,\\
\left[	 \boldsymbol  C_{\boldsymbol {\overline n} }  \right]_{k_1,k_2}   &=&  \rho_A  \mathcal  A\;   \partial_{\overline n_{k_1} }\partial_{\chi_{k_2} }   \mathcal K_{\boldsymbol {\overline n}, \boldsymbol  \chi =0 } (z).
\label{eq:diffusion-phaseSpaceMatrices}
\end{eqnarray}
This equation has to be solved with $ \boldsymbol \Sigma^2 $  of the photonic state in Eq.~\eqref{eq:initialStateMatterSubsystem} as initial condition. Noteworthy, the coefficient matrices $\boldsymbol  D_{\boldsymbol {\overline n} } $ and $\boldsymbol  C_{\boldsymbol {\overline n} } $ depend on the mean photon numbers. 

The matrix $\boldsymbol  D_{\boldsymbol {\overline n} } $ is obtained by a twofold derivation of the dynamical cumulant-generating function with respect to the  counting fields. Using similar methods as in Ref.~\cite{Engelhardt2024b}, it is possible to show that $ \left[	 \boldsymbol  D_{\boldsymbol {\overline n} }   \right]_{k_1,k_2} \geq 0 $, i.e., its contribution enhances the variance. For this reason, we denote this term as the diffusion matrix in the following.

The other coefficient matrix $\boldsymbol  C_{\boldsymbol {\overline n} } $ incorporates the photon number dependence of the creation and annihilation operators into the dynamical equation, and describes the resulting deformation of the photonic  probability distribution. We will denote it as the phase-space matrix in the following. 

(iii) In a similar fashion, one can derive the flow equations for higher-order cumulants. Inspection of Eq.~\eqref{eq:cumulantFlowEquation} shows that the flow equation for a cumulant of order $l = l_1 + \dots l_D$ involves the Taylor expansion up to order $r =l-1$ of the dynamical cumulant-generating function in Eq.~\eqref{eq:dynamicalCumulantGeneratingFunction}. Consequently, a cumulant of order $l$ cannot depend on a higher-order cumulant $l^\prime>l$. In the derivation of the Eq.~\eqref{eq:cumulantFlowEquation} it has been assumed that the laser pulses interact with an ensemble of non-interacting quantum emitters. For this reason, the probability distribution will approach (stay) a Gaussian function for long enough propagation distances $z$. Consequently, the knowledge of the first two cumulant orders is enough to describe realistic measurement statistics.

\section{Spectroscopic quantum sensing}

\label{sec:spectroscopicObservables}

In Sec.~\ref{sec:framework}, we have established a theoretical framework which is capable of predicting the photon exchange between pulses of light and an ensemble of quantum emitters without  further specifying the physical meaning of the photonic operators. In this section, we now link the photonic operators in the Hamiltonian in Eq.~\eqref{eq:hamiltonian:quantum} to relevant observables in spectroscopic quantum sensing, namely the intensity of a laser, the polarization direction of linearly polarized light and the phase shift of the laser-light imposed by the light-matter interaction. In doing so, we provide practical guidelines on how to best integrate the flow equations in Eqs.~\eqref{eq:tra4:momGenFunction}  and \eqref{eq:varianceFlow}, and explain how this can be utilized to evaluate the Fisher information in Eq.~\eqref{eq:def:FisherInformation}.

\subsection{Intensity}
\label{sec:intensity}

A vital observable in spectroscopic experiments is the optical intensity of the laser field, which is  defined as the radiation energy traversing a unit area $\mathcal A$ within a unit  time. In contrast, the PRFT counts the number of photons contained in a laser pulse. The laser intensity of a pulse with frequency $\omega$, duration $t_{\text{P}}$ and cross section $\mathcal A$ can be estimated by
\begin{equation}
	I_{\text{L}} = \frac{\hbar \omega}{t_{\text{P}} \mathcal A }n.
\end{equation}
where $n$ is the measured photon number. 
Thus, from the statistics of the photon number $n$ (as predicted by the PRFT), we can obtain  the  intensity.

To improve the signal to noise ratio of the intensity measurement, one can invoke  longer measurement times $t_{\text{M}}$ by recording the measurement outcomes for the continuous train of  pulses. Quantizing the pulses with the photonic operators $\hat a_k$, we denote the measured photon number  by $n_k$.  Crucially, the $n_k$ can not be regarded as independent random variables. Rather, the photon numbers  are highly correlated because of memory effects of the atoms, which the pulses interact with. For example, the correlation time of the atomic dynamics can be determined by relaxation processes with rate $\gamma$ such that $ t_{\text{P}} \ll 1/\gamma$. 

To estimate the mean intensity of this pulse train, we employ the cumulative photon number $ \sum_k n_k$ as a random variable. Comparing with the definition of the cumulant-generating function in  Eq.~\eqref{eq:def:generatingFcts}, we see that we can simply obtain the statistics of the cumulative photon number by setting all counting-fields equally $\chi_k =\chi$. We then evaluate the flow equations in Sec.~\ref{eq:photonResovledFloquetTheory} for $\chi$  and $t=t_{\text{M}}$ in Eq.~\eqref{eq:def:auxiliaryGenerator}.  The intensity of the radiation field can be then estimated as
\begin{equation}
I_{\text{L}} = \frac{\hbar \omega}{t_{\text{M}} \mathcal A } \sum_{k=1}^{D} n_k,
\end{equation}
 where the statistics of the cumulative photon number $\sum_{k=1}^{D} n_k$ can be obtained using the dynamical cumulant-generator function in Eq.~\eqref{eq:def:auxiliaryGenerator} with integration time $t_{\text{M}}$. Noteworthy, in principle it is possible to construct more precise estimators for the intensity based on the joint photonic measurement probabilities $p_{n_1 \dots n_D}$ using, e.g., a Markov-chain approach. However, this would exceed the scope of this work and will be considered elsewhere.

\subsection{Polarization direction}
\label{sec:polarization}

In this section, we assume that the laser pulse is initially polarized  in the direction
\begin{equation}
		\boldsymbol E_{\text{in} } (\boldsymbol r,t) \propto  \boldsymbol  e_{\text{x} } + \boldsymbol e_{\text{y} } ,
		\label{eq:initialDirectionElectricField}
\end{equation}
where $\boldsymbol  e_{\text{x} }$ and $\boldsymbol  e_{\text{y} }$ denote the unit vectors in x and y directions as shown in the inset of Fig.~\ref{figOverview}(b). Accordingly, we introduce the operators $\hat a_{1 }$  and $\hat a_{2 } $ which quantize the electromagnetic field of  the laser pulse in x and y directions, respectively. Thus, initially, both occupation number operators $\hat n_{\alpha }= \hat a_{\alpha }^\dagger  \hat a_{\alpha } $ with $\alpha=1,2$ have the same mean value,
\begin{equation}
 \left< \hat n_{1 }\right>_{\text{in}} = \left< \hat n_{2 }\right>_{\text{in}}   = \frac{\overline n_{+ }^{(\text{in})}} {2},
 \label{eq:initialOccpuation}
\end{equation}
where  $\overline n_{+}^{(\text{in})} $ is the initial total number of photons in the pulse, which is a function of its shape, duration and frequency. The choice of the basis direction in Eq.~\eqref{eq:initialDirectionElectricField} is in principle arbitrary as long as the photonic field in both directions is macroscopically occupied. Yet, the initial symmetric occupation in both measurement direction is advantageous for the evaluation of the flow equations.

When the light pulse is propagating through an ensemble of quantum emitters, the polarization direction can be changed, e.g., because of the Faraday effect. The change in the polarization direction manifests itself in an imbalance in the photonic occupations of $\hat n_{1} $ and $\hat n_{2} $. The photon numbers can be measured using a polarization beam splitter and two photon detectors as sketched in Fig.~\ref{figOverview}(b), from which we can infer the mean value of the polarization rotation $\overline \theta$  via
\begin{equation}
	\sin \overline \theta = \frac{\sqrt{\overline n_{2 }  }  -\sqrt{\overline n_{1 }  }   }{\sqrt{2\overline n_{+ }}},
	\label{eq:rotationAngel}
\end{equation}
where $\overline n_\alpha$ is the mean photon number  at detector $\alpha= 1,2$, and $\overline n_+ = \overline n_{1} + \overline n_{2} $.

\subsection{Phase}

\label{sec:phase}

Even though the PRFT is a framework which counts the number of photons, it is also capable to predict the statistics of phase measurements. This  assertion becomes clear when recalling that  it is   only possible to  measure intensities in optical experiments. A phase measurement is typically carried out by superposing the probe field  (whose phase is to be measured) with a local oscillator, and then inferring the phase information from the intensity change of the resulting light field. 

The corresponding experimental setup is depicted in Fig.~\ref{figOverview}(c), where the probe field is quantized by $\hat a_0$ and the local oscillator is quantized by $\hat a_{\text{LO}}$. Using a beamsplitter, we create the two superposition output fields $\hat a_1,\hat a_2$, which are related to  the input fields $\hat a_0,\hat a_{\text{LO} }$ via
\begin{eqnarray}
\hat a_1 &=& \frac{1}{\sqrt{2}} \left( \hat a_{\text{LO}}  + i\hat a_0\right) ,\nonumber  \\
\hat a_2 &=& \frac{1}{\sqrt{2}} \left( \hat a_{\text{LO}}  - i\hat a_0\right).
\label{eq:probe-LO-superpositionModes}
\end{eqnarray}
For illustration, let us assume that the probe pulse is in a coherent state with complex  amplitude $\sqrt{\overline n_0}e^{i\overline \varphi_0}$, where $\overline n_0$ is the mean photon number and $\overline \varphi_0$ is the phase to be measured. Likewise, the local oscillator is in a coherent state with complex amplitude $\sqrt{\overline n_{\text{LO}}}e^{i\overline \varphi_{\text{LO}}} $.  Denoting the mean occupations of the operators by $\overline n_\alpha = \left< \hat a_\alpha^\dagger \hat a_\alpha  \right>$ for $\alpha =1,2$, we find  relations for the sum and difference of the mean photon numbers 
\begin{eqnarray}
\overline n_{1} + \overline n_{2}  &=& \overline n_0 + \overline n_{\text{LO}} \nonumber ,\\
\overline n_{1} -  \overline n_{2} &=&  2 \sqrt{\overline n_0 \overline n_{\text{LO}}} \sin \left( \overline \varphi_0 -\overline \varphi_{\text{LO}}\right),
\end{eqnarray}
from which we can infer the  phase of the probe field via
\begin{eqnarray}
\sin \left( \overline \varphi_0 -\overline \varphi_{\text{LO}}\right)  &=& \frac{ \overline n_{1} - \overline n_2 }{2\sqrt{\overline n_0 \overline n_{\text{LO}} }}.
\label{eq:phaseEstimation}
\end{eqnarray}
Thus, we obtain the phase information via a photon number  measurement, which is closely related to the intensity.

Based on these considerations, we can now establish a connection to the PRFT. To this end, we assume that originally the quantum emitters interact with the probe field via the operator $\hat a_0$ in the Hamiltonian. Now we invert Eq.~\eqref{eq:probe-LO-superpositionModes}, i.e.,
\begin{eqnarray}
\hat a_0 &=& \frac{1}{\sqrt{2}i} \left( \hat a_1  - \hat a_2\right)
\label{eq:outputModeReplacement}
\end{eqnarray}
and insert this into the Hamiltonian. As a consequence, the quantum emitters interact now with the two modes $\hat a_1 ,\hat a_2$, for which we can use the formalism in Sec.~\ref{sec:framework} to calculate the projected measurement statistics. The replacement in Eq.~\eqref{eq:outputModeReplacement} is a unitary transformation into the basis of measurement. Thus, we express the Hamiltonian in exactly the basis, in which we will finally measure the photon statistics.

\subsection{Relation between polarization and phase measurements}

\label{sec:rotation-phase-relation}

It is worth noting that the polarization direction measurement in Sec.~\ref{sec:polarization} and the phase-shift measurement  in Sec.~\ref{sec:phase} are intimately related. The mutual  connection becomes clear when recalling that linearly polarized light can be considered as a superposition of two circularly polarized light modes, where the modes have an orthogonal polarization. Using simple geometric considerations, one finds that a phase shift of one circularly polarized mode $\Delta \overline \varphi$ leads to a rotation of the resulting linearly polarized light $\Delta \overline \theta$, and both angles are related by
\begin{equation}
	\Delta \overline \varphi = 2 \Delta \overline \theta.
	\label{eq:rotation-phase-relation}
\end{equation}
This is the origin of the celebrated Faraday effect, in which each circular polarization mode of the light field is subject to a different susceptibility for matter in an external magnetic field.

We can discover this effect also when comparing Eqs.~\eqref{eq:rotationAngel} and \eqref{eq:phaseEstimation}. To this end, we assume  that both modes are initially equally occupied according to Eq.~\eqref{eq:initialOccpuation}. Deriving Eq.~\eqref{eq:rotationAngel} with respect to the position $z$, we find
\begin{eqnarray}
\frac{d}{dz} \overline  \theta 
&=& \frac{\frac{d}{dz} {\overline n}_{1} -\frac{d}{dz} {\overline n}_{2}  } {2\overline n_{+  }} ,
\label{eq:infinitesimalRotation}
\end{eqnarray}
which is the differential change of the rotation axis induced by the light-matter interaction.
Likewise, for the phase shift in Eq.~\eqref{eq:phaseEstimation},  we assume that originally  $\overline n = \overline n_{\text{LO} } $ and $ \overline \varphi =\overline \varphi_{\text{LO}}$. Deriving then  Eq.~\eqref{eq:phaseEstimation} with respect to $z$,  we find
\begin{eqnarray}
\frac{d}{dz} \overline \varphi   &=& \frac{  \frac{d}{dz} \overline n_{1} - \frac{d}{dz} \overline n_2 }{\overline n_{+ }} ,
\label{eq:infinitesimalPhaseShift}
\end{eqnarray}
which describes  the differential change of the measured phase due to the light-matter interaction.

Comparing Eq.~\eqref{eq:infinitesimalRotation} with Eq.~\eqref{eq:infinitesimalPhaseShift}, we readily find
\begin{eqnarray}
\frac{d\overline \varphi }{dz}  =	2\frac{d \overline \theta}{dz}  ,
\end{eqnarray}
which is simply the differential form of  Eq.~\eqref{eq:rotation-phase-relation}. This relation implies  that we can formally map a phase-shift measurement  to a polarization-direction measurement. The later interpretation might be more appealing, as the measured photonic operators already appear in the original Hamiltonian. This more intuitive approach  helps us to represent the phase measurement  in a form, which is more convenient for the numerical and analytical integration, as will be explained in the next section.

\subsection{Rotating measurement frame}

\label{sec:rotatingMeasurementFrame}

For the following explanations, we consider a polarization measurement as introduced in Sec.~\ref{sec:polarization}. The flow equations in Eqs.~\eqref{eq:tra4:momGenFunction}  and \eqref{eq:varianceFlow}  predict the measured photonic probabilities in a basis which is fixed in the laboratory frame. While this setup is intuitive, there are two advantages to consider a measurement basis, which rotates along with the polarization direction of the light pulse:
\begin{enumerate}
	\item[(i)]  The PRFT is only valid if the photonic driving fields are highly occupied. For particular rotation angles, the occupation of one of the modes might vanish, which makes the integration of the flow equations invalid within the PRFT.
	
	\item[(ii)] The matrices $C_{\boldsymbol {\overline n}}$ in Eq.~\eqref{eq:varianceFlow} often dominate the time evolution of the covariance matrix. In a rotating measurement frame, these matrices will be rendered small, which enables a  stable numerical integration and a theoretical analysis.
	
\end{enumerate}
The two different measurement basis setups are sketched in Fig.~\ref{figOverview}(d)  and Fig.~\ref{figOverview}(e), showing the fixed and rotated measurement directions for various positions $z$, respectively. In the rotated basis, the instantaneous direction of the electric field polarization is depicted by the red arrow, and encloses  a $\pi/4$ angle with both rotated basis directions for all positions $z$.

To parameterize the rotation, we utilize the position dependent mean polarization angel $\overline \theta =\overline \theta(z)$ defined in Eq.~\eqref{eq:rotationAngel}. In terms of $\overline \theta$, the rotated photonic operators are defined by
\begin{eqnarray}
\hat a_{\text{rot},1} &=&  \cos \overline \theta  \hat a_{1}  - \sin \overline \theta \hat a_{2}, \nonumber \\
\hat a_{\text{rot}, 2} &=&  \sin \overline \theta \hat a_{1}  +  \cos \overline \theta \hat a_{2}  .
\end{eqnarray}
The angel $\overline\theta$ is chosen such that for all positions $z$ the mean occupations  of the photonic modes $\hat n_{\text{rot},\alpha} = \hat a_{\text{rot},\alpha} ^\dagger \hat a_{\text{rot},\alpha}  $ with $\alpha=1,2$ are equal, i.e.,
\begin{eqnarray}
\left< \hat n_{\text{rot},1}   \right>  =    \left< \hat n_{\text{rot},2}   \right>  \equiv \frac{ \overline n_{+}(z)}{2},
\label{eq:rotationCondition}
\end{eqnarray}
where $\overline n_{+ }(z)$ denotes the total number of photons in both modes.

As shown in Appendix~\ref{app:rotatedMeasurementFrame}, the rotation angle and the total number of photons can be calculated using the following differential relations
\begin{eqnarray}
\frac{d\overline \theta }{dz}  
&=& \frac{\partial_z {\overline n}_{\text{rot}, 2} -\partial_z {\overline n}_{\text{rot}, 1}  } {2\overline n_{+ }} , \nonumber \\
\frac{d{\overline n}_{+ } }{dz}  &=&\partial_z  {\overline n}_{\text{rot}, 2} + \partial_z {\overline  n}_{\text{rot}, 1}.
\label{eq:rotationAngel-photonNumber}
\end{eqnarray}
Thereby,  $\partial_z {\overline n}_{\text{rot}, \alpha }  $ refers to the derivatives of the mean photon number ${\overline n}_{\text{rot}, \alpha } $  in the fixed measurement basis and can be evaluated via
\begin{equation}
\partial_z {\overline n}_{\text{rot}, \alpha }  =\left[   \boldsymbol  I_{\boldsymbol {\overline n_{\text{rot}}}   }\right]_{\alpha},
\end{equation}
where $\boldsymbol  I_{\boldsymbol {\overline n}_{\text{rot}}} $ is the photon flux vector in Eq.~\eqref{eq:fluxEvaluation} for the semiclassical Hamiltonian, in which the photonic mode occupations satisfy the rotation condition in Eq.~\eqref{eq:rotationCondition}. 

The differential equation for the correlation matrix, as it will be measured in the rotated measurement basis,  is given by
\begin{eqnarray}
\frac{d}{dz}  \boldsymbol \Sigma_{\text{rot}}^2  &=& \boldsymbol  D_{\boldsymbol {\overline n}_{\text{rot}}}   +  \tilde {\boldsymbol   C}_{\boldsymbol {\overline n}_{\text{rot}}}    \boldsymbol   \Sigma^2_{\text{rot}}   +\boldsymbol  \Sigma^2_{\text{rot} } \tilde { \boldsymbol  C}_{\boldsymbol {\overline n}_{\text{rot}}} ,  \nonumber \\
 \tilde {\boldsymbol   C}_{\boldsymbol {\overline n}_{\text{rot}}} &=&  {\boldsymbol   C}_{\boldsymbol {\overline n}_{\text{rot}}} -  \frac{d\overline \theta }{dz}  \hat \Xi \boldsymbol  ,
\label{eq:correlationMatrixFlow-rotated}
\end{eqnarray}
where we have introduced $\hat \Xi  = i\hat \sigma_{\text{y}} + \hat \sigma_{\text{z}}  $ ($\hat \sigma_{\alpha}$ with $\alpha= \text{x},\text{y},\text{z}$ denoting the common Pauli matrices). Thereby, $\boldsymbol  D_{\boldsymbol {\overline n}_{\text{rot}}} $ and $\boldsymbol  C_{\boldsymbol {\overline n}_{\text{rot}}} $ refer to the diffusion and phase-space matrices as defined in Eq.~\eqref{eq:diffusion-phaseSpaceMatrices}, but evaluated under the rotation condition in Eq.~\eqref{eq:rotationCondition}. Typically, the entries of $\tilde {\boldsymbol   C}_{\boldsymbol {\overline n}_{\text{rot}}} $ are significantly smaller than the ones of $ {\boldsymbol   C}_{\boldsymbol {\overline n}_{\text{rot}}} $, enabling a more efficient numerical and analytical treatment.

Conversely, in terms of the observables $\overline n_{+}$ and $\overline \theta$, it is also straightforward to retrieve the means and covariances for the measurement in the original measurement basis in the lab frame by evaluating
\begin{eqnarray}
	\left(
	\begin{array}{c}
	\overline n_{1} \\
	\overline n_{2}
	\end{array}
	\right) 
	=
	\overline n_{+}	\left(
	\begin{array}{c}
	\cos^2(\frac{\pi}{4}+ \overline\theta) \\
	\sin^2(\frac{\pi}{4}+ \overline \theta)
	\end{array}
	\right) ,
	\label{eq:meanPhotonNumbersAFOangle}
\end{eqnarray}
and
\begin{equation}
\boldsymbol \Sigma_{\alpha,\beta}^2   =    \frac{2 \sqrt{\overline n_{\alpha}}  \sqrt{\overline n_{\beta}} }{\overline n_{+  }}\left( e^{i\overline \theta  \hat \sigma_{\text{y}}}\boldsymbol \Sigma_{\text{rot}}^2e^{-i\overline \theta \hat \sigma_{\text{y}}} \right)_{\alpha,\beta},
\label{eq:inverseTransformation:rotFrame-labFrame}
\end{equation}
for $\alpha,\beta =1,2$. The detailed  derivation is given in Appendix~\ref{app:rotatedMeasurementFrame}.

Using the relation between rotation and phase measurements in Sec.~\ref{sec:rotation-phase-relation}, one can also transform a phase measurement into an adjusted measurement setup, which simplifies the flow equation. Physically, instead of rotating the measurement basis, we adjust the amplitude and  phase of the local oscillator such that Eq.~\eqref{eq:rotationCondition} is fulfilled. This implies that $\overline n =\overline n_{\text{LO}}$ and $\overline \varphi =\overline \varphi_{\text{LO}}$. Using the correspondence in Eq.~\eqref{eq:rotation-phase-relation}, we can interpret the phase measurement as a \textit{virtual} polarization measurement. After calculating the virtual polarization change, we can then use Eq.~\eqref{eq:rotation-phase-relation} to obtain the mean phase value. Moreover, the measurement statistics of the covariance matrix is the same in the phase and the polarization estimations.

{\subsection{Fisher information}

\label{sec:fisherInformation}

 According to the Cramer-Rao bound in  Eq.~\eqref{eq:cramerRaoBound}, the measurement precision is bounded by the inverse Fisher information.  Under the assumption of a Gaussian probability distribution, the leading-order term of the Fisher information is given by
\begin{eqnarray}
\mathcal I_{X} &=&  \left( \partial_{X} \boldsymbol {\overline n}_{X}\right) \boldsymbol \Sigma_{X}^{-2} \left( \partial_{X} \boldsymbol {\overline n}_{X}\right)^T ,
\label{eq:fisherInfoGaussianState}
\end{eqnarray}
as shown in Appendix~\ref{app:fisherInformationGaussian}. Thus, it can be expressed in terms of the mean values and covariance matrix of the probability distribution of $\boldsymbol  n$. As the photonic probability distribution  approaches a Gaussian distribution when interacting with a macroscopic number of independent quantum emitters, we will use Eq.~\eqref{eq:fisherInfoGaussianState} in the following calculations.

When working in the rotated measurement basis introduced in Sec.~\ref{sec:rotatingMeasurementFrame}, the measurement statistics and the Fisher information can be conveniently expressed in terms of 
\begin{eqnarray}
n_{+ }  = n_1 + n_2 , \nonumber \\
n_{- }  = n_1 - n_2.
\end{eqnarray}
 The total photon number $n_{+ }$ carries information about the total intensity of a laser pulse, while the photon number difference $n_{-}$ is related to the phase shift $ \overline \theta$ via Eq.~\eqref{eq:phaseEstimation}. We recall that, by definition, the mean value  $\overline n_{- } =0 $ in the rotated measurement basis, such that the information about the polarization rotation is expressed by  $\overline \theta$.
In the rotated measurement picture,  $\partial_{X} \boldsymbol {\overline n}_{X} $ can be simply evaluated via
\begin{eqnarray}
\partial_{X} \boldsymbol {\overline n}_{X}
=
\overline n_{+} \frac{d\overline \theta}{dX}	\left(
\begin{array}{c}
1 \\
-1
\end{array}
\right) 
+\frac{d\overline n_{+}  }{dX} 	\left(
\begin{array}{c}
1 \\
1
\end{array}
\right) ,
\end{eqnarray}
as can be shown using Eq.~\eqref{eq:meanPhotonNumbersAFOangle} for $\overline \theta =  0$.

\section{Tavis-Cummings  model}

\label{sec:tavisCummingsModel}

\begin{figure*}
	\includegraphics[width=\linewidth]{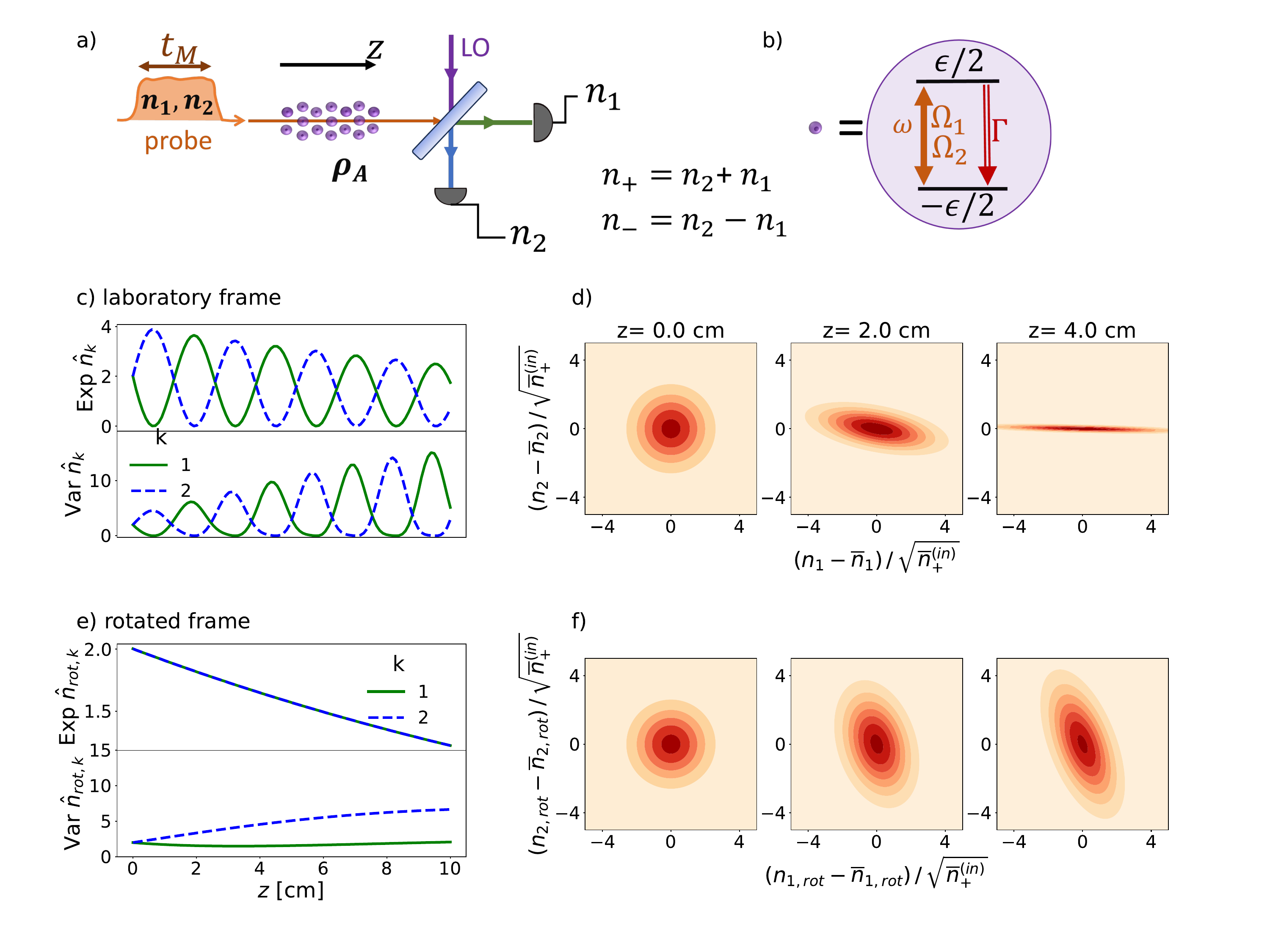}
	\caption{
		(a) Experimental setup designed for a phase measurement.  (b)  Dissipative two-level system. (c) Mean photon number and the variance as a function of propagation distance $z$ for the mode $k=1,2$ in the laboratory measurement frame. 
 	(d) Probability distribution for three selected distances. (e) and (f) depict the same as (c) and (d) but in the rotated measurement basis. The dynamics in the rotated frame is  obtained by integrating the flow equations in Eqs.~\eqref{eq:rotationAngel-photonNumber} and \eqref{eq:correlationMatrixFlow-rotated}  using the coefficients in Eqs.~\eqref{eq:phaseShiftAngle-photonNumber}, \eqref{eq:tavisCummingsModel:LeadingphotonFluctuations} and \eqref{eq:tavisCummingsModel:rotationCoefficient}.  The values in the laboratory frame can be obtained using Eqs.~\eqref{eq:meanPhotonNumbersAFOangle} and \eqref{eq:inverseTransformation:rotFrame-labFrame}. Parameters are $\gamma = 1\,\text{Mhz}$, $z_{\text{max} }=10\text{cm}$, $P = 100\,\text{mW}$, $\mathcal A = 1\,\text{cm}^2$, $\rho_A=10^{14} \text{m}^{-3}$, $t_{\text{M}} = 1\,\text{s}$ and atomic dipole moment $d= 5ea_0$. This gives rise to an overall Rabi frequency of $\Omega=60\,\text{MHz}$.  }
	\label{figFlowdynamics}
\end{figure*}

To demonstrate the theoretical concepts in Sec.~\ref{sec:framework} and Sec.~\ref{sec:spectroscopicObservables} for a concrete example, we consider here a dissipative version of the celebrated Tavis-Cummings model, whose  Hamiltonian reads
\begin{equation}
	\hat H  = \hat H_{\text{TC}} +  \hat H_{\text{Bath}}.
	\label{eq:ham:dissipativeTCmodel}
\end{equation}
Thereby, the closed Tavis-Cumming model is given by
\begin{eqnarray}
\hat H_{\text{TC}} &=&  \omega  \hat a_0^\dagger  \hat a_0 +\sum_{m} \frac{ \epsilon }{2} \hat \sigma_{m,\text{z}}    \nonumber \\
 &+&    \sum_{m}   g_{m}(t)  \left( \hat \sigma_{m,-}  \hat a_0^\dagger   +\hat \sigma_{m,+}  \hat a_0 \right) ,	
\label{eq:hamiltonian:JaynesCummings}
\end{eqnarray}
where the photonic operators $\hat a_0, \hat a_0^\dagger$ quantizes a laser pulse with frequency $\omega$, $\hat \sigma_{m,\text{z}} $ describes the two levels of atom $m$, which has level splitting $\epsilon$. The coupling operators of the quantum emitters are given by the Pauli operators $\hat \sigma_{m,-} ,\hat \sigma_{m,+} $, whose time-dependent coupling to the photonic field is parameterized by $g_{m}(t)$. We recall that the time-dependence of  $g_{m}(t)$ is determined by the  pulse shape and  quantum emitter location.

Moreover, each quantum emitter is coupled to its own bath which gives rise to dissipation. The bath and the coupling operators are described by
\begin{eqnarray}
\hat H_{\text{Bath}} &=&  \sum_{m,q}  \omega_{q}  \hat b_{m,q}^\dagger  \hat b_{m,q} 
+    \sum_{m,q}   v_{q}  \left( \hat \sigma_{m,-}  b_{m,q}^\dagger   +\text{h.c.} \right) , \nonumber \\	
\end{eqnarray}
where the bosonic bath modes $\hat b_{m,q}$ with frequency $\omega_{q}$ are coupled to the quantum emitters with strength $v_{q}$. We assume that the baths are in a thermal state with temperature $k_B T \ll \hbar \omega$.

In the rest of this section, we will use the PRFT to estimate the counting statistics for the phase-shift measurement of the probe pulse quantized by $\hat a_{0}$. A sketch of the corresponding experimental setup and the dissipative two-level systems can be found in Fig.~\ref{figFlowdynamics}(a) and  Fig.~\ref{figFlowdynamics}(b), respectively.

\subsection{Cumulant-generating function}
\label{sec:jaynesCummingsModel}

The first step in the framework is to get an expression for the dynamical cumulant-generating function defined in Eq.~\eqref{eq:def:auxiliaryGenerator}. Using Eq.~\eqref{eq:outputModeReplacement} we represent the Hamiltonian in the measurement basis, which reads
\begin{eqnarray}
\hat H_{\text{TC}} &=& \sum_{k=1,2} \omega  \hat a_k^\dagger  \hat a_k +\sum_{m} \frac{ \epsilon }{2} \hat \sigma_{m,\text{z}}    \nonumber \\
&+&    \sum_{m, k=1,2}\left[ \frac{ g_{m}(t)}{\sqrt{2}}i   \hat \sigma_{m,-}  \hat a_k^\dagger  +\text{ h.c.}  \right] .	
\label{eq:hamiltonian:TavisCummings}
\end{eqnarray}
Next, we consider the Hamiltonian in the semiclassical limit by replacing $\hat a_k \rightarrow \sqrt{\overline n_k} e^{i\overline \varphi_k}$, such that
 \begin{equation}
 	\hat {\mathcal H}(t)  = \sum_{m }\hat {\mathcal H}_{m,\boldsymbol \varphi} (t) + \hat H_{\text{Bath}} ,
 	\label{eq:tavisCummingsModel-semiclassical}
 \end{equation}
where the semiclassical light-matter interaction of two-level system $m$ is described by
\begin{equation}
\hat {\mathcal H}_{m,\boldsymbol \varphi} (t) =  \frac{ \epsilon}{2} \hat \sigma_{m,\text{z}}  +  \sum_{k=1,2} \left[ \frac{\Omega_{m,k} (t)}{2}e^{i(\omega t - \varphi_k) }\hat \sigma_{m,-} +\text{ h.c.} \right],
\label{eq:semiclassicalJaymesCummingsModel}
\end{equation}
in which the effective Rabi frequencies are given by $\Omega_{m,k} (t)  = \sqrt{2\overline n_k}i g_{m,k} (t)  $.
In the following, we will omit the index $m$, as the quantum emitters are formally decoupled in the semiclassical Hamiltonian and assumed to be identical.  Of note, the Tavis-Cummings model  describes the light-matter interaction within a rotating-wave approximation, which allows us to derive analyical expression of the measurement statistics. Yet, we emphasize that the methods developed in this work are also applicable for models beyond the rotating-wave approximation.

As each quantum emitter is coupled to a continuum of bath modes, we cannot exactly construct the semiclassical time-evolution operator in Eq.~\eqref{eq:timeEvolutionOperator:semiclassical}. Instead, we use an open quantum system approach of the PRFT to construct the dynamical moment-generating function in Eq.~\eqref{eq:def:auxiliaryGenerator}. To this end, we deploy a generalized quantum master equation, which describes the system-bath coupling in second-order perturbation theory as introduced in Ref.~\cite{Engelhardt2024b}. To obtain the dynamical moment-generating function, we first have to integrate the generalized master equation
\begin{eqnarray}
	\frac{d}{dt}\rho_{\boldsymbol \chi}  = \mathcal L_{\overline{\boldsymbol n},\boldsymbol \chi} \rho_{\boldsymbol \chi},
\end{eqnarray}
and then use the generalized density matrix of the quantum-emitter system $\rho_{\boldsymbol \chi}$ to construct the dynamical moment-generating function via
\begin{eqnarray}
	\mathcal K_{\overline{\boldsymbol n}, \boldsymbol\chi} (z)  = \log \text{tr} \left[ \rho_{\boldsymbol \chi}(t_{\text{M} })  \right] \rightarrow \lambda_{\overline{\boldsymbol n},\boldsymbol \chi} t_{\text{M}}	.
	\label{eq:auxGenFkt-longTimeLimit}
\end{eqnarray}
The information about the laser pulse properties (i.e., the occupations of the photonic modes) is expressed by the subscript $\overline{\boldsymbol n}$ of the generalized Liouvillian. For long enough measurement times $t_{\text{M}}$, the generating function becomes proportional to the dominating eigenvalue of the Liouvillian, i.e., the eigenvalue which is $\lambda_{\overline{\boldsymbol n},\boldsymbol \chi}\rightarrow 0$ for $\boldsymbol \chi \rightarrow 0$~\cite{Engelhardt2024b}, which leads to the simple expression for the generating function in Eq.~\eqref{eq:auxGenFkt-longTimeLimit}.

For the Hamiltonian in Eq.~\eqref{eq:semiclassicalJaymesCummingsModel} under the influence of the disspative bath modes, the generalized Liouvillian is time-independent and given by
\begin{eqnarray}
	\mathcal L_{\overline{\boldsymbol n},\boldsymbol \chi}\rho &=& -i \left[\hat {\mathcal H}_{\boldsymbol \varphi +\frac{\boldsymbol \chi}{2}  }  \rho -\rho  \hat {\mathcal H}_{\boldsymbol \varphi-\frac{\boldsymbol \chi}{2} } \right]  \nonumber \\
	&+& \gamma \left[ \hat \sigma_{-}  \rho \hat \sigma_{+}  -\frac{1}{2} \left\lbrace   \rho, \hat \sigma_{+} \hat \sigma_{-} \right  \rbrace \right] ,
	\label{eq:liouvillian:twoLevelSys}
\end{eqnarray}
where $\gamma$ parameterizes the coupling strength of the atoms to the bath modes, and the Hamiltonian acting on the quantum emitters reads as
\begin{equation}
	\hat {\mathcal H}_{\boldsymbol \varphi}  =  \frac{ \epsilon_{\Delta}}{2} \hat \sigma_{\text{z}}  +  \sum_{k=1,2} \left[ \frac{\Omega_{k}}{2} e^{i \varphi_k }\hat \sigma_{-} + \text{ h.c.} \right],
	\label{eq:semiclassicalTavisCummingsModel}
\end{equation}
where $\epsilon_{\Delta} = \epsilon -\omega$ denotes the laser detuning. For simplicity, we consider a  Rabi frequency which is constant during the pulse duration $t_{\text{M}}$.

\subsection{Flow equations}

\label{sec:tavisCummings:flowEquations}

We are now in  position to construct the flow equations for the mean and the covariance matrix in Eqs.~\eqref{eq:tra4:momGenFunction}  and \eqref{eq:varianceFlow}, respectively,  using the generating function in Eq.~\eqref{eq:auxGenFkt-longTimeLimit}. The analytical calculation of the low-order derivatives with respect to the counting fields is demonstrated in Appendix~\ref{app:coeefficientsFlowEquationTwolevelSystem}. The following formulas are represented in a vector notation in the basis $\left(n_1 , n_2\right)$.

The photon flux coefficients defined in Eq.~\eqref{eq:fluxEvaluation} read
\begin{multline}
\boldsymbol  I_{\boldsymbol {\overline n}    }   =  \frac{t_{\text{M}} \rho_A  \mathcal  A  }{4 \epsilon_{\Delta}^{2} + 2 \Omega_{\tilde \varphi} ^2+\gamma  } \left[ 
	   \Omega_1 \Omega_2 \epsilon_{\Delta}  \sin\left(\tilde \varphi \right)   \left( \begin{array}{c} 1\\-1 \end{array}\right)  \right. \\\left.
	-\gamma \left( \begin{array}{c} \Omega_{1}^2 \\\Omega_{2}^2 \end{array}\right) 
	-\gamma \Omega_{1} \Omega_{2} \cos\left(\tilde \varphi \right)\left( \begin{array}{c} 1 \\1  \end{array}\right) 
	\right] ,
\label{eq:photonFluxExact}
\end{multline}
where we have introduced
\begin{equation}
\Omega_{\tilde \varphi} ^2=    \Omega_{1}^{2} + 2 \Omega_{1} \Omega_{2} \cos\left(\tilde \varphi \right) +  \Omega_{2}^{2}  .
\end{equation}
This is the effective Rabi frequency which the two-level experiences due to the driving of the two modes. Moreover, we have defined $\tilde \varphi =\overline \varphi_1 -\overline \varphi_2$. The term proportional to $\epsilon_{\Delta} $ gives rise to a mean-photon number imbalance between  modes $k=1$ and $k=2$, which can be associated with a phase shift according to the explanations in Eq.~\eqref{eq:phaseEstimation}. The terms proportional to $\gamma$ describe  absorption of photons by the dissipative two-level systems.

The diffusive matrix defined in Eq.~\eqref{eq:diffusion-phaseSpaceMatrices}  can be represented as
\begin{eqnarray}
\boldsymbol  D_{\boldsymbol {\overline n} }     &\approx&  \rho_A  \mathcal  A t_{\text{M}}
\frac{32  \Omega_{1}^{2} \Omega_{2}^{2} \Omega_{\tilde \varphi} ^4 \sin^{2}\left(\tilde \varphi \right) }{\gamma \left[4\epsilon_{\Delta}^{2} + 2 \Omega_{\tilde \varphi} ^2   \right]^{3}} \begin{pmatrix}
1 & -1\\-1 & 1
\end{pmatrix} 
+ \mathcal O \left( \gamma^0 \right) , \nonumber \\
\label{eq:LeadingphotonFluctuations}
\end{eqnarray}
which we have expanded for small dissipation $\gamma$. The full expression can be found in Eq.~\eqref{eq:fullExpressionFlowEquationCoeff}. Interestingly, the lowest order contribution scales with $\gamma^{-1}$. The matrix has two eigenvectors. The first one is $(1,1)$ and has a vanishing eigenvalue, and thus does  not contribute to the diffusive dynamics. The second  one is $(1,-1)$ and has a finite eigenvalue and thus contributes to the diffusion. As the eigenvector $(1,-1)$  is proportional to the first line in Eq.~\eqref{eq:photonFluxExact}, and the leading order of the diffusion matrix is propotional to $\sin^{2}\left(\tilde \varphi \right)$ we conclude that the leading order of $\boldsymbol  D_{\boldsymbol {\overline n} }  $ describes phase diffusion.

Finally, the phase-space matrix defined in Eq.~\eqref{eq:diffusion-phaseSpaceMatrices} can be calculated by deriving the photon fluxes with respect to the mean photon numbers $\overline n_1 ,\overline n_2$. In doing so, we obtain
\begin{multline}
\boldsymbol  C_{\boldsymbol {\overline n} }     =  
\frac{\rho_A  \mathcal  A t_{\text{M}}  }{4 \epsilon_{\Delta}^{2} + 2\Omega_{\tilde \varphi} ^2+\gamma }
\left[ 
\frac{\Omega_1 \Omega_2\epsilon_{\Delta}  }{2}\sin\left(\tilde \varphi \right) 
\left(
\begin{array}{cc}
\frac{1}{\overline n_1}   &  \frac{1}{\overline n_2}  \\
-\frac{1}{\overline n_1}   &  -\frac{1}{\overline n_2}  \\
\end{array}
\right)  \right. \\
- \gamma  
\left(
\begin{array}{cc}
\frac{\Omega_1^2}{\overline n_1}   &  0  \\
0    &  \frac{\Omega_2^2}{\overline n_2}  \\
\end{array}
\right)
-\frac{\Omega_1 \Omega_2 \gamma}{2}  \cos\left(\tilde \varphi \right)
\left(
\begin{array}{cc}
\frac{1}{\overline n_1}   &  \frac{1}{\overline n_2}  \\
\frac{1}{\overline n_1}   &  \frac{1}{\overline n_2}  \\
\end{array}
\right)
  \\ 
-  \left. \frac{ \boldsymbol  I_{\boldsymbol {\overline n}    }}{\rho_A  \mathcal  A t_{\text{M}} }   \otimes \left(\frac{d \Omega_{\tilde \varphi}^2}{d\overline n_1}  , \frac{d \Omega_{\tilde \varphi}^2}{d\overline n_2} \right)\right].
\label{eq:phaseCoefficients}
\end{multline}
where $\otimes$ denotes the tensor product of two vectors producing a matrix.
The first three terms in the rectangular bracket are the derivatives of the first three terms in the rectangular bracket in Eq.~\eqref{eq:photonFluxExact}. The fourth term in the bracket is generated by the prefactor in Eq.~\eqref{eq:photonFluxExact}, and has not been explicitly written down for brevity.

The  mean-photon numbers and the corresponding variances as a function of position $z$ are depicted in Fig.~\ref{figFlowdynamics}(c).  For small distances $z$, we observe that the mean value of $\hat n_{1}$ decreases while the mean value of $\hat n_{2}$ increases. This means that the polarization direction rotates towards the $\text{y}$ direction. Likewise, the variance of $\hat n_{1}$ decreases while the variance of $\hat n_{2 }$ increases. This is a consequence of the phase space terms $C_{\boldsymbol {\overline n}}$ in Eq.~\eqref{eq:phaseCoefficients} which describe the deformation of the probability distribution due the photon-number dependence of the creation and annihilation operators. The effect of the diffusion matrix $D_{\boldsymbol {\overline n}}$ leads to an overall increase of the variance for larger $z$.

In Fig.~\ref{figFlowdynamics}(d), we depict the probability distribution for three propagation distances. For $z=2\,\text{cm}$, the distribution is noticeably deformed due to the action of the phase space terms $C_{\boldsymbol {\overline n}}$. This deformation is even clearer for $z=4\,\text{cm}$ for which the $\text{Var}\, \hat n_{2}$ almost vanishes. At this point,   $\overline n_2 \approx 0$, meaning that the measurement direction is almost orthogonal to the polarization of the laser field such that the corresponding photon mode is close to its vacuum state.

\subsection{Rotating frame}

\label{sec:tavisCummingsModel:rotFrame}

For the reasons explained in Sec.~\ref{sec:rotatingMeasurementFrame}, it is beneficial to  integrate the flow equations in a rotating frame, in which $\Omega_1= \Omega_2 \equiv \Omega/\sqrt{2}$ and $\tilde \varphi = \pi/2$. In doing so, the coefficients in Sec.~\eqref{sec:tavisCummings:flowEquations} and the equations of motion simplify drastically, enabling  a theoretical analysis in limiting cases. 

The photon-flux coefficients in Eq.~\eqref{eq:photonFluxExact} read
\begin{multline}
	\boldsymbol  I_{\boldsymbol {\overline n_{\text{rot}}}   }   =  \frac{\rho_A  \mathcal  A t_{\text{M}} }{ 4\epsilon_{\Delta}^{2} +  2\Omega^2+\gamma^2  } \left[ 
	\epsilon_{\Delta}   \Omega^2    \left( \begin{array}{c} 1\\-1 \end{array}\right)  
	-\gamma \Omega^2  \left( \begin{array}{c} 1 \\1  \end{array}\right) 
	\right] ,\\
	\label{eq:photonFluxRotated}
\end{multline}
which allows us to express the equations for the phase shift and the total number of photons according to Sec.~\eqref{sec:rotatingMeasurementFrame} as
\begin{eqnarray}
\frac{d\overline \varphi }{dz}  
&=&\frac{ \rho_A  \mathcal  A t_{\text{M}} }{{\overline n}_{+  }  }  \frac{\epsilon_{\Delta}   \Omega^2   }{ 4\epsilon_{\Delta}^{2} + 2 \Omega^2+ \gamma^2  }    , \nonumber \\
\frac{d\overline n_{+  } }{dz}  &=&-\rho_A  \mathcal  A t_{\text{M}}  \frac{\gamma   \Omega^2   }{ 4\epsilon_{\Delta}^{2} + 2 \Omega^2 + \gamma^2  }  .
\label{eq:phaseShiftAngle-photonNumber}
\end{eqnarray}
Notably, the integration of the phase shift equation is trivial for a vanishing $\gamma$, for which $\overline n_{+}$  becomes constant: Noting  that in the rotated basis $\Omega = g \sqrt{\overline n_{+ }}$, we see that the right-hand side of the phase equation becomes independent of $z$.

Likewise, the diffusion matrix simplifies  and now reads
\begin{eqnarray}
\boldsymbol  D_{\boldsymbol {\overline n}_{\text{rot}} }     &\approx&  
 \frac{8\rho_A  \mathcal  A t_{\text{M}}    \Omega ^8  }{\gamma \left[4 \epsilon_{\Delta}^{2} + 2\Omega ^2 \right]^{3}} \begin{pmatrix}
1 & -1\\-1 & 1
\end{pmatrix} 
+ \mathcal O \left( \gamma^0 \right).
\label{eq:tavisCummingsModel:LeadingphotonFluctuations}
\end{eqnarray}
As for the phase-shift equation \eqref{eq:phaseShiftAngle-photonNumber}, the diffusion matrix becomes independent of position for a small dissipation $\gamma$, allowing for an analytical integration of the variance equation.

A drastic simplification appears for the phase-space matrix, which in the rotated frame becomes
\begin{multline}
\tilde {\boldsymbol  C}_{\boldsymbol {\overline n_{\text{rot}}}   }     = 
\frac{t_{\text{M}}   \rho_A  \mathcal  A  }{\overline n_{+ } }\frac{\Omega^2}{4 \epsilon_{\Delta}^{2} + 2\Omega ^2+\gamma^2 } \\  \times \left[ 
 -4\gamma 
\left(
\begin{array}{cc}
1   &  0  \\
0    & 1 \\
\end{array}
\right)
 \right. 
- \left. \frac{ \boldsymbol  I_{\boldsymbol {\overline n_{\text{rot}}}   }  }{\rho_A  \mathcal  A t_{\text{M}} }  \otimes \left( \begin{array}{c} 1 \\1  \end{array}\right)  \right],
\label{eq:tavisCummingsModel:rotationCoefficient}
\end{multline}
i.e., the first and the third terms in Eq.~\eqref{eq:phaseCoefficients}  disappear. The first term in the bracket corresponds to the dissipative dynamics, which results in a contraction of the covariance matrix. The second term describes the photon-number  dependence of the Rabi frequency.

 In Fig.~\ref{figFlowdynamics}(c), we depict the means and variances of the photonic modes $k=1,2$ in the rotated measurement frame. By definition, $\overline n_{\text{rot},1} = \overline n_{\text{rot} ,1} = \overline n_{+}/2$ for all positions. The total mean photon number is declining because of the   absorption by the atoms. The rapid oscillations of the variance, which are visible in the laboratory frame, disappear in the rotated measurement frame. 

Interestingly, we find that the variances of both modes are not equal. While the variance of mode $k=1$ decreases, the variance of mode $k=2$ significantly increases. This asymmetry is a consequence of  a distortion of the probability distribution as depicted for selected values of $z$ in Fig.~\ref{figFlowdynamics}(d). 
 The distortion is driven by  the second term  of the phase-space coefficient in Eq.~\eqref{eq:tavisCummingsModel:rotationCoefficient}:
The vector $(1,1)$ acts on  $\boldsymbol \Sigma$ in the direction of $\overline n_{+}$, and distorts the covariance matrix  controlled by $\boldsymbol  I_{\boldsymbol {\overline n}    }$.

We interpret this distortion as a result of the  dependence of the energy level splitting in the two-level on the photon number  (i.e., the laser intensity):  The larger the light intensity, the larger the level splitting. This relation leads to a reduced photon transfer between the atoms and the photonic modes for larger photon numbers. As the variance of the photonic probability distribution is on the order $\sqrt{ \overline n_k}$, and the level splitting is a function of the Rabi frequencies $\Omega_k\propto \sqrt{ \overline n_k}$, i.e., the variance and Rabi frequencies both change on the same order of magnitude, the level splitting dependence thus results in a significant distortion of the probability distribution.

\subsection{Mean and variance}

\label{sec:tavisCummingsModel:analysis}

\begin{figure*}
	\includegraphics[width=\linewidth]{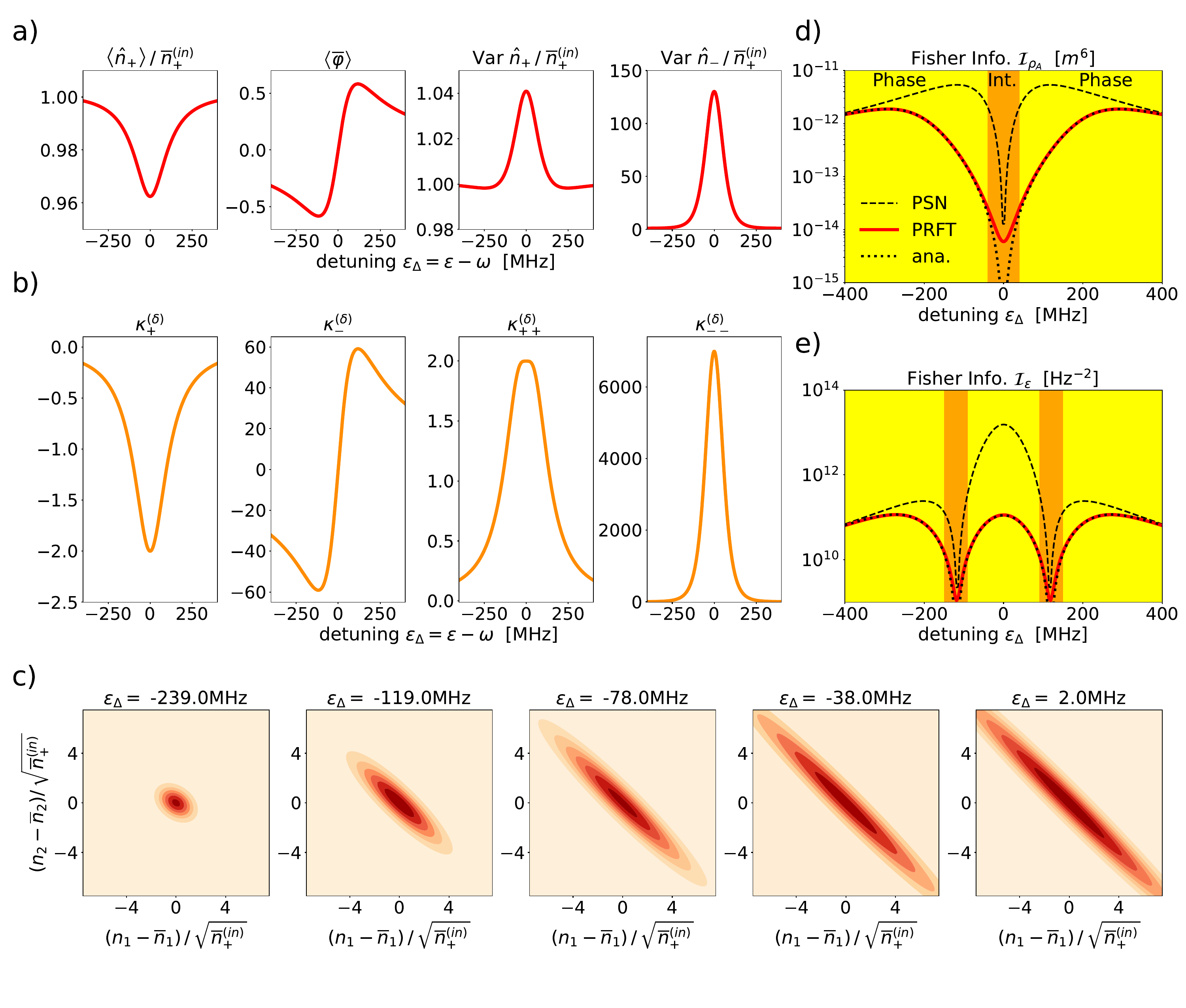}
	\caption{Measurement statistics  for the dissipative Tavis-Cummings model  in Eq.~\eqref{eq:ham:dissipativeTCmodel}.  (a) Means and variances for the observed photon numbers $n_+$ and $n_-$ obtained by integrating the flow equations in Eqs.~\eqref{eq:rotationAngel-photonNumber} and \eqref{eq:correlationMatrixFlow-rotated}  using the coefficients in Eqs.~\eqref{eq:phaseShiftAngle-photonNumber}, \eqref{eq:tavisCummingsModel:LeadingphotonFluctuations} and \eqref{eq:tavisCummingsModel:rotationCoefficient}.  
		(b) Corresponding cumulant aptitudes at $z=0$ given in Eq.~\eqref{eq:tavisCummingsModel:aptitudes}. (c) Probability distributions for selected detunings.   (d) Fisher information for the atom density as a function of detuning. The dashed line shows the Fisher information  where the photonic variance has been estimated with the initial photon-shot noise (PSN). (e) is the same as (d), but for the Fisher information for the detuning. Overall parameters are the same as in Fig.~\ref{figFlowdynamics}.  }
	\label{fig:tavisCummingsModel}
\end{figure*}

Using the coefficients in Sec.~\ref{sec:tavisCummingsModel:rotFrame}, we can  integrate Eq.~\eqref{eq:correlationMatrixFlow-rotated} and analyze the  measurement statistics after the laser pulses have propagated through the atomic cloud at $z=z_{\text{max}}$. The results are depicted in Fig.~\ref{fig:tavisCummingsModel}. 

 In Fig.~\ref{fig:tavisCummingsModel}(a), we depict the mean photon number $\overline n_{+ }$ and the phase shift $\overline \varphi_0$ along with their corresponding variances as a  function of the detuning $\epsilon_{\Delta}$. Note that the variance associated with the $\overline \varphi_0$ measurement is given by the variance of $n_{-}$.  The mean total photon number $\overline n_{+ }$ exhibits the common absorption dip at resonance $\epsilon_{\Delta} = 0$, while the mean phase shift $\overline \varphi_0$ resembles the well-known dispersion-relation curve. The variance of $n_{+ }$ and $n_{- }$ both exhibit a single peak at resonance. While the variance of  $n_{+ }$ hardly overcomes the photon-shot-noise, the variance of $n_{- }$  exceeds the photon-shot noise by more than two orders of magnitude.

To analyze this effect theoretically, we first note that the photon number statistics of $n_{+ }$ and $n_{- }$ can be obtained by expressing the cumulant-generating function in Eq.~\eqref{eq:def:generatingFcts} it terms of the new counting fields
$\chi_{+ }  =   \left( \chi_1 +\chi_2 \right)/2 $ and
$ \chi_{-}  =  \left( \chi_1 - \chi_2 \right)/2. $
For a theoretical purpose, we introduce the coefficients
\begin{eqnarray}
\kappa^{(\delta)}_{\alpha} &=& \frac{i}{t_{\text{M}}}\partial_{ \chi_{\alpha} }      \mathcal K_{\boldsymbol {\overline n}_{\text{rot}},\boldsymbol \chi =0  } , \nonumber  \\
	\kappa^{(\delta)}_{\alpha\beta} &=&- \frac{1}{t_{\text{M}}}\partial_{ \chi_{\alpha} }\partial_{\chi_{\beta} }       \mathcal K_{\boldsymbol {\overline n}_{\text{rot}},\boldsymbol \chi =0  } ,
	\label{eq:def:cumulantAptitudes}
\end{eqnarray}
with $\alpha,\beta= +,-$, which we denote as cumulant aptitudes in the following. They describe the intrinsic ability of a single atom to modify the photonic statistics of their driving fields. The dynamical cumulant-generating function is thereby evaluated in the rotating frame (as expressed by $\boldsymbol {\overline n}_{\text{rot}}$) for which $\overline n_{\text{rot},- } = 0$. 

Analytically, the evaluation of the cumulant aptitudes is closely related to the calculation of the flow equation coefficient matrices in the rotated measurement basis in Sec.~\ref{sec:tavisCummingsModel:rotFrame}, and we obtain
\begin{eqnarray} 
	\kappa^{(\delta)}_{+}  &=&   \frac{\gamma \Omega^2  }{ 4\epsilon_{\Delta}^{2} + 2 \Omega^2+\gamma  }   ,\nonumber \\  %
	\kappa^{(\delta)}_{-}  &=&    \frac{2\epsilon_{\Delta} \Omega^2  }{4 \epsilon_{\Delta}^{2} + 2 \Omega^2+\gamma  } , \nonumber \\  %
	\kappa^{(\delta)}_{++} &=&    \frac{\gamma\Omega^2}{ 4\epsilon_{\Delta}^2 +2\Omega^2 + \gamma^2 } 
	+\frac{\Omega^4\gamma \left( 8\epsilon_{\Delta}^2 -6 \gamma^{2} \right) }{  \left(4\epsilon_{\Delta}^2 +2\Omega^2 + \gamma^2 \right)^3},\nonumber \\  %
	\kappa^{(\delta)}_{--} &=&   \frac{2\Omega^4  +     \gamma^2 \Omega^2  }{\gamma \left(4\epsilon_{\Delta}^2 +2\Omega^2 + \gamma^2 \right)} 
	- \frac{8 \epsilon_{\Delta}^2 \Omega^4 \left( 4\Omega^2  +  4\epsilon_{\Delta}^{2} + 5 \gamma^{2} \right) }{\gamma \left(4\epsilon_{\Delta}^2 +2\Omega^2 + \gamma^2 \right)^3}  ,\nonumber \\  %
	\label{eq:tavisCummingsModel:aptitudes}
\end{eqnarray}
as shown in Appendix~\ref{app:cumulantAptitudes}.
The cumulant  aptitudes are depicted in Fig.~\ref{fig:tavisCummingsModel}(b) below the corresponding measured cumulants. We see that the aptitudes closely resemble the corresponding  cumulants, which shows the basic importance of the aptitudes to describe the response of the matter system to an external driving field. In particular, from Eq.~\eqref{eq:tavisCummingsModel:aptitudes} we learn that the phase variance aptitude $\kappa^{(\delta)}_{--} $ is proportional to $1/\gamma$, which is the essential driver of the phase fluctuations. As investigated in Ref.~\cite{Engelhardt2024b} and also explained in Appendix~\ref{app:weakCouplingRegimeDynamics}, the $1/\gamma$ dependency originates from incoherent jumps  between the two eigenstates of the two-level system, each of which featuring distinct optical properties, leading to diverging fluctuations.

To further illustrate the large fluctuations of $n_{-}$, we depict the photonic probability distribution shown in Fig.~\ref{fig:tavisCummingsModel}(c)  as a function of the photon numbers $n_1$ and $n_2$ for various detunings. The closer the detuning is to the resonance $\epsilon_{\Delta}=0$, the more the probability ellipsoid extends across the diagonal, which is in agreement with the variances in Fig.~\ref{fig:tavisCummingsModel}(a) and the cumulant aptitude $\kappa^{(\delta)}_{--} $ in Fig.~\ref{fig:tavisCummingsModel}(b). This effect describes the tremendous fluctuations which can appear in phase measurements. 

\subsection{Fisher information}

We continue to discuss the spectroscopic setup sketched in Fig.~\ref{figFlowdynamics}(a) for its potential in spectroscopic quantum sensing. In the following, we will focus on the estimation of two quantities, namely  the atomic density $\rho_A$, and the level splitting $\epsilon$, which could be a function of an unknown external stimulus. 

  Using  Eq.~\eqref{eq:fisherInfoGaussianState}, which we evaluate using the mean values and covariance matrix obtained by the numerical integration of the flow equations, we can conveniently  estimate  the Fisher information for vast parameter regimes.
In Fig.~\ref{fig:tavisCummingsModel}(d) and Fig.~\ref{fig:tavisCummingsModel}(e) we depict the Fisher information $\mathcal I_X$ for the estimation of the density $X=\rho_A$ and the energy splitting $X=\epsilon$ as a function of the detuning by a red solid line. Both Fisher informations exhibit a non-trivial dependence on the detuning. Thus, the methods introduced in this work constitute a flexible methodology to quickly find experimental parameters which optimize the sensitivity of the measurement apparatus.

 To emphasize the importance of taking the added fluctuation predicted by the variance flow equation in Eq.~\eqref{eq:correlationMatrixFlow-rotated} into account, we depict also a simplistic variance estimate by a dashed black line, which approximates the noise by the photon-shot noise of the incoming laser pulse. As we observe in Fig.~\ref{fig:tavisCummingsModel}(d) and Fig.~\ref{fig:tavisCummingsModel}(e), this oversimplified estimate drastically fails to accurately predict the actual Fisher information by several orders of magnitudes. Only far away from resonance, where the variance aptitude is strongly suppressed [see Fig.~\ref{fig:tavisCummingsModel}(d)], and very close to the resonance condition, where the Fisher information is mainly carried by the total photon-number change, the shot-noise estimate succeeds to be a reliable measure. 

\subsection{ Small dissipation}

\begin{figure*}
	\includegraphics[width=\linewidth]{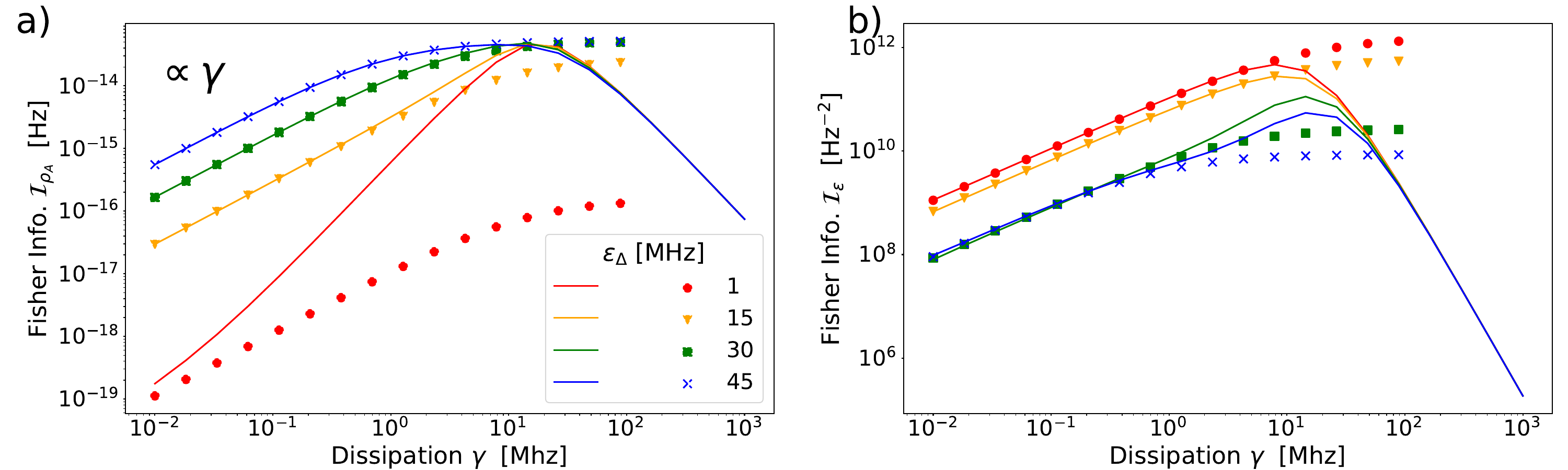}
	\caption{Fisher information of the atom density $\mathcal I_{\rho_A}$  and the detuning $\mathcal I_{\epsilon}$ for the Tavis-Cummings model as function of dissipation rate $\gamma$ depicted in panels (a) and (b), respectively. The solid lines show the numerical results obtained by integrating the flow equations using the coefficients in Eqs.~\eqref{eq:photonFluxRotated}, \eqref{eq:tavisCummingsModel:LeadingphotonFluctuations}, and \eqref{eq:tavisCummingsModel:rotationCoefficient}. The symbols show the analytical calculations in Eq.~\eqref{eq:tavisCummings:FisherInfo:density} and \eqref{eq:tavisCummings:FisherInfo:energy}.  The overall parameters are the same as in Fig.~\ref{figFlowdynamics}. }
	\label{fig:fisherInfoDissipation}
\end{figure*}

To better understand the functional dependence of the Fisher information on the experimental parameters, we investigate the Fisher information in the small dissipation regime $\gamma\ll \Omega$. In this limit, it is safe to assume that $\overline n_{+}$ is constant, for which the integration of the flow equations in the rotating measurement frame becomes trivial. Integrating Eq.~\eqref{eq:phaseShiftAngle-photonNumber}, the total phase shift becomes
\begin{equation}
\overline \varphi_0  = \frac{ t_{\text{M}}  \rho_A \mathcal  A   z_{\text{max}} }{\overline n_{+}  }   \frac{ 2\epsilon_{\Delta}	\Omega^2    }{ 4\epsilon_{\Delta}^{2} +  2\Omega^2 } +  \mathcal O\left(\gamma^0 \right)  ,
\label{eq:tavisCummingsModel:varphi:weakDissipation}
\end{equation}
where we have neglected terms of order $\gamma^0$. Noteworthy, if $\Omega\ll \epsilon_\Delta$, we find that $\overline \varphi$ is independent of $\overline n_{+} $ as $ \Omega^2 \propto \overline n_{+} $. In the opposite regime $\Omega\gg \epsilon_\Delta$, the phase shift is suppressed by a factor $\overline n_{+}^{-1}$.

Integrating Eq.~\eqref{eq:correlationMatrixFlow-rotated} using Eq.~\eqref{eq:tavisCummingsModel:LeadingphotonFluctuations} to obtain the covariance matrix in the weak dissipation regime, we find
\begin{multline}	
	\boldsymbol \Sigma^2  =  \left(
	\begin{array}{cc}
	  \frac{\overline n_{+ }} {2}  &  0  \\
	0    &  \frac{\overline n_{+}} {2}\\
	\end{array}
	\right) \\
	  +  8 t_{\text{M}} \rho_A  \mathcal  A     z_{\text{max}}  \frac{   \Omega ^8 }{\gamma \left(4\epsilon_{\Delta}^{2} +2 \Omega ^2 \right)^{3}}   \begin{pmatrix}
	1 & -1\\-1 & 1\end{pmatrix}  \\
	  +\mathcal O\left(\gamma^0,\overline n_{+}^0\right) ,
\end{multline}
where we have neglected terms scaling with $\gamma^0$ and $\overline n_{+}^0$. The first term is the photon-shot noise, while the second terms describes the noise added by the light-matter interaction. According to the explanations in Sec.~\ref{sec:tavisCummingsModel:analysis}, it describes the fluctuations appearing in the phase measurement.  The product $N_A =\rho_A  \mathcal  A     z_{\text{max}}$ is the total number of two-level systems interacting with the laser pulse.

Using now Eq.~\eqref{eq:fisherInfoGaussianState} to calculate the Fisher information for the atom density, we obtain
\begin{eqnarray}
\mathcal I_{\rho_A} &\approx&     
\frac{ \left(t_{\text{M}}   \mathcal  A     z_{\text{max}}  \frac{ 2\epsilon_{\Delta}	\Omega^2    }{ 4\epsilon_{\Delta}^{2} + 2 \Omega^2 }  \right)^2  }{\overline n_{+ }  + 8 t_{\text{M}} N_{A}   \frac{   \Omega ^8 }{\gamma \left(4 \epsilon_{\Delta}^{2} +2 \Omega ^2 \right)^{3}}  },
\label{eq:tavisCummings:FisherInfo:density}
\end{eqnarray}
which is valid in the weak dissipation limit.  Likewise, we find  that the Fisher information for the energy splitting is given by
\begin{eqnarray}
\mathcal I_{\epsilon } &\approx&     
\frac{
	 \left[  t_{\text{M}} N_{A}   \frac{ 	\Omega^2 \left(  4\epsilon_{\Delta}^{2} -2 \Omega ^2  \right) }{  \left( 4\epsilon_{\Delta}^{2} + 2 \Omega^2 \right)^2 }  \right]^2 
 }
{\overline n_{+}  + 8 t_{\text{M}} N_{A}  \frac{   \Omega ^8 }{\gamma \left(4\epsilon_{\Delta}^{2} + 2\Omega ^2 \right)^{3}}  }.
\label{eq:tavisCummings:FisherInfo:energy}
\end{eqnarray}
For comparison, these analytical expression are depicted in Fig.~\ref{fig:tavisCummingsModel}(d) and Fig.~\ref{fig:tavisCummingsModel}(e), respectively, with dotted lines, which agree with the actual Fisher information for vast parameter regimes. For the detunings for which we observe deviations, the total photon number measurement $n_+$ contributes significantly to the Fisher information. Overall, we observe that the Fisher information contained in the phase shift is considerably larger than the one in the total photon number (i.e., intensity). For this reason, it is sufficient to optimize the Fisher information related to the phase information, that is  approximately given by Eqs.~\eqref{eq:tavisCummings:FisherInfo:density} and \eqref{eq:tavisCummings:FisherInfo:energy}.

In Fig.~\ref{fig:tavisCummingsModel}(d) we observe that the Fisher information has a minimum at $\epsilon_{\Delta}=0$, which is a consequence of the vanishing derivative of $\overline \varphi_0$ in Eq.~\eqref{eq:tavisCummingsModel:varphi:weakDissipation} with respect to $\rho_A$. Similarly, we find a Fisher information minimum in Fig.~\ref{fig:tavisCummingsModel}(e), which is caused by the vanishing derivative of $\overline \varphi_0$ with respect to $\epsilon$. Away from the resonance, the Fisher information $\mathcal I_{\rho_A}$ increases and finally exhibits a turnover: For large detunings $\epsilon_{\Delta}$, the added noise [second term in the denominator of Eq.~\eqref{eq:tavisCummings:FisherInfo:density}] vanishes and only the photon-shot noise remains. In this parameter regime, the predictions of the PRFT and the naive photon-shot-noise approximation agree with each other. The decreasing  derivative of $\overline \varphi_0$ in Eq.~\eqref{eq:tavisCummingsModel:varphi:weakDissipation} for large detunings is eventually reflected in a decreasing Fisher information $\mathcal I_{\rho_A}$. A similar explanation holds for $\mathcal I_{\epsilon}$.

\subsection{Dissipation dependence}

\label{sec:dissipationDependence}

In Fig.~\ref{fig:fisherInfoDissipation},  we investigate the dependence of the Fisher information  on the dissipation rate $\gamma$. Intriguingly, we observe that both $\mathcal I_{\rho_A}$ and $\mathcal I_{\epsilon}$ exhibit a turnover as a function of $\gamma$, i.e., they increase for small $\gamma$ and decrease for large $\gamma$.  

In the small $\gamma$ regime, we can readily explain the enhancement of the Fisher information using Eqs.~\eqref{eq:tavisCummings:FisherInfo:density} and ~\eqref{eq:tavisCummings:FisherInfo:energy}. If $\gamma \rightarrow 0$, we can neglect the photon-shot-noise term in the denominator, such that $\mathcal I_{\rho_A} \propto \gamma$ and $\mathcal I_{\epsilon}\propto \gamma$. From this considerations we conclude that the increase of the Fisher information is driven by an decrease of the phase fluctuation in the measurement, while the  differential change of the mean phase  remains almost unchanged.

In the opposite regime of strong dissipation, we learn from Eq.~\eqref{eq:photonFluxRotated} that the mean photon flux becomes increasingly suppressed for larger $\gamma$. This is consequently reflected in an decreasing Fisher information. In the large dissipation regime, $\gamma$ dominates the atomic dynamics, such that all curves in Fig.~\ref{fig:fisherInfoDissipation} collapse on each other. 

Worthwhile, similar turnovers as a function of system-bath coupling are well known from exciton and energy transport in molecular systems~\cite{Xu2016} and polariton systems~\cite{Engelhardt2022}.  Thereby, the transport efficiency is optimized by an interplay of coherent and incoherent dynamics. Surprisingly, here we observe a similar turnover, yet, in a genuine quantity of information science. Similar to transport in molecular systems, the optimal Fisher information requires a fine tuned  coherent dynamics, which can be controlled by an appropriately adjusted dissipation rate $\gamma$.

\begin{figure*}
	\includegraphics[width=\linewidth]{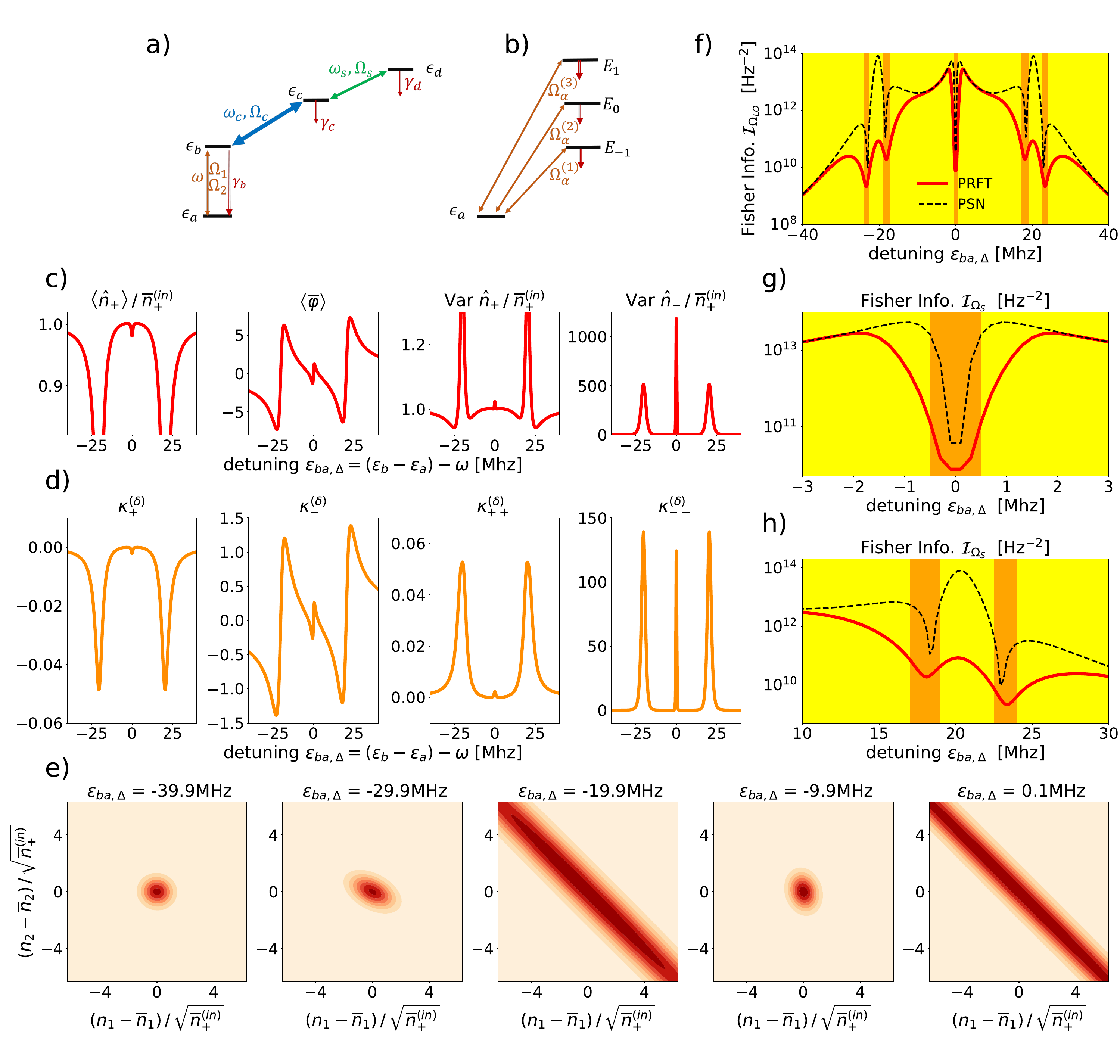}
	\caption{Analysis of the measurement statistics for the four-level system in Eq.~\eqref{eq:fourLevelSystem:ham}  which is sketched in the local basis in (a), and in the partially diagonalized basis in (b). (c) Mean values and variances of the photon measurement statistics, which are calculated  in the same way as for  Fig.~\ref{fig:tavisCummingsModel}.  (d) Corresponding cumulant aptitudes at $z=0$. (e) Photonic probability distribution for selected detunings. (f)-(h) Fisher information $\mathcal I_{\Omega_{\text{S}} }$.  Parameters of the probe field are  $P = 0.1 \,\text{mW}$, $\mathcal A = 1\,\text{cm}^2$,  $t_{\text{P} } = 1\,\text{s}$, $\lambda = 500\,\text{nm}$, and atomic dipole moment $d = 5 e a_0$, which gives rise to an  Rabi frequency of $\Omega_{\text{P}}=1.9\,\text{MHz}$. Other parameters are $\Omega_{\text{C}} =20\,\text{MHz}  $,  $\Omega_{\text{S}} =3 \,\text{MHz}  $ $\gamma_{b} = 0.1\,\text{MHz}$, $\gamma_{c} = 0.01\,\text{Mhz}$ = $\gamma_{d} = 0.01\,\text{MHz}$ , $z_{ \text{max} }= 10 \text{cm}$, $\rho_A=5\cdot 10^{13} \text{m}^{-3}$, and $t_{\text{M}} = 1\text{s}$.  }
	\label{fig:fourLevelSystem}
\end{figure*}

\section{Four-level system}

\label{sec:fourLevelSystem}

As a more sophisticated application, we consider an ensemble of  four-level systems (i.e., atoms), whose level structure  is sketched in Fig.~\ref{fig:fourLevelSystem}(a). Thereby, the four energy levels $j=a,b,c,d$ are coupled by three lasers fields, namely, the probe, coupling, and signal lasers. The Hamiltonian of a single atom coupled to the light field is given by
	\begin{multline}
\hat H(t) 
= \sum_{j= a,b,c,d} \epsilon_{j} \ketbra{j}{j}   + \sum_{k=\text{P,C,S}} \omega_{k} \hat a_{\text{k}}^\dagger \hat a_{\text{k}} \\
+  \left[  g_{\text{P} } \hat a_{\text{P}} \ketbra{b}{a} + g_{\text{C} } \hat a_{\text{C}}  \ketbra{c}{b}   +g_{\text{S}}  \hat a_{S}\ketbra{d}{c} +\text{h.c.}  \right], \\
\label{eq:fourLevelSystem:ham}
\end{multline}
where $ \epsilon_{j}$ and $\ket{j}$   denote the energies and the eigenstates of the atom. The probe, coupling, and signal fields are quantized by  $ \hat a_{\text{P}} $,  $ \hat a_{\text{C}} $, and $ \hat a_{\text{S}} $. The corresponding frequencies and light-matter coupling strengths are denoted by $\omega_{k}$ and  $g_{k}$, respectively.

This and similar protocols find application in the detection of electric fields in the radiofrequency regime using Rydberg states~\cite{Holloway2017,Cox2018,Meyer2020,Jing2020,Meyer2021,Holloway2022,Liu2022,Borowka2024,Kumar2023}, and has been suggestion for the detection of axion dark matter in the galatic halo~\cite{Engelhardt2024a}. The signal to be estimated is proportional to the Rabi frequency  $\Omega_{\text{S}}= g_{\text{S}} \left<  \hat a_{\text{S}}\right>$ and typically in the range $\omega_{\text{S}}<20\,\text{GHz}$, which couples two Rydberg states $c$ and $d$. Here, we carry out a theoretical investigation of its sensitivity  by estimating the Fisher information $\mathcal I_{X }$ for $X= \Omega_{\text{S}}$ utilizing the methods developed in this work.

\subsection{Semiclassical equations of motion}

To estimate the measurement statistics for the four-level system, we follow the same steps introduced in Sec.~\ref{sec:tavisCummingsModel} for the two-level systems. We first express the probe field  in the basis to be measured
\begin{eqnarray}
\hat a_{\text{P} }  &=& \frac{1}{\sqrt{2}i} \left( \hat a_1 -\hat a_2\right)
\end{eqnarray}
according to Eq.~\eqref{eq:outputModeReplacement}. Next, we transform the Hamiltonian into an interaction frame to remove  the second term of the Hamiltonian in Eq.~\eqref{eq:fourLevelSystem:ham}. The corresponding  semiclassical Hamiltonian thus becomes
\begin{multline}
\hat {\mathcal H}_{\boldsymbol \varphi } 
= \sum_{j= a,b,c,d} \epsilon_{j,\Delta} \ketbra{j}{j}    \\
+  \left[ \frac{ \Omega_{\text{P},\boldsymbol  \varphi} }{2} \ketbra{b}{a} + \frac{\Omega_{\text{C}}}{2}  \ketbra{c}{b}   + \frac{\Omega_{\text{S} }}{2}\ketbra{d}{c} + \text{h.c.}  \right], \\
\end{multline} 
where $\epsilon_{a,\Delta}  = \epsilon_{a }  $, $\epsilon_{b,\Delta}  = \epsilon_{b } -\omega_{\text{P}} $,  $\epsilon_{c,\Delta}  = \epsilon_{c } -\omega_{\text{P}} -\omega_{\text{C}}$ and $\epsilon_{d,\Delta}  = \epsilon_{d } -\omega_{\text{P}} -\omega_{\text{C}}-\omega_{S}$. The Rabi frequencies of the coupling and signal fields are denoted by $\Omega_{\text{C}}$ and $\Omega_{\text{S}}$. We choose the corresponding driving phases to be zero.  We express  the Rabi frequency of the probe field in terms of the Rabi frequencies of the measured  fields $\Omega_1$ and $\Omega_2$ as
\begin{equation}
	\Omega_{\text{P},\boldsymbol  \varphi}  =\frac{\sqrt{2}}{i} \left(  \Omega_1e^{i\varphi_1}- \Omega_2e^{i\varphi_2} \right),
\end{equation}
where $\varphi_1$ and $\varphi_2$ are the respective phases.

To construct the dynamical cumulant-generating function, we follow the same procedure as for the two-level system in Sec.~\ref{sec:jaynesCummingsModel}. To model dissipative effects, we solve the generalized Liouvillian equation
\begin{equation}
\frac{d}{dt}\rho_{\boldsymbol \chi} = -i\left[ \hat {\mathcal H}_{\boldsymbol \varphi + \frac{\boldsymbol \chi}{2} }  \rho_{\boldsymbol \chi} -\rho_{\boldsymbol \chi} \hat{ \mathcal H}_{\boldsymbol \varphi - \frac{\boldsymbol \chi}{2} } \right]  
 +\sum_{j=b,c,d} \gamma_j D_{\ketbra{a}{j} } \rho_{\boldsymbol \chi}  ,  
\label{eq:blochEquationLabFrame}
\end{equation}
where $\rho_{\boldsymbol \chi}(t)$ denotes the reduced density matrix of a single atom. The  dynamical cumulant-generating function is then equal to the trace of $\rho_{\boldsymbol \chi}(t)$. In Eq.~\eqref{eq:blochEquationLabFrame}, we have phenomenologically added the decay from the excited states to the ground states with dissipation rate $\gamma_j$, where the dissipator is defined by
\begin{equation}
D_ {\hat O  } \rho =  \hat O \rho \hat O^\dagger  - \frac{1}{2}\left( \hat O^\dagger \hat O \rho + \rho \hat O^\dagger \hat O \right) .
\end{equation}
The corresponding decay rates are denoted by $\gamma_j$.

\subsection{Findings}

 The numerically evaluated findings for the mean and variances of the measurement statistics of the total photon number and the phase estimation are depicted in Fig.~\ref{fig:fourLevelSystem}(c).  In contrast to the findings in Fig.~\ref{fig:tavisCummingsModel}(a), the four-level system exhibit three resonances, two of which feature broad absorption dips for $\overline n_+$  around $\epsilon_{ba,\Delta}\approx \pm 20 \text{MHz}$, and one narrow dip for $\epsilon_{ba,\Delta}=0$. Accordingly, the dispersion of $\overline \varphi$ consists of three characteristics phase steps. The variance of $\hat  n_+$ hardly deviates from the initial photon-shot noise. In contrast the variance of $\hat n_-$ exceeds the initial photon-shot noise by several orders of magnitude for all three resonances. This is thus reminiscent to the findings in Fig.~\ref{fig:tavisCummingsModel}(a), where the variance of $\hat n_-$ is significantly more pronounced than that of $\hat n_+$.

The corresponding cumulant aptitudes are depicted in Fig.~\ref{fig:fourLevelSystem}(d), which exhibit a close resemblance with the means and the variances in Fig.~\ref{fig:fourLevelSystem}(c). We conclude that the main information about the measured photon statistics is directly determined by  these quantities. In Fig.~\ref{fig:fourLevelSystem}(e), we depict the probability distribution for selected detunings, which illustrates the dominance of the $\hat n_-$ noise over the $\hat n_+$ noise for parameters close to the resonance. 

In Fig.~\ref{fig:fourLevelSystem}(f), we depict  the estimate for the Fisher information $\mathcal I_{\Omega_S}$. As the system is more sophisticated than the two-level system in Sec.~\ref{sec:tavisCummingsModel}, the overall dependence  of all investigated quantities on the system parameters is more complicated. We compare the Fisher information predicted by the PRFT with the photon-shot-noise estimation as we did in Fig.~\ref{fig:tavisCummingsModel}. In doing so, we find that the shot-noise estimation would drastically overestimate the Fisher information by more than three orders of magnitude.

In Fig.~\ref{fig:fourLevelSystem}(g), and Fig.~\ref{fig:fourLevelSystem}(h) we depict the Fisher information in  two special detuning intervals $\epsilon_{ba,\Delta} \in \left[-3\,\text{MHz} , 3\,\text{MHz} \right]$ and  $\epsilon_{ba,\Delta} \in \left[10\,\text{MHz} , 30\,\text{MHz} \right]$. Notably,  Fig.~\ref{fig:fourLevelSystem}(g) and Fig.~\ref{fig:fourLevelSystem}(h) closely resemble  Fig.~\ref{fig:tavisCummingsModel}(d) and Fig.~\ref{fig:tavisCummingsModel}(e), respectively,  investigating the two-level system. This close resemblance will be explained in the next section.

\subsection{Analysis}

The more advanced four-level system does not allow for a simple non-perturbative treatment as the two-level system in Sec.~\ref{sec:tavisCummingsModel}. Nevertheless, here we give some analytical arguments which explain why Fig.~\ref{fig:tavisCummingsModel}(d) and Fig.~\ref{fig:tavisCummingsModel}(e) closely resemble  Fig.~\ref{fig:fourLevelSystem}(g) and Fig.~\ref{fig:fourLevelSystem}(h).

To this end, we assume that the probe-field Rabi frequency is significantly smaller than the Rabi frequencies of the coupling and the signal fields. In this regime, it is beneficial to represent the three excited states in their  diagonal basis as sketched in Fig.~\ref{fig:fourLevelSystem}(b). In this basis, the Hamiltonian reads
\begin{multline}
\hat {\mathcal H}_{\boldsymbol \varphi } 
= \sum_{j= a,0,\pm 1}E_{j,\Delta} \ketbra{j}{j}    
+   \sum_{j= 0,\pm 1} \left[ \frac{\Omega_{j,\boldsymbol \varphi }}{2} \ketbra{j}{0} + \text{h.c.} \right] ,  
\label{eq:fourLevelSystem:effectiveHamiltonian}
\end{multline}
where $E_{a,\Delta} =\epsilon_a$,
\begin{eqnarray}
E_{0,\Delta} &=& \epsilon_b \nonumber, \\
E_{\pm 1,\Delta} &=& \epsilon_b \pm \frac{1}{2}\sqrt{\Omega_{\text{C}}^2 +\Omega_{\text{S}}^2  },
\label{eq:fourLevelSystem:effectiveExcitationEnergies}
\end{eqnarray}
denote the  level energies, and
\begin{eqnarray}
\Omega_{0,\boldsymbol \varphi} &=&  \frac{\Omega_{\text{S}}}{2\sqrt{\Omega_{\text{C}}^2 +\Omega_{\text{S}}^2  }} \Omega_{\text{P},\boldsymbol  \varphi}, \nonumber \\
\Omega_{\pm 1,\boldsymbol \varphi} &=& \pm  \frac{\Omega_{\text{C}}}{\sqrt{\Omega_{\text{C}}^2 +\Omega_{\text{S}}^2  }} \Omega_{\text{P},\boldsymbol \varphi}
\label{eq:fourLevelSystem:effectiveRabiFreq}
\end{eqnarray}
 are the effective Rabi frequencies in the modified basis.

Based on this representation, we can interpret the ensemble of  four-level systems as consisting of three ensembles of  two-level systems with distinct physical properties. The respective excitation energies  and Rabi frequencies of each two-level systems are given by Eqs.~\eqref{eq:fourLevelSystem:effectiveExcitationEnergies} and \eqref{eq:fourLevelSystem:effectiveRabiFreq}. Each ensemble of two-level systems contributes independently to the measurement properties in Fig.~\ref{fig:fourLevelSystem}(c). For this reason, the means and variances af the total photon number measurement and the phase measurement appear to be a concatenation of three two-level-system measurements, each with a different excitation frequency. 

 We use this interpretation to explain the respective similarity of Fig.~\ref{fig:tavisCummingsModel}(d) and Fig.~\ref{fig:tavisCummingsModel}(e) with Fig.~\ref{fig:fourLevelSystem}(g) and Fig.~\ref{fig:fourLevelSystem}(h).  The derivation of the mean photon numbers $\overline n_k$ with respect to the signal Rabi frequency $\Omega_{\text{S}}$ is closely related to the derivations of the Hamiltonian parameters in Eqs.~\eqref{eq:fourLevelSystem:effectiveExcitationEnergies} and \eqref{eq:fourLevelSystem:effectiveExcitationEnergies} with respect to $\Omega_{\text{S}}$: From Eq.~\eqref{eq:fourLevelSystem:effectiveExcitationEnergies} we find that only $E_{\pm 1,\Delta}$ have a functional dependence on $\Omega_{\text{S}}$, while from Eq.~\eqref{eq:fourLevelSystem:effectiveRabiFreq}, we see that $\Omega_{0,\boldsymbol \varphi}$ depends more sensitively on $\Omega_{\text{S}}$ than   $\Omega_{\pm 1,\boldsymbol \varphi}$. In other words, a variation of $\Omega_{\text{S}}$ leads thus mainly to variations of $\Omega_{0,\boldsymbol \varphi}$ and $E_{\pm 1,\Delta}$.

 A variation of $\Omega_{0,\boldsymbol \varphi}$ is closely related to a variation of the atomic density, as the  probe field experiences  an effectively modified impact of  the atomic ensemble. As $\Omega_{0,\boldsymbol \varphi}$ is prevalent for detunings close to $\epsilon_{ba,\Delta}=0$, the Fisher information in Fig.~\ref{fig:tavisCummingsModel}(d) for the density estimation resembles the Fisher information  in Fig.~\ref{fig:fourLevelSystem}(g). Likewise, a variation of $E_{\pm 1,\Delta}$ corresponds to a variation of the level splitting in a two-level system. As the $E_{\pm 1,\Delta}$ variation impact is prevalent for detunings close to $\epsilon_{ba,\Delta}=\pm 20 \text{MHz}$, the Fisher information in Fig.~\ref{fig:tavisCummingsModel}(e) for $\epsilon $ resembles the Fisher information in Fig.~\ref{fig:fourLevelSystem}(h). 
 
 Our analysis in terms of the partially diagonalized Hamiltonian in Eq.~\eqref{eq:fourLevelSystem:effectiveHamiltonian} is thus capable to explain the overall structure of the Fisher information in Fig.~\ref{fig:fourLevelSystem}(f) in the parameter regime investigated here. The investigation of arbitrary parameter regimes of this four-level model would require a more comprehensive numerical study and a more sophisticated analytical treatment, which exceeds the scope of the present work.

\section{Conclusions}
\label{sec:conclusions}

We have developed a comprehensive framework of light-matter interaction which is capable of predicting the spectroscopic measurement statistics by evaluating the semiclassical dynamics of quantum emitters, such as atoms and molecules. This allows us to predict the sensitivity of spectroscopic quantum sensing protocol in terms of the Fisher information. The framework constructs flow equations for the mean values and the covariance matrix of  relevant spectroscopic observables. In this work, we have focused on the measurement statistics related to the  intensity, the polarization direction and the phase shift. The application to dissipative two-level systems and four-level systems has demonstrated the flexibility of the developed framework.

We conjecture that the spectroscopic analysis deploying a full-counting-statistics approach has plenty  of application potential. In particular, it can optimize the operational configuration of currently existing spectroscopic quantum sensing devices, which deploy complicated level structures or periodic driving protocols~\cite{Meyer2020,Jing2020,Bloch2022}. As our approach is non-perturbative, it allows us to predict and interpret the statistical features close to resonance conditions, for which both the signal and the noise are strongly enhanced. This might show advantageous when extending the current methods to investigate the dynamics of laser systems, which are naturally closely related to spectroscopy. Laser systems, which typically work close to resonance conditions to enhance the gain, often suffer from phase diffusion, which can be investigated with virtually the same methods as introduced here. Moreover, the simplicity of the semiclassical approach enables us to investigate the measurement statistics of interacting many-body system, and their potential in quantum sensing. 

Furthermore, there are several relevant topics which have not been addressed in this work. In the cases studied in Secs.~\ref{sec:tavisCummingsModel} and \ref{sec:fourLevelSystem}, we have focused on the stationary state. When assuming short  pulse durations $t_{\text{P}}$, one must integrate the Schr\"odinger equation (or Liouville equation) to obtain the correct dynamical cumulant-generating function. This will be important for transient dynamics, for example, in the read-out of atomic interferometers. In addition, we have not considered some other effects, such as the Doppler shift or a broken rotational symmetry of the quantum emitters, for which the integration of the flow equations will become more elaborate.

Moreover, in our studies we have assumed that the measurement time is relatively long. Modern measurement devices allow to record  spectroscopic signals with a high temporal  resolution. As the PRFT counts the number of accumulated photons, it accounts for the time-averaged signal. A theoretical prediction of the precise joint probability distribution for the time-resolved measurement statistics will give a higher estimate for the Fisher information. As such, the Fisher information  predicted by the PRFT can be interpreted as a lower bound for the Fisher information of the joint probability distribution. Given that the PRFT can also construct Kraus operators related to  specific measurement outcomes~\cite{Engelhardt2024b}, it is also possible  to calculate the detailed joint probability distribution. These   questions will be addressed in future work.

\section*{Acknowledgments}

G.E. acknowledges the support by the National Natural
Science Foundation of China (Grant No. W2432004). Z.Z acknowledges the  Excellent Young Scientists Fund by National Science Foundation of China (No. 9240172), the General Fund by National Science Foundation of China (No. 12474364), and the National Science Foundation of China/RGC Collaborative Research Scheme (No. 9054901).

\appendix

\section{Light-matter Hamiltonian}
\label{sec:lightMatterHamiltonian}

\subsection{Microscopic Hamiltonian}

We consider a system consisting of an ensemble of  $N$ quantum emitters, e.g., atoms or molecules, which interacts with the light field. For concreteness, we will refer to atoms in the following. To describe the system in a microscopic fashion, we deploy the following light-matter Hamiltonian
\begin{equation}
	\hat H = \hat  H_{\text{M}}  +\hat H_{\text{L}} + \hat H_{\text{LM}},
	\label{eq:microscopic-hamiltonian}
\end{equation}
where $\hat  H_{\text{M}}$ acts on the matter system. The free radiation Hamiltonian  and the light-matter interaction Hamiltonians are given by
\begin{eqnarray}
\hat H_{\text{L}}  
&=& \sum_{\boldsymbol k ,\lambda} \omega_{\boldsymbol k } \hat d_{\boldsymbol k , \lambda}^\dagger \hat d_{\boldsymbol k , \lambda}  \label{eq:ham:freeRad} ,\\
\hat H_{\text{LM}} &=& \sum_{m=1}^{N}\int d\boldsymbol r  \hat {\boldsymbol E}(\boldsymbol r) \cdot \hat {\boldsymbol P}_m(\boldsymbol r),
\end{eqnarray}
 respectively.
 
  In terms of the photonic creation and annihilation operators, $\hat d_{\boldsymbol k , \lambda}^\dagger $ and $ \hat d_{\boldsymbol k}$ , the  radiation Hamiltonian  is diagonal with photonic mode frequencies $\omega_{\boldsymbol k }$, which is given by the dispersion relation of the radiation field in  the cloud of atoms. 
The  electric field operator  $\hat {\boldsymbol E}(\boldsymbol r)$  can be expressed as
\begin{equation}
	\hat {\boldsymbol E}(\boldsymbol r)   = \sum_{\boldsymbol k ,\lambda}i \epsilon_{\boldsymbol k} \boldsymbol e_{\boldsymbol k , \lambda}\left( \hat d_{\boldsymbol k , \lambda}^\dagger   e^{-i\boldsymbol k\cdot \boldsymbol r } -  \hat d_{\boldsymbol k , \lambda} e^{i\boldsymbol k\cdot \boldsymbol r }  \right),
\end{equation}
where $\boldsymbol k$ denotes the wavevector of the plain electromagnetic waves, and $\lambda=1,2$ labels the polarization. The  corresponding polarization vectors are given by $\boldsymbol e_{\boldsymbol k , \lambda}$. The polarization of atom $m$ is denoted by $\hat {\boldsymbol P}_m(\boldsymbol r)$.

\subsection{Definition of the laser pulses}

Before proceeding, we  first define the spectroscopic observables of interest on a sound theoretical basis.  In realistic experiments, it is not possible to  measure the occupation of the  photonic operators $ \hat d_{\boldsymbol k , \lambda}$ describing plain waves, as their mode function extends over the full space. Instead, in optical spectroscopy a laser pulse  transverses a cloud of atoms and is then measured using a photon-multiplier. Alternatively, continuous wave lasers are being deployed, which can be likewise interpreted as a continuous train of laser pulses.

For this reason, we establish a theoretical definition of laser pulses in terms of the windowed Fourier transformation~\cite{Clerk2010}. For simplicity, we reduce the three dimensional system to the single dimension of propagation of the laser field, i.e., we  consider a (quasi) one-dimensional system  extended in z direction.  To this end, we first define the directed photonic field operators in position space by
\begin{eqnarray}
	\hat  d_{ \lambda}^{\rightarrow }(r) =\sum_{ k>0 ,\lambda}  \hat d_{ k , \lambda} e^{i kr } , \nonumber \\
	 \hat  d_{ \lambda}^{\leftarrow }(r) =\sum_{ k<0 ,\lambda}  \hat d_{ k , \lambda} e^{i kr }.
	 \label{eq:directedFieldOperator}
\end{eqnarray}
The commutation relations of these operators converge to the common photonic commutation relations if the quantization volume of the $\hat d_{ k , \lambda}$ approaches  infinity. In terms of the directed operators, we now introduce the windowed Fourier transformation. Thereby, we distribute the one-dimensional space into intervals of length $L$, such that the $j$-th interval starts at $r_j = jL$ and ends at $r_{j+1}= (j+1)L$. For each interval, we now define a set of Fourier transformed operators
\begin{eqnarray}
	\hat a_{\lambda,j, M }^{\rightarrow} \equiv \int_{r_j}^{r_{j+1}}  \hat  d_{ \lambda}^{\rightarrow }(r) e^{-ik_M r } dr,
	\label{eq:def:pulseOpeators}
\end{eqnarray} 
which can be regarded as the operators quantizing the laser pulses. Each index combination ($\lambda,j, M$)  labels one specific photonic operator, which can be associated with a laser pulse.   Thereby, $\lambda= 1,2$ denotes the polarization, the integer $j$ refers to the spatial position, and  the integer $M$ corresponding to the (windowed) wave number $k_M = \frac{2\pi}{L} M$. The pulse length is given by $L$, while the bandwidth of the pulse, defined by $\Delta k =k_{M+1} -k_M  $, is $\Delta k =2\pi/L$. Conversely, the directed electric field operators can be expressed as
\begin{equation}
	  d_{ \lambda}^{\rightarrow }(r)  = \tilde \theta(r_j,r_{j+1}, r)  \sum_{M} 	\hat a_{\lambda,j, _M }^{\rightarrow}e^{ ik_M r } ,
	  \label{eq:inversePulseTransformation}
\end{equation}
where we have introduced a modified Heayside function function $\tilde \theta(x,y, z)  $
\begin{equation}
	\tilde \theta(x,y, z)   =
	\begin{cases}
1&\text{if }	x<z\leq y ,\\
0 & \text{else }
	\end{cases}.
\end{equation}
to simplify the notation.

With evolving time, the directed field operators in Eq.~\eqref{eq:directedFieldOperator} and the pulse operators defined in Eq.~\eqref{eq:def:pulseOpeators} propagate along the z direction. We describe this propagation in an interaction picture defined by
\begin{eqnarray}
	\hat U_{\text{prop} } (t) = \exp \left(-i t\sum_{k,\lambda} \omega_k \hat d_{ k , \lambda}^\dagger \hat d_{ k , \lambda}  \right),
\end{eqnarray}
where we recall that $\omega_k$ denotes the dispersion relation of light within the cloud of atoms, and not the free dispersion relation. To proceed, we assume that the laser pulses   have a small wavelength $\lambda = 2\pi/ k_{M_0}$ as compared to the pulse length $\lambda \ll L$. In this regime, we can approximate the time-evolved directed photonic operators as
\begin{eqnarray}
		\hat  d_{ \lambda}^{\rightarrow }(r,t) &\equiv&  \hat U_{\text{prop} }^\dagger  (t)	\hat  d_{ \lambda}^{\rightarrow }(r) \hat U_{\text{prop} } (t)\nonumber  \\
		&=& \sum_{ k>0 ,\lambda}  \hat d_{ k , \lambda} e^{i (kr -\omega_k ) t } \nonumber  \\
		&\approx& \sum_{ k>0 ,\lambda}  \hat d_{ k , \lambda} e^{i \left[ kr -\omega_{k_M}t  -  \frac{d\omega_{k_M}} {d k}(k - k_M ) t\right] }   \nonumber  \\
		&=& \hat  d_{ \lambda}^{\rightarrow }(r - c_{G} t) e^{-i ( \omega_{k_M} -c_G k_M )  t   } .
\end{eqnarray}
In the third line, we have approximated the dispersion relation $\omega_k$ by its Taylor expansion around $k_M$. To this end, we have introduced $c_{\text{G} } =  \frac{d\omega_{k_M}} {d k}$, which can be interpreted as the group velocity of the laser pulse. In doing so, the time-evolved directed photonic operators are shifted in position, and additionally acquire a time-dependent phase $ ( \omega_{k_M} -c_{ \text{G} } k_M )  t $.

Alternatively, we express the time-evolved electric field operators using Eq.~\eqref{eq:inversePulseTransformation} as
\begin{eqnarray}
d_{ \lambda}^{\rightarrow }(r,t)  &=& \sum_{j} \tilde \theta(r_j,r_{j+1}, r-c_Gt) \nonumber   \\
  &\times&   \sum_{M} 	\hat a_{\lambda,j, _M }^{\rightarrow}(0) e^{ ik_M r }  e^{-i  \omega_{k_M}   t   } ,
\end{eqnarray}
where the $\hat a_{\lambda,j, _M }^{\rightarrow}(0) $ are the initial pulse operators. Using this expression, we find that the Hamiltonian in the interaction picture defined by Eq.~\eqref{eq:microscopic-hamiltonian} becomes
\begin{equation}
\hat H(t) = \hat  H_{\text{M}} +  H_{\text{LM}}(t),
\end{equation}
where $\hat  H_{\text{M}}$ acts on the matter system, and the light-matter Hamiltonian is given by
\begin{multline}
\hat H_{\text{LM}} = \sum_{m=1}^{N}\sum_{\lambda, j,M}  \int dr   \hat {\boldsymbol P}_m(r)\cdot \boldsymbol  e_{\lambda}\\ 
\times 
 \tilde \theta(r_j,r_{j+1}, r-c_Gt)   	\hat a_{\lambda,j, _M }^{\rightarrow}  e^{ ik_M r }  e^{-i  \omega_{k_M}   t   } + \text{h.c. }.  
\end{multline}
For a notational reason, we now introduce the short-cuts $\hat a_{k} = 	\hat a_{\lambda,j, _M }^{\rightarrow}  $ and  $\omega_k  = \omega_{k_M}$ for the photonic operators and frequencies. Likewise, we use
\begin{equation}
	g_{m,k}(t) \hat V_m = \int dr  \tilde \theta(r_j,r_{j+1}, r-c_Gt)   \hat {\boldsymbol P}_m(r)\cdot \boldsymbol  e_{\lambda}e^{ ik  r } 
	\label{eq:timedependentInteractionParameter}
\end{equation}
to write the light-matter Hamiltonian in a more compact fashion, where $g_{m,k}(t) $ is a scalar number and $\hat V_m$ is an operator acting on atom $m$.
In doing so, we eventually arrive at the time-dependent quantum-optical Hamiltonian
\begin{equation}
\hat H(t)   = \sum_{j=1}^{N} \hat H_{\text{M},m}   +  \sum_{m,k} g_{m,k}(t) \hat V_m \left(  \hat a_k^\dagger  e^{i \omega_k t} + \hat a_k e^{-i \omega_k t}\right),
\label{eq:hamiltonian:quantum:rotatingFrame}
\end{equation}
which is the same as in Eq.~\eqref{eq:hamiltonian:quantum} but in a frame rotating with the frequencies $\omega_k$.

\section{Derivation of the cumulant flow equation}

\label{app:derivation:CumulantFlowEquation}

\subsection{Exact moment-generating function}

Here, we derive an expression for the moment-generating function of the laser pulse operators $\hat a_k$ appearing in the Hamiltonian in Eq.~\eqref{eq:hamiltonian:quantum}. For later purpose, we assume a more generic light-matter initial state than the one in Eq.~\eqref{eq:InitalState}, namely
\begin{equation}
\rho_{\text{tot} }(t=0) =  \rho^{(\text{M})}(0) \otimes \rho^{(\text{L})}(0),
\label{eq:InitalStateMatterGeneric}
\end{equation}
i.e., where the light field is described by a mixed state instead of a pure state. The moment-generating function of the photonic probability distribution is defined by
\begin{eqnarray}
	M_{\boldsymbol\chi}(t) &\equiv&  \text{tr} \left[ \hat U^{(N)\dagger} (t) e^{-i \sum_k  \chi_k \hat a_k^\dagger \hat a_k   } \hat U^{(N)}  (t) \rho_{\text{tot}}(0)   \right] \nonumber ,
	\label{eq:def:momentGeneratingFunction} \newline\\
\end{eqnarray}
where $ \hat U^{(N)}  (t)$ denotes the time-evolution operator corresponding to the Hamiltonian in Eq.~\eqref{eq:hamiltonian:quantum}. The index $N$ refers thereby to the number of quantum emitters. 
  In an interaction picture defined by $U^{(\text{free})}=  \exp(-i H_{\text{L} }t)$ with $H_{\text{L}}=\sum_{k}  \omega_k  \hat a_k^\dagger  \hat a_k $, the time-evolution operators can be written as  
\begin{equation}
	\hat U^{(N)}  (t) = \hat U^{(\text{free})}(t) \hat U^{(N, \text{int})} (t),
\end{equation}
where $ \hat U^{(N, \text{int})} (t)$ denotes the time-evolution operator in the interaction picture.
Evaluating the time-evolution operator in the Fock basis, we find
\begin{eqnarray}
\left[ \hat  U^{(N) } (t) \right]_{ \boldsymbol  n,\boldsymbol  n_1  } &=& \left<\boldsymbol  n  \right| \hat  U^{(N) } (t)   \left|\boldsymbol  n_1 \right> \nonumber \\
&=&  e^{-i \boldsymbol\omega  \cdot \boldsymbol n}\left[ \hat  U^{(N, \text{int} ) } (t) \right]_{ \boldsymbol  n,\boldsymbol  n_1  }  .
\label{eq:timeEvolutionOperatorRelation}
\end{eqnarray}
We can now evaluate the moment-generating function of the photonic probability distribution 
\begin{widetext}
	\begin{eqnarray}
	M_{\boldsymbol\chi}(t) &=& \sum_{\boldsymbol n, \boldsymbol n_1,  \boldsymbol n_2 }   \text{tr}\left    \lbrace  \left[  \hat  U^{(N) \dagger} (t) \right]_{ \boldsymbol  n-\boldsymbol  n_1,\boldsymbol  n  } e^{-i \boldsymbol \chi  \cdot \boldsymbol n   }    \left[\hat  U^{(N)}  (t) \right]_{\boldsymbol  n, \boldsymbol  n-\boldsymbol  n_2 }    \rho^{(\text{M})}(0) \right\rbrace  
   \rho^{(\text{L} )}_{\boldsymbol n- \boldsymbol n_{1},\boldsymbol n-\boldsymbol n_{2}}   \nonumber  \\
	&=& \sum_{\boldsymbol n, \boldsymbol n_1,  \boldsymbol n_2 }   \left<      \left[ \hat  U^{(N,\text{int}) \dagger} (t) \right]_{ \boldsymbol  n-\boldsymbol  n_1,\boldsymbol  n  } e^{-i \boldsymbol \chi  \cdot \boldsymbol n   }    \left[\hat  U^{(N,\text{int})}  (t) \right]_{\boldsymbol  n, \boldsymbol  n-\boldsymbol  n_2 }    \right> _0 
\rho^{(\text{L} )}_{\boldsymbol n- \boldsymbol n_{1},\boldsymbol n-\boldsymbol n_{2}}   \nonumber  \\
		&=& \sum_{\boldsymbol n, \boldsymbol n_1,  \boldsymbol n_2 }   \left<      \left[ \hat  U^{(N,\text{int}) \dagger} (t) \right]_{\overline{\boldsymbol  n} + \boldsymbol  n-\boldsymbol  n_1,\boldsymbol  n  }   \left[\hat  U^{(N,\text{int})}  (t) \right]_{\overline{\boldsymbol  n} + \boldsymbol  n, \boldsymbol  n-\boldsymbol  n_2 }    \right> _0  e^{-i \boldsymbol \chi  \cdot( \overline{\boldsymbol  n}+ \boldsymbol n )  }  
		\rho^{(\text{L} )}_{\overline{\boldsymbol  n} +\boldsymbol n- \boldsymbol n_{1},\overline{\boldsymbol  n} + \boldsymbol n-\boldsymbol n_{2}}   .
\end{eqnarray}		
In the first equality, we have expressed Eq.~\eqref{eq:def:momentGeneratingFunction}  in the Fock basis. 
In the second equality, we have introduced the notation $\left< \bullet \right>_{0} = \text{tr} \left[ \bullet \rho^{(\text{M} )}(0)  \right]$ and deployed Eq.~\eqref{eq:timeEvolutionOperatorRelation}.		In the third equality, we have shifted the dummy index $\boldsymbol n $ by $\overline{\boldsymbol n} $.

Next, we carry out a Taylor series of the matter-system expectation value as a function of photon number around the mean-photon number of the system $\overline {\boldsymbol n}$, guided by the intuition that the photon-number dependence of the photonic creation and annihilation operators depends only weakly on the photon number in the high-occupation regime. In doing so, we obtain
\begin{eqnarray}		
	M_{\boldsymbol\chi}(t)
	&= & \sum_{r=0}^{\infty} \sum_{\iota_1+ \dots+\iota_{ D }=r } \left\lbrace \partial_{\overline n_1}^{\iota_1 }\cdots  \partial_{\overline n_\textD }^{\iota_\textD }    \left<     \left[  \hat  U ^{(N,\text{int}) \dagger} (t) \right]_{\overline { \boldsymbol  n}-\boldsymbol  n_1,\overline{\boldsymbol  n}  }    \left[\hat  U^{(N,\text{int})}   (t) \right]_{\overline{\boldsymbol  n}, \overline{\boldsymbol  n}-\boldsymbol  n_2 }   \right>_{0}     \right\rbrace    \frac{n_1^{\iota_1} \cdots n_\textD^{\iota_\textD}   }{\iota_1 !\cdots \iota_\textD! } \nonumber \\
	 && \qquad \times e^{-i \boldsymbol \chi  \cdot( \overline{\boldsymbol  n}+ \boldsymbol n )  }  
\rho^{(\text{L} )}_{\overline{\boldsymbol  n} +\boldsymbol n- \boldsymbol n_{1},\overline{\boldsymbol  n} + \boldsymbol n-\boldsymbol n_{2}} ,
	\end{eqnarray}
where $r$ denotes the Taylor expansion order. The subscript of the second summation ($\iota_1+ \dots+\iota_{N_\textD }=r$) means that we have to sum over all combinations $\iota_1, \dots,\iota_{N_\textD }$ whose index sum equals  $r$. The derivatives with respect to the mean photon numbers $\overline n_k$ are denoted by $\partial_{\overline n_k}^{i}$, where $i$ denotes the order of the derivation.

We continue to  carry out  Fourier transformations from the photon-numbers $\boldsymbol n$ to their conjugated phase $\boldsymbol  \varphi$, which are explicitly defined by
\begin{eqnarray}
\left[\hat  U^{(N,\text{int})}   (t) \right]_{\overline{\boldsymbol  n}, \overline{\boldsymbol  n}-\boldsymbol  n_1 } &=&  \frac{1}{(2\pi)^{D} } \int_{-\pi}^{\pi}  d\boldsymbol \varphi    \hat  U^ {(N) \dagger}  _{\overline { \boldsymbol  n},\boldsymbol  \varphi   }   (t) e^{-i \boldsymbol  \varphi   \cdot \boldsymbol  n_1 },  \nonumber \\ \nonumber \\
\rho^{(\text{L} )}_{\boldsymbol n_{1}, \boldsymbol n_{2}}  &=&  \frac{1}{(2\pi)^{ D} } \iint_{-\pi}^{\pi}  d\boldsymbol  \varphi_1    d\boldsymbol  \varphi_2  	 \rho^{(\text{L} )}_{\boldsymbol  \varphi_1   ,  \boldsymbol   \varphi_2  } e^{-i ( \boldsymbol  \varphi_1   \cdot \boldsymbol  n_1 -\boldsymbol  \varphi_2   \cdot \boldsymbol  n_2  ) }    \nonumber .
\end{eqnarray}
Inserting these Fourier transformations, the moment-generating function reads
\begin{eqnarray}		
M_{\boldsymbol\chi}(t)
&= & \sum_{r=0}^{\infty} \sum_{\iota_1+ \dots\iota_{N_\textD } } \frac{1}{(2\pi)^{3D} }  \iiiint \boldsymbol  \varphi_1 \boldsymbol  \varphi_2 \boldsymbol  \varphi_3 \boldsymbol  \varphi_4 \sum_{\boldsymbol n, \boldsymbol n_1,  \boldsymbol n_2 }  \left\lbrace \partial_{\overline n_1}^{\iota_1 }\cdots  \partial_{\overline n_D }^{\iota_\textD }  
\left<     \hat  U^ {(N) \dagger}  _{\overline { \boldsymbol  n},\boldsymbol  \varphi_1  }   (t)  \hat  U^{(N)}_{\overline{\boldsymbol  n},  \boldsymbol  \varphi_2   } (t)   \right>_{0}  
    \right\rbrace    \frac{n_1^{\iota_1} \cdots n_\textD^{\iota_D}   }{\iota_1 !\cdots \iota_D! } \nonumber  \\
   &&\qquad \times  \rho^{(\text{L} )}_{\boldsymbol  \varphi_3   ,  \boldsymbol   \varphi_4  }    e^{i\boldsymbol n\cdot ( \boldsymbol \varphi_4 - \boldsymbol \chi   +\boldsymbol \varphi_3 ) + i \boldsymbol n_1\cdot (\boldsymbol \varphi_3 -\boldsymbol \varphi_1)  - i \boldsymbol n_2\cdot  (\boldsymbol \varphi_4 -\boldsymbol \varphi_2)  + i\overline{\boldsymbol n } \cdot ( \boldsymbol \varphi_4 - \boldsymbol \chi   +\boldsymbol \varphi_3 ) }  . \nonumber  \\
\label{eq:momentGeneratingFunctionFourier}
\end{eqnarray}
The ongoing strategy is to  deploy the identity
\begin{equation}
\frac{1}{2\pi} \sum_{-\infty}^{\infty} e^{-i \varphi n} = \delta(\varphi)
\label{eq:deltaIdentity}
\end{equation}
to identify delta functions  $ \delta(\varphi)$ in Eq.~\eqref{eq:momentGeneratingFunctionFourier}, which can then be used to evaluate some of the phase integrals. However, the photon numbers $n_k^{\iota_k}$ in Eq.~\eqref{eq:momentGeneratingFunctionFourier} prevent us from identifying the delta functions so far. For this reason, we express the photon numbers using derivatives with respect to the counting fields in the following way
\begin{eqnarray}		
M_{\boldsymbol\chi}(t)
&= & \sum_{r=0}^{\infty} \sum_{\iota_1+ \dots+\iota_{ D } =r } \frac{1}{(2\pi)^{3 D} } \iiiint \boldsymbol  \varphi_1 \boldsymbol  \varphi_2 \boldsymbol  \varphi_3 \boldsymbol  \varphi_4 \sum_{\boldsymbol n, \boldsymbol n_1,  \boldsymbol n_2 }  \left\lbrace \partial_{\overline n_1}^{\iota_1 }\cdots  \partial_{\overline n_D }^{\iota_D }  
\left<     \hat  U^ {(N) \dagger}  _{\overline { \boldsymbol  n},\boldsymbol  \varphi_1  }   (t)  \hat  U^{(N)}_{\overline{\boldsymbol  n},  \boldsymbol  \varphi_2   } (t)   \right>_{0}  
\right\rbrace    \nonumber  \\
&&\qquad \times    \rho^{(\text{L})}_{\boldsymbol  \varphi_3   ,  \boldsymbol   \varphi_4  }    e^{i\boldsymbol n\cdot ( \boldsymbol \varphi_4    +\boldsymbol \varphi_3 ) + i \boldsymbol n_1\cdot (\boldsymbol \varphi_3 -\boldsymbol \varphi_1)  - i \boldsymbol n_2\cdot  (\boldsymbol \varphi_4 -\boldsymbol \varphi_2)  + i\overline{\boldsymbol n } \cdot ( \boldsymbol \varphi_4 - \boldsymbol \chi   +\boldsymbol \varphi_3 ) } \frac{\partial_{-i\chi_1}^{\iota_1 }\cdots  \partial_{-i\chi_D }^{\iota_D }     }{\iota_1 !\cdots \iota_D! }  e^{-i \boldsymbol n \cdot \boldsymbol \chi}  \nonumber  \\ \nonumber \\\nonumber \\
&= & \sum_{r=0}^{\infty} \sum_{\iota_1+ \dots+\iota_{D } =r} \frac{1}{(2\pi)^{3D} }   e^{- i\overline{\boldsymbol n }\cdot \boldsymbol \chi }  \frac{\partial_{-i\chi_1}^{\iota_1 }\cdots  \partial_{-i\chi_D }^{\iota_D }     }{\iota_1 !\cdots \iota_D! }\iiiint \boldsymbol  \varphi_1 \boldsymbol  \varphi_2 \boldsymbol  \varphi_3 \boldsymbol  \varphi_4 \sum_{\boldsymbol n, \boldsymbol n_1,  \boldsymbol n_2 }  \left\lbrace \partial_{\overline n_1}^{\iota_1 }\cdots  \partial_{\overline n_D }^{\iota_D }  
\left<     \hat  U^ {(N) \dagger}  _{\overline { \boldsymbol  n},\boldsymbol  \varphi_1  }   (t)  \hat  U^{(N)}_{\overline{\boldsymbol  n},  \boldsymbol  \varphi_2   } (t)   \right>_{0}  
\right\rbrace    \nonumber  \\
&&\qquad \times    \rho^{(\text{L} )}_{\boldsymbol  \varphi_3   ,  \boldsymbol   \varphi_4  }    e^{i\boldsymbol n\cdot ( \boldsymbol \varphi_4 - \boldsymbol \chi   +\boldsymbol \varphi_3 ) + i \boldsymbol n_1\cdot (\boldsymbol \varphi_3 -\boldsymbol \varphi_1)  - i \boldsymbol n_2\cdot  (\boldsymbol \varphi_4 -\boldsymbol \varphi_2)  + i\overline{\boldsymbol n } \cdot ( \boldsymbol \varphi_4    +\boldsymbol \varphi_3 ) }   . \nonumber  \\
\label{eq:tra1:momGenFunction}
\end{eqnarray}

In this form, we can now use the identity in Eq.~\eqref{eq:deltaIdentity}  to simplify the above expression. By evaluating the delta functions, we can cancel three of the four integrals over the phase $\boldsymbol \varphi_{\alpha}$, such that we obtain
\begin{eqnarray}		
M_{\boldsymbol\chi}(t)
&= & \sum_{r=0}^{\infty} \sum_{\iota_1+ \dots+\iota_{D }=r }  e^{- i\overline{\boldsymbol n }\cdot \boldsymbol \chi }  \frac{\partial_{-i\chi_1}^{\iota_1 }\cdots  \partial_{-i\chi_D }^{\iota_D }     }{\iota_1 !\cdots \iota_D! } \int d \boldsymbol  \varphi  \left\lbrace \partial_{\overline n_1}^{\iota_1 }\cdots  \partial_{\overline n_D }^{\iota_D }  
\left<     \hat  U^ {(N) \dagger}  _{\overline { \boldsymbol  n},\boldsymbol  \varphi -\frac{\boldsymbol  \chi}{2}  }   (t)  \hat  U^{(N)}_{\overline{\boldsymbol  n},  \boldsymbol  \varphi + \frac{\boldsymbol  \chi}{2}   } (t)   \right>_{0}  
\right\rbrace      \rho^{(\text{L} )}_{\boldsymbol  \varphi +\frac{\boldsymbol  \chi}{2}   ,  \boldsymbol   \varphi - \frac{\boldsymbol  \chi}{2}  }    e^{ i\overline{\boldsymbol n } \cdot  \boldsymbol \chi  }  .  \nonumber  \\
\label{eq:momGenFct:Exact}
\end{eqnarray}
We emphasize that this is  still an exact expression for the moment-generating function.

\subsection{Semiclassical approximation}

While exact, the expression in Eq.~\eqref{eq:momGenFct:Exact} is impossible to evaluate. At this point, we recall the objective of this work to establish a theoretical framework that is capable to predict the measurement statistics of spectroscopic experiments using only semiclassical methods. Motivated by this aim,  we carry out a  semiclassical approximation in Eq.~\eqref{eq:momGenFct:Exact}. This semiclassical approach has a two-fold effect: 

(i)  we replace the quantum optical time-evolution operators in Eq.~\eqref{eq:momGenFct:Exact} by their semiclassical counterparts, i.e.,
\begin{eqnarray}
\hat  { U }^{(N)}_{\overline{\boldsymbol  n}, \boldsymbol  \varphi }  &\rightarrow& \hat  { \mathcal U }^{(N)}   _{\overline{\boldsymbol  n}, \boldsymbol  \varphi  },
\end{eqnarray}
which is defined in Eq.~\eqref{eq:timeEvolutionOperator:semiclassical}.  We note that this replacement is exact, when representing the Hamiltonian in Eq.~\eqref{eq:hamiltonian:quantum} in the Sambe space as shown in Ref.~\cite{Engelhardt2024b}. The Sambe space is closely related to the Fock space, but with the common photonic operators replaced by $\hat a_k^\dagger  \hat a_k \rightarrow \sum_{n_k} n_k \ketbra{n_k}{n_k}$ and $ \hat a_k \rightarrow \sum_{n_k} \sqrt{\overline n_k} \ketbra{n_k}{n_k+1}$, such that this approximation is well motivated in the large photon-number regime.

(ii) The second approximation  concerns the integral over the phase variables in Eq.~\eqref{eq:momGenFct:Exact}. As we work in the semiclassical limit, we assume that the phase of the photonic field should be uniquely defined. Consequently, it is reasonable to assume that the photonic state $\rho^{(\text{L} )}_{\boldsymbol  \varphi +\frac{\boldsymbol  \chi}{2}   ,  \boldsymbol   \varphi - \frac{\boldsymbol  \chi}{2}  } $ is strongly peaked around a specific mean phase $\overline  { \boldsymbol   \varphi } $, at which we evaluate the integral. 

When applying the semiclassical approximation, the moment-generating function in the semiclassical approximation becomes
\begin{eqnarray}		
M_{\boldsymbol\chi}(t)
&= & \sum_{r=0}^{\infty} \sum_{\iota_1+ \dots\iota_{D } }  e^{- i\overline{\boldsymbol n }\cdot \boldsymbol \chi }  \frac{\partial_{-i\chi_1}^{\iota_1 }\cdots  \partial_{-i\chi_D }^{\iota_D }     }{\iota_1 !\cdots \iota_D! }   \left\lbrace \partial_{\overline n_1}^{\iota_1 }\cdots  \partial_{\overline n_D }^{\iota_\textD }  
\left<     \hat  {\mathcal U}^ {(N) \dagger}  _{\overline { \boldsymbol  n},\overline  { \boldsymbol   \varphi } -\frac{\boldsymbol  \chi}{2}  }   (t)  \hat  {\mathcal U}^{(N)}_{\overline{\boldsymbol  n}, \overline  { \boldsymbol   \varphi } + \frac{\boldsymbol  \chi}{2}   } (t)   \right>_{0}  
\right\rbrace  e^{ i\overline{\boldsymbol n } \cdot  \boldsymbol \chi  }  \int d \boldsymbol  \varphi   \rho^{(\text{L} )}_{ \boldsymbol   \varphi  +\frac{\boldsymbol  \chi}{2}   ,   \boldsymbol   \varphi  - \frac{\boldsymbol  \chi}{2}  }     ,\nonumber \\
	\label{eq:intStep:momentGenFct:semiclassialApproximation}
\end{eqnarray}
in which we can   identify
\begin{equation}
	M_{\boldsymbol\chi}(t=0) =   \int d \boldsymbol  \varphi   \rho^{(\text{L} )}_{\overline  { \boldsymbol   \varphi } +\frac{\boldsymbol  \chi}{2}   ,  \overline  { \boldsymbol   \varphi } - \frac{\boldsymbol  \chi}{2}  }  .
	\label{eq::momentGenFct:initial}
\end{equation} 
Noteworthy, this relation becomes exact  when the initial state is the pure Gaussian state given in Eq.~\eqref{eq:initialStateMatterSubsystem}. The photonic probability distribution in this case has been studied in details in Ref.~\cite{Engelhardt2024c}.

To make progress, we use the defining relation between cumulant and moment-generating functions,
\begin{equation}
M_{\boldsymbol\chi}(0) = \exp \left[ K_{\boldsymbol\chi}(0) \right].
\end{equation}
Moreover, we introduce the dynamical cumulant-generating function describing the action of the ensemble of $N$ atoms on the photonic fields
\begin{eqnarray}
	\mathcal K_{\boldsymbol\chi}^{(N)}(t) &\equiv&
  \log  \left<     \hat  {\mathcal U}^ {(N) \dagger}  _{\overline { \boldsymbol  n},\overline  { \boldsymbol   \varphi } -\frac{\boldsymbol  \chi}{2}  }   (t)  \hat  {\mathcal U}^{(N)}_{\overline{\boldsymbol  n}, \overline  { \boldsymbol   \varphi } + \frac{\boldsymbol  \chi}{2}   } (t)   \right>_{0} \nonumber \\
		&=& \log    \left<     \hat  {\mathcal U}^ {\dagger}  _{\overline { \boldsymbol  n},\overline  { \boldsymbol   \varphi } -\frac{\boldsymbol  \chi}{2}  }   (t)  \hat  {\mathcal U}_{\overline{\boldsymbol  n}, \overline  { \boldsymbol   \varphi } + \frac{\boldsymbol  \chi}{2}   } (t)   \right>_{0}^N\nonumber  \\
		&=& N \log    \left<     \hat  {\mathcal U}^ {\dagger}  _{\overline { \boldsymbol  n},\overline  { \boldsymbol   \varphi } -\frac{\boldsymbol  \chi}{2}  }   (t)  \hat  {\mathcal U}_{\overline{\boldsymbol  n}, \overline  { \boldsymbol   \varphi } + \frac{\boldsymbol  \chi}{2}   } (t)   \right>_{0}\nonumber  \\
	 &=& N	\mathcal K_{\boldsymbol\chi} (t).
\end{eqnarray}
In the second line, we have assumed that the atoms are independent of each other, such that the expectation value of the ensemble is just the $N$-th power of the expectation value for an individual atom. In the last equality, we introduced $G_{\boldsymbol\chi} (t)$, which describes the impact of a single atom on the photonic modes.

In the following, we assume that these $N$ atoms are confined in a volume with area $\mathcal A$ and infinitesimal length $dz$, i.e., $N = \rho_A \mathcal A dz$. Using these definitions, the moment-generating function  in Eq.~\eqref{eq:intStep:momentGenFct:semiclassialApproximation}  becomes
\begin{eqnarray}		
M_{\boldsymbol\chi}(t)
&= & \sum_{r=0}^{\infty} \sum_{\iota_1+ \dots\iota_{ D } }  e^{- i\overline{\boldsymbol n }\cdot \boldsymbol \chi }  \frac{\partial_{-i\chi_1}^{\iota_1 }\cdots  \partial_{-i\chi_D }^{\iota_D }     }{\iota_1 !\cdots \iota_D! }   \left\lbrace \partial_{\overline n_1}^{\iota_1 }\cdots  \partial_{\overline n_D }^{\iota_D }  
\exp \left[ \rho_A \mathcal A dz 	\mathcal K_{\boldsymbol\chi}(t) \right]
\right\rbrace      e^{  K_{\boldsymbol\chi}(0)  +i\overline{\boldsymbol n } \cdot  \boldsymbol \chi  }    .
\label{eq:momentGenFct:semiclassical:dynCGF}
\end{eqnarray}
The following scaling analysis will show how this expression can be simplified. 

To this end, we first have to specify in which fashion we approach the semiclassical limit. As one can see from Eq.~\eqref{eq:hamiltonian:semiclassical},  the bare light-matter interaction parameters $g_{j,k}$ and mean photon numbers $\overline n_k$ appear only as the products $\Omega_{j,k} =  g_{j,k}\sqrt{\overline n_k} $ (i.e., the Rabi frequencies) in the semiclassical Hamiltonian. Thus, the dynamical moment-generating function $\mathcal K_{ \boldsymbol\chi}(t)$ is only  a function of the Rabi frequencies. To reach the semiclassical limit in an unambiguous fashion, we thus require that
\begin{eqnarray}
	\overline n_k &\rightarrow& \infty \nonumber, \\
	g_{j,k} &\rightarrow& 0 \nonumber, \\
	\text{s.t.} \quad g_{j,k}\sqrt{\overline n_k} &=& \Omega_{j,k} =\text{const}.
	\label{eq:scalingAssumptions}
\end{eqnarray}
Physically, the condition can be met by increasing the quantization volume of a pulse in the directions transversal to the propagation.

First, we investigate the scaling properties of the factor 
\begin{equation}
 \partial_{\overline n_1}^{\iota_1 }\cdots  \partial_{\overline n_D }^{\iota_D }  
\exp \left[ \rho_A \mathcal A dz 	\mathcal K_{\boldsymbol\chi}(t) \right]  \equiv  \partial_{\overline n_1}^{\iota_1 }\cdots  \partial_{\overline n_D }^{\iota_D }    	g_{\boldsymbol\chi}(t) ,
\end{equation}
when the mean photon numbers  approach infinity $\overline n_k\rightarrow \infty$. 
Using the scaling assumptions in Eq.~\eqref{eq:scalingAssumptions}, we can thus conclude that
\begin{eqnarray}
	\partial_{\overline n_1}^{\iota_1 }\cdots  \partial_{\overline n_D }^{\iota_D }  
g_{\boldsymbol\chi}(t)  &=& 
		\partial_{\Omega_1}^{\iota_1 }\cdots  \partial_{\Omega_D }^{\iota_D }  
g_{\boldsymbol\chi}(t)  (\partial_{\overline n_1}^{\iota_1 }\Omega_1 ) \cdots  (\partial_{\overline n_D }^{\iota_D }   \Omega_{ D}) \nonumber \\
	&\propto& \frac{g^r}{\sqrt{ \overline n^{r }}}   \propto \frac{1}{\overline n^{r}}, \qquad r= \iota_1+ \dots\iota_{N_D }.
\end{eqnarray}
In the first equality, we have used the chain rule of differentiation.
In the second line,  we utilized $g$ and $ \overline n$ as measures for the $g_{j,k}$ and the $\overline n_k$.  Importantly, note that the derivatives of $g_{\boldsymbol\chi} $ with respect to $\Omega_k$ are independent of the photon number , i.e., $g_{\boldsymbol\chi} \propto \overline n^0 $ under the scaling assumptions in Eq.~\eqref{eq:scalingAssumptions}, as $g_{\boldsymbol\chi}$ is only a function of the Rabi frequencies. Thus, the scaling properties are solely determined by the derivatives of the Rabi frequencies with respect to the mean-photon numbers, showing the terms vanish with  $\overline n^{-r}$ in the Taylor order $r$.

Consequently, in order for the terms of order $r$ to remain in Eq.~\eqref{eq:momentGenFct:semiclassical:dynCGF}, we require additional factors which must jointly scale with $\overline n^r$. These factors can be delivered by the factor $ K_{\boldsymbol\chi}(0)$ in the exponential appearing in Eq.~\eqref{eq:momentGenFct:semiclassical:dynCGF}. To this end, we make the reasonable assumption that
\begin{equation}
	K_{\boldsymbol\chi}(0) \propto \overline n ,
\end{equation}  
which implies that all initial  cumulants  scale as $\propto \overline n $. This is definitely fulfilled for Gaussian states or states which are similar to Gaussian states such as coherent squeezed states.  More precisely, we have to repeatably bring down the factor $ K_{\boldsymbol\chi}(0)$ from the exponent by deriving $r$ times to produce the scaling $\propto \overline n^r$.

Taking all these scaling relations into account, we can express the moment-generating function in Eq.~\eqref{eq:intStep:momentGenFct:semiclassialApproximation} as
\begin{eqnarray}		
M_{\boldsymbol\chi}(t)
&= &  M_{\text{aux},\boldsymbol\chi}(t) M_{\boldsymbol\chi}(0) ,
\label{eq:momentGenFct:semiclassical:final}
\end{eqnarray}
where
\begin{eqnarray}		
M_{\text{aux},\boldsymbol\chi} (t)
&= & \left\lbrace 1+ \rho_A \mathcal A \tilde  K_{\text{aux}, \boldsymbol\chi}  (t)  dz  + \mathcal O\left[ (\rho_A \mathcal Adz)^2,\frac{1}{\overline n}\right] \right\rbrace 
 e^{ \rho_A \mathcal A \mathcal K_{ \boldsymbol\chi}(t)dz  }  , \nonumber  \\ \nonumber  \\
\tilde K_{\text{aux}, \boldsymbol\chi}(t) &= & \sum_{r=1}^{\infty}  \sum_{\iota_1+ \dots\iota_{D } =r }     \left\lbrace \frac{\partial_{\overline n_1}^{\iota_1 }\cdots  \partial_{\overline n_D }^{\iota_D }  }{\iota_1 !\cdots \iota_D! }  \mathcal K_{ \boldsymbol\chi}(t) 
\right\rbrace    
\left\lbrace
\left[ \partial_{-i\chi_1} K_{\boldsymbol\chi}(0) -\overline n_1\right] ^{\iota_1 }\cdots  \left[ \partial_{-i\chi_D } K_{\boldsymbol\chi}(0) -\overline n_D  \right] ^{\iota_D }
\right \rbrace
\label{eq:momentGenFct:semiclassical:step}
\end{eqnarray}
will be denoted as the auxiliary moment-generating function in the following. Within the auxiliary moment-generating function, we have ordered the terms according to their powers of $dz$ and $1/\overline n$. The terms with $1/\overline n^r$ with $r\geq 1$ will vanish in the semiclassical limit $\overline n \rightarrow \infty$. The terms $dz^r$ with $r\geq 2$ will become irrelevant when construction the cumulant flow equations in the next subsection.

\end{widetext}

\subsection{Cumulant flow equation}

In the previous subsection, we have derived an expression for the moment-generating function describing the occupation of the laser pulse modes after they have interacted with an ensemble of $N= \rho_A \mathcal A dz$ atoms for a time duration $t$. As we have seen in the derivation of the pulse Hamiltonian in Sec.~\ref{sec:lightMatterHamiltonian}, the interaction of a pulse with each atom is described by the time-dependent light-matter interaction parameters $g_{m,k}(t)$. For the following considerations, we assume that the train of pulses has a total duration of $t_{\text{M} }$, which we identify  as the time of the light-matter interaction $t=t_{\text{M}}$ in Eq.~\eqref{eq:momentGenFct:semiclassical:final}.

More precisely, we describe the propagation of a  laser pulse (train)   trough the cloud of atoms as a sequential interaction of the  pulse with each layer of the of the atom cloud as sketch in Fig.~\ref{figOverview}(a). After each layer of interaction, the moment-generating function describing the probability distribution is slightly modified according to Eq.~\eqref{eq:momentGenFct:semiclassical:final}. For brevity, we consider the interaction duraction $t=t_{\text{M}}$ as a fixed parameter and consider the moment and cumulant-generating functions as a function of position $z$, i.e., $	K_{\boldsymbol\chi}(z)  =  \log M_{\boldsymbol\chi}(z)$ from now on.

We  can thus interpret  Eq.~\eqref{eq:momentGenFct:semiclassical:final}  as the moment-generating function after interaction with the $N$ atoms in the volume limited by $z$ and $z+dz$. Taking the logarithm on both sides, we thus can formally write
\begin{eqnarray}
	K_{\overline{\boldsymbol n} }(z+dz)  =  \log\left[  M_{\text{aux}, \boldsymbol\chi}  \right]  +   K_{\overline{\boldsymbol n}} (z),
\end{eqnarray}
where $ M_{\boldsymbol\chi}(0) $ in Eq.~\eqref{eq:momentGenFct:semiclassical:final} corresponds to $K_{\overline{\boldsymbol n}} (z)$ and denotes the cumulant-generating function before the interaction with the next atom layer. To derive an effective equation of motion, we  expand $\log\left[  M_{\text{aux}, \boldsymbol\chi}  \right]  $ up to the first order in $dz$. In doing so, we finally obtain
\begin{eqnarray}
	\frac{d}{dz}K_{\overline{\boldsymbol n}} (z) = \rho_A \mathcal A  \left[\mathcal K_{\boldsymbol\chi}(z)  + \tilde K_{\text{aux}, \boldsymbol\chi}(z)    \right],
	\label{eq:flowEquation:nonSimplified}
\end{eqnarray}
which is equivalent to the equation of motion  in Eq.~\eqref{eq:cumulantFlowEquation}. We note that $ \tilde K_{\text{aux}}$ defined in Eq.~\eqref{eq:momentGenFct:semiclassical:step} and $  K_{\text{aux}}$ in  Eq.~\eqref{eq:dynamicalCumulantGeneratingFunction} differ only in the start of the Taylor expansion.

\section{Spectroscopic observables}

\subsection{Rotated measurement frame}
\label{app:rotatedMeasurementFrame}

Here we provide detailed derivations regarding the representation of the flow equation in a frame, which rotates along with the polarization direction of the electromagnetic field, as explained  in Sec.~\ref{sec:rotatingMeasurementFrame}. To this end, we consider the dynamics in terms of the variables
\begin{eqnarray}
	A_{\alpha} &=& \sqrt{n_{\alpha }  } 
	\label{eq:amplitudeDefinition}
\end{eqnarray}
 for $\alpha =1,2  $,  where $n_{\alpha } $  is  the photon numbers of the  occupations operators $\hat a_{\alpha} ^\dagger \hat a_{\alpha} $.  The observables $A_{1} $  and $A_{2} $  are proportional to the electromagnetic fields in $\text{x}$ and $\text{y}$ directions. Besides for a notational reason, we have introduced the variables $A_{\alpha}$ as they have more convenient transformation properties under rotations than the $n_{\alpha }$.  Next, we distribute these variables  in their mean and fluctuations parts, i.e.,  $A_\alpha = \overline A_\alpha + dA_\alpha$. The same separation is also applied to  $n_\alpha = \overline n_\alpha + dn_\alpha$.

First, we consider the mean values of the $A_{\alpha}$. We transform them into a rotated measurement frame defined by the rotation angle of the electric field polarization $\overline \theta = \overline \theta(z)$, such that
\begin{eqnarray}
\left( 
\begin{array}{c}
\overline A_{\text{rot}, 1} \\
\overline A_{\text{rot},2}
\end{array}
\right)&=&
e^{i\overline \theta  \hat \sigma_{\text{y}}}
\left( 
\begin{array}{c}
\overline A_{1} \\
\overline A_{2}
\end{array}
\right).
\label{eq:rotatedFrame:Amplitudes}
\end{eqnarray}
Clearly, the condition in Eq.~\eqref{eq:rotationCondition} is equivalent to the relation
\begin{eqnarray}
 \overline A_{\text{rot}, 1}  &=&  \overline  A_{\text{rot}, 2}
 \label{eq:rotatingFrame:condition}
\end{eqnarray}
for all positions, meaning that the electric field in the rotating frame encloses always an angle $\pi/4$ with both measurement axes.  

Using Eq.~\eqref{eq:rotatedFrame:Amplitudes}, we find that the time evolution  in the rotated frame is given by 
\begin{eqnarray}
\frac{d}{dz}
\left( 
\begin{array}{c}
\overline A_{\text{rot}, 1} \\
\overline A_{\text{rot},2}
\end{array}
\right)&=&
i  \frac{d\overline\theta}{dz}  \hat \sigma_{\text{y}}
\left( 
\begin{array}{c}
\overline A_{\text{rot},1} \\
\overline  A_{\text{rot},2}
\end{array}
\right)
+
\left( 
\begin{array}{c}
 \dot{ \overline A}_{\text{rot},1}  \\
 \dot {\overline A}_{\text{rot},2}
\end{array}
\right), \nonumber \\
\label{eq:rotatedFields}
\end{eqnarray}
where $\dot{ \overline A}_{\text{rot},\alpha}$ denotes the position derivative in the fixed measurement frame. Using now Eqs.~\eqref{eq:amplitudeDefinition}, \eqref{eq:rotatingFrame:condition} and $\overline n_{+} = \overline n_{1}  + \overline n_{2}$, we find 
\begin{eqnarray}
\frac{d\overline\theta}{dz}     &=& \frac{\dot { \overline A}_{\text{rot}, 2}  -  \dot{\overline  A}_{\text{rot}, 1} }   { \overline A_{\text{rot},2}  +   \overline A_{\text{rot}, 1 }   } ,\nonumber\\
\dot{\overline n}_{+} &=&2\overline A_{\text{rot}, 2}   \dot { \overline  A}_{\text{rot}, 2}  +2  \overline A_{\text{rot}, 1}   \dot{\tilde A}_{\text{rot}, 1} ,
\end{eqnarray}
which is equivalent to Eq.~\eqref{eq:rotationAngel-photonNumber}.

Next, we consider the covariance matrix $\boldsymbol \Sigma^2$, which can be expressed in terms of the fluctuations $dn_\alpha$. In the rotated measurement basis, its differential fulfills
\begin{equation}
 d n_{\text{rot},\alpha }  =  2 \overline A_{\text{rot},\alpha}d A_{\text{rot},\alpha},
\end{equation}
where the amplitude fluctuations are related to fluctuations in the fixed measurement basis by
\begin{eqnarray}
\left( 
\begin{array}{c}
dA_{\text{rot}, 1} \\
d A_{\text{rot},2}
\end{array}
\right)&=&
e^{i\theta_z \hat \sigma_{\text{y}}}
\left( 
\begin{array}{c}
d A_{1 } \\
d A_{2 }
\end{array}
\right).
\label{eq:amplitudeFluctuationTransformation}
\end{eqnarray}
Consequently, the amplitude fluctuations fulfill the same differential equations as their corresponding mean values, i.e., 
\begin{eqnarray}
\frac{d}{dz}
\left( 
\begin{array}{c}
dA_{\text{rot}, 1} \\
d A_{\text{rot},2}
\end{array}
\right)&=&
i\frac{d\overline\theta}{dz}  \hat \sigma_{\text{y}}
\left( 
\begin{array}{c}
d A_{\text{rot},1} \\
d A_{\text{rot},2}
\end{array}
\right)
+
\left( 
\begin{array}{c}
d \dot A_{\text{rot},1}  \\
d \dot A_{\text{rot},2}
\end{array}
\right)\nonumber .\\
\label{eq:rotatedFluctuations}
\end{eqnarray}
Using now Eqs.~\eqref{eq:rotatedFields} and \eqref{eq:rotatedFluctuations}, we can construct the flow equation in the rotated measurement frame for $\boldsymbol \Sigma^2$. For instance, we find
\begin{eqnarray}
\frac{d}{dz}\left[ \boldsymbol \Sigma_{\text{rot}}^2\right]_{1,1 }&=& \frac{d}{dz} \left< 	d n_{\text{rot},1 }  d n_{\text{rot},1 }   \right> \nonumber  \\
&= &  \left<   \left( \frac{d}{dz}	d n_{\text{rot},1 }  \right)  d n_{\text{rot},1 }  \right> \nonumber \\
 &+& \left< 	d n_{\text{rot},1 }   \left( \frac{d}{dz}   d n_{\text{rot},1 }   \right) \right> \nonumber  \\
&=& \left<  \dot {\overline  A}_{\text{rot},1} \overline A_{\text{rot},1} d A_{\text{rot},1}d A_{\text{rot},1} \right>\nonumber  \\
 &+&\left<  \overline A_{\text{rot},1} \dot {\overline A}_{\text{rot},1} d A_{\text{rot},1}d A_{\text{rot},1}\right> \nonumber \\
 &+& \left< \overline A_{\text{rot},1}\overline A_{\text{rot},1} d \dot A_{\text{rot},1}d A_{\text{rot},1}\right> \nonumber \\
 &+&\left<  \overline A_{\text{rot},1}\overline A_{\text{rot},1} dA_{\text{rot},1}d  \dot  A_{\text{rot},1} \right> \nonumber \\
 &+&\frac{d\overline\theta}{dz}   \left< \overline  A_{\text{rot},2} \overline A_{\text{rot},1} d A_{\text{rot},1}d A_{\text{rot},1} \right>  \nonumber  \\
 &+&\frac{d\overline\theta}{dz}   \left< \overline A_{\text{rot},1} \overline A_{\text{rot},2} d A_{\text{rot},1}d A_{\text{rot},1} \right>  \nonumber \\
 &+&\frac{d\overline\theta}{dz}    \left< \overline A_{\text{rot},1}\overline A_{\text{rot},1} d A_{\text{rot},2}d A_{\text{rot},1} \right> \nonumber \\
 &+&\frac{d\overline\theta}{dz}   \left<  \overline A_{\text{rot},1}\overline A_{\text{rot},1} d A_{\text{rot},1}d A_{\text{rot},2}\right> .
\end{eqnarray}
The first four terms represent the derivative of $\boldsymbol \Sigma^2$ in the fixed measurement basis. The second four terms can be simplified. Taking the fifth term as example, we find
\begin{eqnarray}
&&\left< \overline  A_{\text{rot},2} \overline A_{\text{rot},1} d A_{\text{rot},1}d A_{\text{rot},1}\right>\nonumber \\
&&=\left< \overline  A_{\text{rot},1} \overline A_{\text{rot},1} d A_{\text{rot},1}d A_{\text{rot},1}  \right> \nonumber \\
&& =\left< 	d n_{\text{rot},1 }  d n_{\text{rot},1 }   \right>   =  \left[ \boldsymbol \Sigma_{\text{rot}}^2\right] _{1,1 },
\end{eqnarray}
where we have used Eq.~\eqref{eq:rotatingFrame:condition}. Putting now everything together, we finally obtain
\begin{eqnarray}
\frac{d}{dz}  \boldsymbol \Sigma_{\text{rot}}^2  &=& \dot{ \boldsymbol \Sigma}^2 _{\text{rot}}\nonumber\\
&& +\frac{d\overline\theta }{dz}   \left[ \left( 
\begin{array}{cc}
1&1 \\
-1&-1
\end{array}
\right) \boldsymbol  \Sigma_{\text{rot}}^2  +  \boldsymbol \Sigma_{\text{rot}}^2 \left( 
\begin{array}{cc}
1&-1 \\
1&-1
\end{array}
\right)   \right], \nonumber \\
\end{eqnarray}
where $ \dot{ \boldsymbol \Sigma}^2 _{\text{rot}}$ denotes the derivative of the correlation matrix in the instantaneous  fixed measurement basis. This expression in equivalent to Eq.~\eqref{eq:correlationMatrixFlow-rotated}.

Finally, we explain how to transform the measurement statistic found in the rotated basis back into the fixed laboratory basis. Using that the mean amplitudes in the fixed measurement basis fulfill
\begin{eqnarray}
\left( 
\begin{array}{c}
\overline A_{ 1} \\
\overline A_{2}
\end{array}
\right)&=&
\sqrt{\frac{\overline n_{+ }}{2}}
e^{i\overline\theta\hat \sigma_{\text{y}}}
\left( 
\begin{array}{c}
1 \\
1
\end{array}
\right),
\label{eq:inverseAmplitudeTransformation}
\end{eqnarray}
 it is straightforward to show that the mean photon numbers in the fixed measurement basis are given by Eq.~\eqref{eq:meanPhotonNumbersAFOangle}.
To find the relation between $\boldsymbol \Sigma^2_{\text{rot}}$ and $\boldsymbol \Sigma^2$ we use that the differential of $n_{\alpha}$ is given by $ d n_{\alpha }  =  2 \overline A_{\alpha}d A_{\alpha}$ for $\alpha = 1,2$. Using the inverse relation of Eq.~\eqref{eq:amplitudeFluctuationTransformation}, we can thus find, for example,
\begin{eqnarray}
\left[ \boldsymbol \Sigma^2\right]_{1,1  }&=&  \left< 	d n_{\text{x} }  d n_{\text{x}  }   \right> \nonumber  \\
%
%
%
&+&\cos(\overline\theta )  \cos(\overline\theta ) \overline  A_{\text{x}} \overline A_{\text{x}}  \left<d A_{\text{rot},1}d A_{\text{rot},1} \right>  \nonumber  \\
&+&\sin(\overline\theta )  \cos(\overline\theta ) \overline A_{\text{x}} \overline A_{\text{x}} \left< d A_{\text{rot},1}d A_{\text{rot},2} \right>  \nonumber \\
&-&\cos(\overline\theta)  \sin(\overline\theta)    \overline A_{\text{x}}\overline A_{\text{x}}\left< d A_{\text{rot},2}d A_{\text{rot},1} \right> \nonumber \\
&+&\sin(\overline\theta)  \sin(\overline\theta)  \overline A_{\text{x}}\overline A_{\text{x}} \left<  d A_{\text{rot},2}d A_{\text{rot},2}\right> \nonumber, \\
\end{eqnarray}
and similarly for the other entries of $\boldsymbol \Sigma^2$. Using Eq.~\eqref{eq:inverseAmplitudeTransformation}, we can  carry out the inverse transformation from the rotated frame to the fixed laboratory by calculating
\begin{equation}
\boldsymbol \Sigma_{\alpha,\beta}^2   =    \frac{2 \overline A_{\alpha} \overline A_{\beta} }{\overline n_{+ }}\left( e^{i\overline\theta \hat \sigma_{\text{y}}}\boldsymbol \Sigma_{\text{rot}}^2 e^{-i\overline\theta  \hat \sigma_{\text{y}}} \right)_{\alpha,\beta}.
\end{equation}
This is equivalent to the expression in Eq.~\eqref{eq:inverseTransformation:rotFrame-labFrame}.

\subsection{Phase measurement}

Here,  we provide background information regarding the statistics of phase measurements in Sec.~\ref{sec:phase}. Using the beamsplitter relation of the photonic operators in Eq.~\eqref{eq:probe-LO-superpositionModes}, we find
\begin{eqnarray}
\hat a_1^\dagger \hat a_1  + \hat a_2^\dagger \hat a_2  &=& \hat a_0^\dagger \hat a_0  + \hat a_{\text{LO} }^\dagger \hat a_{\text{LO}} ,\nonumber \\
\hat a_1^\dagger \hat a_1  - \hat a_2^\dagger \hat a_2  &=&  i\left(  \hat a_0^\dagger \hat a_{\text{LO}}  - \hat a_{\text{LO}}^\dagger \hat a_0  \right),
\end{eqnarray}
which tells us that (i) the total number of photons is preserved in the beamsplitter transformation, (ii) the measured photon number difference contains information about the phase relation of  $\hat a_0 $ and $\hat a_{\text{LO}}$.

Assuming that the mean photon number $\overline n_{\text{LO}}$ and the mean phase $\overline \varphi_{\text{LO}}$ are  known, we can define the following estimators for the total photon number $n_0$  and the phase $\varphi_0 $ of the photonic mode $\hat a_0$ via
\begin{eqnarray}
 n_{0}  &=& n_{1} + n_{2} -\overline n_{\text{LO} } \nonumber  ,\\
	\sin \left(  \varphi_0 - \overline \varphi_{\text{LO}} \right)  &=& \frac{  n_{1} -  n_2 }{2\sqrt{ n_{0} \overline n_{\text{LO}} }},
\end{eqnarray}
from which we trivially see that the average photon number and phase are predicted by Eq.~\eqref{eq:phaseEstimation}.

In the considerations of Sec.~\ref{sec:rotation-phase-relation}, we have assumed that $\overline  n_{0}  = \overline n_{\text{LO}} $, as this configuration is directly related to the case of the polarization measurement. For completeness, we sketch here how to calculate the measurement statistics if one  employs a local oscillator with an arbitrary amplitude. In particular, the task is, given our knowledge of $\left< dn_\alpha  d n_\beta \right>$ in the $\overline  n_{0}  = \overline n_{\text{LO}} $  configuration, to calculate  $\left< d n_{\text{mod} ,\alpha} d n_{\text{mod} ,\beta} \right>$ for a modified local oscillator.

To this end, we consider a similar  approach as in Appendix~\ref{app:rotatedMeasurementFrame} by considering the   electric field amplitudes
\begin{eqnarray}
	A_1&=& \frac{1}{\sqrt{2} } \left( A_{\text{LO}} + i A_0\right), \nonumber \\
	A_2&=& \frac{1}{\sqrt{2} } \left( A_{\text{LO}} - i A_0\right) ,
	\label{eq:outputFieldrandomRelation}
\end{eqnarray}
where $A_1$, $A_2$, $A_{\text{LO}}$, and $A_0$  are the electric fields related to the operators $\hat a_1$, $\hat a_2$, $\hat a_{\text{LO}}$, and $\hat a_0$.  We note that these random variables are complex valued, and can be distributed into a mean-field part and a fluctuations part according to  $A_{\alpha} =\overline A_{\alpha} + d A_{\alpha}$.
When modifying the local oscillators by replacing  $A_{\text{LO}} \rightarrow  A_{\text{mod},\text{LO}} $, the corresponding measured fields simply become
\begin{eqnarray}
 A_{\text{mod},1}&=& \frac{1}{\sqrt{2} } \left( A_{\text{mod},\text{LO}} + i  A_0\right), \nonumber  \\
 A_{\text{mod},2}&=& \frac{1}{\sqrt{2} } \left( A_{\text{mod},\text{LO} } - i   A_0 \right). 
 \label{eq:outputFieldrandomRelationModified}
\end{eqnarray}
for which we can determine the statistical properties of $n_{\text{mod},\alpha} = A_{\text{mod},\alpha}^*   A_{\text{mod},\alpha}$.

We assume  that many of the electric field fluctuations correlations vanish, except for  
\begin{eqnarray}
\left< d A_\alpha^*   d A_\beta \right> &=& \left< d A_\beta   d A_\alpha^* \right> \neq 0 ,\nonumber \\
\left< d A_{\text{mod} ,\alpha}^*    d A_{\text{mod} ,\beta} \right> &=& \left< d A_{\text{mod},\alpha}   d A_{\text{mod},\beta}^* \right> \neq 0
\end{eqnarray}
for $\alpha =1,2$, as the electric fields before passing the beamsplitter can be considered to be uncorrrelated and correlations like $\left< d A_\alpha^*   d A_\beta^*  \right>$ vanish because of a random phase relation. 

Resolving Eq.~\eqref{eq:outputFieldrandomRelation} for $A_0$ and inserting into Eq.~\eqref{eq:outputFieldrandomRelationModified}, we find 
\begin{eqnarray}
A_{\text{mod},1}&=&  \frac{1}{\sqrt{2} }  A_{\text{mod},\text{LO}}   + \frac{1}{2} \left( A_1  - A_2  \right),   \\
A_{\text{mod},2}&=&  \frac{1}{\sqrt{2} } A_{\text{mod},\text{LO}}  -  \frac{1}{2 } \left(  A_1  - A_2  \right) ,
\end{eqnarray}
from which we can calculate the modified entries of the covariance matrix  of the photon-number measurement via
\begin{eqnarray}
 \left[ \boldsymbol \Sigma_{\text{mod} } \right]_{\alpha,\beta} &=& \left<dn_{\text{mod},\alpha} ,dn_{\text{mod} , \beta} \right>    \nonumber \\
 &=&A_{\text{mod},\alpha}^*A_{\text{mod},\beta}\left< d A_{\text{mod},\alpha} dA_{\text{mod},\beta}^*\right> \nonumber  \\
 &+& A_{\text{mod},\alpha}A_{\text{mod},\beta}^*  \left< d A_{\text{mod},\alpha}^* dA_{\text{mod},\beta}\right>  .
\end{eqnarray}
By evaluating these terms, one can express $ \left[ \boldsymbol \Sigma_{\text{mod}} \right]_{\alpha,\beta}$  in terms of the $ \left[ \boldsymbol \Sigma \right]_{\alpha,\beta}$ of the balanced $\overline  n_{0}  = \overline n_{\text{LO} } $ measurement configuration.

\section{Fisher information for  Gaussian distributions}

\label{app:fisherInformationGaussian}

In this Appendix, we evaluate the leading-order contributions for the Fisher information under the assumption of a Gaussian probability distribution, which is parameterized by  
\begin{equation}
p_{ X,\boldsymbol  n } = \frac{1}{(2\pi )^{\frac{D }{2} } \sqrt{\det \boldsymbol \Sigma^2} } e^{-\frac{1}{2}\left(\boldsymbol n-\boldsymbol {\overline n} \right) \boldsymbol \Sigma^{-2}  \left(\boldsymbol n-\mathbf {\overline n}  \right)^T   } ,
\label{eq:def:gaussDistribution}
\end{equation}
where $X$ is the parameter which is to be estimated.  To simplify the following calculation, we thereby assume that the entries of $\boldsymbol  n$ are continuous variables. The influence of the parameter $X$ on  the probability distribution enters via the dependence of the mean photon numbers and the covariance matrix, i.e., $\overline{\boldsymbol n} =  \overline{\boldsymbol n}_X  $ and $ \boldsymbol \Sigma^2  =  \boldsymbol \Sigma^2 _X   $.

The Fisher score of the probability distribution regarding the parameter $X$ is defined by 
\begin{eqnarray}
	g(X,\boldsymbol  n  ) &\equiv& -  \partial_X \ln p_{X,  \boldsymbol  n }\nonumber \\
	&=& \frac{1}{2} (\partial_X  \boldsymbol {\overline n} )  \boldsymbol \Sigma^{-2}  \left(\boldsymbol n-\mathbf {\overline n}  \right)^T 	+\frac{1}{2} \left(\boldsymbol n-\mathbf {\overline n} \right)\boldsymbol \Sigma^{-2}  (\partial_X \boldsymbol {\overline n})^T\nonumber \\
	&+& \left(\boldsymbol n-\boldsymbol {\overline n} \right)  \boldsymbol \Sigma^{-2} ( \partial_X  {\boldsymbol\Sigma^{2}}) \boldsymbol  \Sigma^{-2} \left(\boldsymbol n-\mathbf {\overline n}  \right)^T \nonumber \\
	&+& \frac{1}{2} \text{tr}\left[\partial_{X} \ln  \boldsymbol \Sigma^2 \right].
\end{eqnarray}
Using the Fisher score, we can now evaluate the Fisher information
\begin{eqnarray}
I_{\theta=0} &=& \int d\boldsymbol n p_{ \boldsymbol  n }^{(0)} g(X ,\boldsymbol  n  )^2 \nonumber \\
&=&\int  d\boldsymbol n\mathcal N_0 e^{-\frac{1}{2} \boldsymbol n  \boldsymbol \Sigma^{-2}  \boldsymbol  n ^T   }\left[ (\partial_X    \boldsymbol {\overline n} ) \boldsymbol \Sigma^{-2} \boldsymbol  n^T \nonumber \right. \\
&+& \boldsymbol  n  \boldsymbol \Sigma^{-2} (\partial_X   \boldsymbol\Sigma^{2}) \boldsymbol  \Sigma^{-2} \boldsymbol n^T \nonumber \\
&+&\left. \frac{1}{2}  \text{tr}\left(\partial_{X}\ln  \boldsymbol \Sigma^2 \right) \right]^2 \nonumber \\
&=&(\partial_X  \boldsymbol {\overline n} )\boldsymbol \Sigma^{-2}  (\partial_X   \boldsymbol {\overline n} )^T \nonumber  \\
&+& \text{tr} \left( \boldsymbol \Sigma^{-2}\partial_X   \boldsymbol\Sigma^{2} \right)^2 + 2\text{tr} \left( \boldsymbol \Sigma^{-2} \partial_X  \boldsymbol\Sigma^{2} \boldsymbol \Sigma^{-2} \partial_X  \boldsymbol\Sigma^{2} \right) \nonumber  \\
&+& \,\text{tr} \left( \boldsymbol \Sigma^{-2} \partial_X  \boldsymbol\Sigma^{2} \right) \text{tr}\left(\partial_{X} \ln  \boldsymbol \Sigma^2 \right)  \nonumber \\
&+&\frac{1}{4}  \text{tr}\left(\partial_{X}\ln  \boldsymbol \Sigma^2 \right)^2 ,
\end{eqnarray}
where $\mathcal N_0$ is the normalization of the Gauss distribution in Eq.~\eqref{eq:def:gaussDistribution}.
To further simplify this, we assume the scaling relation $\boldsymbol {\overline n} \propto z  $ and $\boldsymbol \Sigma^2 \propto z  $, where $z$ is the propagation distance of the laser pulse within the atom cloud.  Making a simple scaling analysis of the Fisher information, we find that only the first term scales with $\propto z$, while the other terms scale only  with $\propto z^0$. For this reason, the Fisher information is approximately given by Eq.~\eqref{eq:fisherInfoGaussianState} for a Gaussian probability distribution.

\section{Truncated polynomial approach}

\label{app:truncatedPolApproach}

To analytically evaluate the low-order cumulants, we take advantage of the truncated polynomial approach  of Ref.~\cite{Engelhardt2024b}. The starting point is the characteristic polynomial of the Liouvillian, which can be formally expanded as
\begin{equation}
\mathcal P_{\boldsymbol \chi}(z) =  \sum_{j =0}^{N} a_{j }(\boldsymbol   \chi) z^{j },
\end{equation}
where the expansion coefficients $a_{j }( \chi)$ depend on the counting field.  When intending to evaluate a $l$-th order cumulant, it is sufficient to consider the truncated polynomial
\begin{equation}
\mathcal P^{(l)} _{\boldsymbol  \chi}(z) =  \sum_{j =0}^{l} a_{j }( \boldsymbol  \chi) z^{j },
\end{equation}
as the derivatives of the dominating root of the full polynomial $\lambda_{0;\boldsymbol  \chi} $ and of the truncated polynomial $\lambda_{0;\boldsymbol  \chi}^{(l)}  $ are equal, e.g.,
\begin{equation}
\frac{d^{l }}{d(-i\boldsymbol  \chi_\alpha)^{l} } \lambda_{0;\boldsymbol  \chi = 0 }= \frac{d^{l}}{d(-i\boldsymbol  \chi_\alpha)^{l} } \lambda_{0;\boldsymbol  \chi=0}^{(l)} .
\end{equation}
The same relations hold also true for mixed derivatives.

As we  intent to calculate the first two cumulants, we use the dominant eigenvalue of the corresponding truncating polynomial, which reads
\begin{eqnarray}
\lambda^{(2)}_{0;\boldsymbol  \chi} =  \frac{1}{2a_2}\left(-a_1 + \sqrt{a_1^2 - 4 a_0 a_2}  \right) \nonumber ,
\end{eqnarray}
where we have suppressed the dependence of the expansion coefficients on the counting fields. 
For brevity, we define the following short-hand notation for the derivatives of the polynomial coefficient
\begin{eqnarray}
a_j^{(\alpha)}  &=&  \left. \frac{d}{d\boldsymbol  \chi_{\alpha}  }a_j  \right|_{\boldsymbol  \chi=0} \nonumber,\\
a_j^{(\alpha\beta)}  &=&  \left. \frac{d}{d\boldsymbol  \chi_{\alpha}}\frac{d}{d\boldsymbol  \chi_{\beta}}a_j  \right|_{\boldsymbol  \chi=0} .
\end{eqnarray}
After some tedious but straightforward calculations, we find that the derivaties of the dominating eigenvalue with respect to the counting fields can be expressed as
\begin{eqnarray}
\lambda^{(2)({\alpha }) }_{0;\boldsymbol  \chi=0}  &=&   -\frac{a_0^{(\alpha)} }{a_1} \nonumber \\
\lambda^{(2)({\alpha \beta}) }_{0;\boldsymbol  \chi=0}  &=&     -  \frac{  a_0^{(\alpha \beta)}  }{ a_1}  -  \frac{     a_0^{(\alpha)}  a_2^{(\beta)}+  a_0^{(\beta)}  a_2^{(\alpha)}  } {a_2 a_1 }    \nonumber \\
&+&    \frac{  a_0^{(\alpha)}   a_1^{(\beta)}  + a_0^{(\beta)}  a_1^{(\alpha)}  }{ a_1^2}  -  \frac{ 2 a_2 a_0^{(\alpha) }  a_0^{(\beta) } }{ a_1^3}  ,
\label{eq:secondDerivativeLambda}
\end{eqnarray}
which can used to directly evaluate the cumulants of first and second orders by just determining the expansion coefficients of the characteristic polynomial. 

\section{Analytical calculations for the two-level system}

\subsection{Coefficients of the flow equations}

\label{app:coeefficientsFlowEquationTwolevelSystem}

In this Appendix, we  calculate the coefficients of the flow equation in Eq.~\eqref{eq:fluxEvaluation} and Eq.~\eqref{eq:diffusion-phaseSpaceMatrices} for the two-level system Liouvillian  in Eq.~\eqref{eq:liouvillian:twoLevelSys}. To this end, we use the truncated polynomial approach introduced in Appendix~\ref{app:truncatedPolApproach}. The coefficients of the characteristic polynomial are given by
\begin{eqnarray}
a_4 &=&  1,\nonumber \\
a_3  &=& 2  \gamma,\nonumber \\
a_2  &=&  \frac{1}{2} \mathcal A_{\boldsymbol  \chi } +  \epsilon_{\Delta}^{2} + \frac{5}{4} \gamma^{2},\nonumber \\
a_1 &=& \frac{\gamma}{2} \left[\mathcal A_{\boldsymbol  \chi }  - \mathcal C_{\chi } \right]+ \gamma \epsilon_{\Delta}^{2} + \frac{\gamma^3}{4}  , \nonumber \\
a_0 &=& \frac{1}{16} \mathcal B_{\boldsymbol  \chi }^2 + \frac{\gamma^2}{8} \left(\mathcal A_{\boldsymbol  \chi } -2 \mathcal C_{\boldsymbol  \chi } \right)- i \frac{\epsilon_{\Delta} \gamma}{4}\mathcal B_{\boldsymbol  \chi } ,
\label{eq:expansionCoefficientPolTLS}
\end{eqnarray}
in which we have introduced
\begin{eqnarray}
\mathcal A_{\boldsymbol  \chi }&=&   2\Omega_1^2 +2 \Omega_2^2 + 4 \Omega_1 \Omega_2 \cos\left( \overline \varphi_1 - \overline \varphi_2\right)  \cos\left(  \frac{\chi_1}{2} -\frac{\chi_2}{2}  \right) \nonumber, \\
\mathcal B_{\boldsymbol \chi }&=&-4 \Omega_1 \Omega_2 \sin\left( \overline \varphi_1 -\overline \varphi_2\right)  \sin\left(  \frac{\chi_1}{2} -\frac{\chi_2}{2}  \right), \nonumber \\
\mathcal C_{\boldsymbol \chi }&=& \Omega_1^2 e^{i\chi_1} + \Omega_2^2 e^{i\chi_2} + 2\Omega_1 \Omega_2 e^{i\frac{\chi_1+\chi_2}{2} } \cos \left(\overline\varphi_1 -\overline \varphi_2 \right)\nonumber. \\
\end{eqnarray}
To evaluate the cumulants using Eq.~\eqref{eq:secondDerivativeLambda}, we  calculate the first and second derivatives of the expansion coefficients in Eq.~\eqref{eq:expansionCoefficientPolTLS}. A straightforward calculation gives
\begin{eqnarray}
a_0&=& 0 ,\nonumber \\
a_0^{(1)}&=& - i\frac{\gamma^2}{4} \left[\Omega_1^2+ \Omega_1 \Omega_2  \cos \left(\tilde \varphi\right)\right]      + i \frac{\epsilon_{\Delta} \gamma}{4}2 \Omega_1 \Omega_2 \sin\left( \tilde \varphi \right)  , \nonumber \\
a_0^{(2)}&=&  - i\frac{\gamma^2}{4} \left[\Omega_2^2+ \Omega_1 \Omega_2  \cos \left(\tilde \varphi\right)\right]      - i \frac{\epsilon_{\Delta} \gamma}{4}2 \Omega_1 \Omega_2 \sin\left( \tilde \varphi \right),\nonumber \\
a_0^{(11)}&=& \frac{1}{16} \left[ 2 \Omega_1 \Omega_2 \sin\left( \tilde \varphi \right)  \right]^2  , \nonumber \\
&+& \frac{\gamma^2}{8} \left[-\Omega_1 \Omega_2 \cos(\tilde \varphi) + 2\Omega_1^2 + \Omega_1 \Omega_2 \cos(\tilde \varphi) \right] ,\nonumber \\
a_0^{(22)}&=& \frac{1}{16} \left[ 2 \Omega_1 \Omega_2 \sin\left( \tilde \varphi \right)  \right]^2 ,\nonumber \\
 &+& \frac{\gamma^2}{8} \left[-\Omega_1 \Omega_2 \cos(\tilde \varphi) +  2\Omega_2^2 + \Omega_1 \Omega_2 \cos(\tilde \varphi) \right] , \nonumber \\
a_0^{(12)}&=& - \frac{1}{16} \left[ 2 \Omega_1 \Omega_2 \sin\left( \tilde \varphi \right)  \right]^2 , \nonumber \\
&+& \frac{\gamma^2}{8} \left[\Omega_1 \Omega_2 \cos(\tilde \varphi) +    \Omega_1 \Omega_2 \cos(\tilde \varphi) \right] , \nonumber \\
a_0^{(21)}&=& a_0^{(12)},\nonumber \\\nonumber \\
a_1&=& \frac{\gamma}{2} \left[ \Omega_1^2 +\Omega_2^2 + 2\Omega_1 \Omega_2 \cos(\tilde \varphi) \right]+ \gamma \epsilon_{\Delta}^{2} + \frac{\gamma^3}{4},       \nonumber \\
a_1^{(1)}&=&  -i\frac{\gamma}{2} \left[ \Omega_1^2 +\Omega_1\Omega_2 \cos (\tilde \varphi ) \right]  , \nonumber \\
a_1^{(2)}&=& -i \frac{\gamma}{2} \left[ \Omega_2^2 +\Omega_1\Omega_2 \cos (\tilde \varphi ) \right] , \nonumber \\
a_2&=&  \Omega_1^2 + \Omega_2^2 + 2 \Omega_1 \Omega_2 \cos\left( \tilde \varphi\right) +  \epsilon_{\Delta}^{2} + \frac{5}{4} \gamma^{2} ,   \nonumber \\
a_2^{(1)}&=&  0,\nonumber \\
a_2^{(2)}&=&  0 ,
\end{eqnarray}
where we have defined $\tilde \varphi = \overline \varphi_1  -\overline \varphi_2$. Consequently, the coefficients of the flow equations in Eq.~\eqref{eq:fluxEvaluation} and Eq.~\eqref{eq:diffusion-phaseSpaceMatrices}  are given by
\begin{eqnarray}
\left[ \boldsymbol  I_{\boldsymbol {\overline n}    } \right]_{k_\alpha} & &= -i \rho_A  \mathcal  A  t_{\text{M}} \frac{a_0^{(\alpha)} }{a_1 } , \nonumber \\
\left[	 \boldsymbol  D_{\boldsymbol {\overline n} }   \right]_{k_\alpha,k_\beta}&& = \rho_A  \mathcal  A     t_{\text{M}} \nonumber \\ 
& \times&     \left(  \frac{  a_0^{(\alpha \beta)}  }{ a_1}    -    \frac{  a_0^{(\alpha)}   a_1^{(\beta )} + a_0^{(\beta)}  a_1^{(\alpha)}  }{ a_1^2}  +  \frac{ 2 a_2 a_0^{(\alpha) }  a_0^{(\beta) } }{ a_1^3} \right),\nonumber \\
\label{eq:fullExpressionFlowEquationCoeff}
\end{eqnarray}
where we recall that $\rho_A$ denotes the atom density and $\mathcal A$ is the effective laser cross section.

\subsection{Cumulant aptitudes}

\label{app:cumulantAptitudes}

In a similar fashion we can also evaluate the cumulant aptitudes introduced in Eq.~\eqref{eq:def:cumulantAptitudes}. To this end, we first express the characteristic polynomial in the rotated measurement basis, in which $\Omega_1 =\Omega_2 =\frac{\Omega}{\sqrt{2}}$ with $\Omega$ being the Rabi frequency experienced by the atoms. In doing so, the first three expansion coefficients become
\begin{eqnarray}
a_2  &=&  \Omega^2  + \frac{1}{4} \epsilon_{\Delta}^{2} + \frac{5}{4} \gamma^{2}\nonumber, \\
a_1 &=& \frac{\gamma}{2} \left[2 \Omega^2     - \Omega^2 e^{i\chi_+} \cos\left(  \chi_-   \right)   \right]+ \frac{\gamma}{4} \epsilon_{\Delta} ^{2} + \frac{\gamma^3}{4}  , \nonumber \\
a_0 &=& \frac{1}{4} \left[ \Omega^2   \sin\left(  \chi_-  \right)\right]^2 + \frac{\gamma^2}{4} \left[ \Omega^2  - \Omega^2 e^{i\chi_+} \cos\left( \chi_-   \right) \right]  \nonumber \\
&+& i \frac{\epsilon_{\Delta} \gamma}{4}\Omega^2   \sin\left(  \chi_-   \right),
\end{eqnarray}
where we have used the symmetric and anti-symmetric representation of the counting fields $\chi_{+ }  = \frac{1}{2}\left(  \chi_1 +\chi_2\right) $ and
$ \chi_{-}  = \frac{1}{2}\left(  \chi_1 - \chi_2\right)$. Calculating the derivatives with respect to the counting fields $\chi_+$ and $\chi_-$, we obtain
\begin{eqnarray}
a_0^{(+)}&=&  -i \frac{\gamma^2}{4} \Omega^2 , \nonumber \\
a_0^{(-)}&=&   i \frac{\epsilon_{\Delta} \gamma}{2}\Omega^2 , \nonumber \\
a_0^{(++)}&=&  \frac{\gamma^2}{4} \Omega^2, \nonumber \\
a_0^{(--)}&=&  \frac{1}{2} \Omega^4  +    \frac{1}{4} \gamma^2 \Omega^2 ,\nonumber \\
a_0^{(+-)}&=&  0, \nonumber \\
a_0^{(-+)}&=& 0 ,\nonumber \\\nonumber \\
a_1&=&   \frac{\gamma}{2}  \Omega^2  + \gamma \epsilon_{\Delta}^{2} + \frac{\gamma^3}{4}  ,\nonumber \\
a_1^{(+)}&=&   -i \frac{ \gamma}{2} \Omega^2 ,\nonumber \\
a_1^{(-)}&=&  0 , \nonumber \\
a_2&=& \Omega^2  +  \epsilon_{\Delta}^{2} + \frac{5}{4} \gamma^{2}   . 
\end{eqnarray}
Using now Eq.~\eqref{eq:secondDerivativeLambda}, we can readily obtain closed-form expressions for the cumulant aptitudes 
\begin{eqnarray}
\kappa^{(\delta)}_{+} &=&  -\frac{\gamma\Omega^2}{ 4\epsilon_{\Delta}^2 +2\Omega^2 + \gamma^2 },\nonumber \\
\kappa^{(\delta)}_{-} &=& \frac{2\epsilon_{\Delta}\Omega^2}{ 4\epsilon_{\Delta}^2 +2\Omega^2 + \gamma^2 }  , \nonumber \\
\kappa^{(\delta)}_{++} &=& \frac{a_{0}^{(++)}  }{a_1}- \frac{2 a_{0}^{(+)} a_{1}^{(+)} }{a_1^2} + \frac{2a_2 a_0^{(+)} a_0^{(+)} }{a_1^3 } \nonumber  \\
&=&  \frac{\gamma\Omega^2}{ \left(4\epsilon_{\Delta}^2 +2\Omega^2 + \gamma^2 \right)} 
+\frac{\Omega^4\gamma \left( 8\epsilon_{\Delta}^2 -6 \gamma^{2} \right) }{  \left(4\epsilon_{\Delta}^2 +2\Omega^2 + \gamma^2 \right)^3},\nonumber  \\
\kappa^{(\delta)}_{--} &=&     \frac{  a_0^{(--)}  }{ a_1}         +  \frac{2 a_2 a_0^{(- )}  a_0^{(- )} }{ a_1^3}   \nonumber  \\
&=&  \frac{2\Omega^4  +     \gamma^2 \Omega^2  }{\gamma \left(4\epsilon_{\Delta}^2 +2\Omega^2 + \gamma^2 \right)} 
- \frac{8 \epsilon_{\Delta}^2 \Omega^4 \left( 4\Omega^2  +  4\epsilon_{\Delta}^{2} + 5 \gamma^{2} \right) }{\gamma \left(4\epsilon_{\Delta}^2 +2\Omega^2 + \gamma^2 \right)^3},\nonumber  \\
\end{eqnarray}

which are the same as in Eq.~\eqref{eq:tavisCummingsModel:aptitudes}.

\section{Weak coupling regime}

\label{app:weakCouplingRegimeDynamics}

In this Appendix, we explain the appearance of  the diverging fluctuations in the weak dissipation regime which we have shortly discussed in Sec.~\ref{sec:dissipationDependence} for the two-level system. To show that this a generic effect, we consider the  quantum master equation
 \begin{eqnarray}
\frac{d}{dt } \rho &=& -i \left[\hat {\mathcal H}_{\boldsymbol \varphi +\frac{\boldsymbol \chi}{2}  }  \rho -\rho  \hat {\mathcal H}_{\boldsymbol \varphi-\frac{\boldsymbol \chi}{2} } \right]  + \sum_{j} \gamma_{j} D_j \left[ S_j\right]  \rho ,\nonumber\\
\end{eqnarray}
without further specifying the Hamiltonian $\hat {\mathcal H}_{\boldsymbol \varphi   } $ or the dissipators $ D \left[ S_j\right]$, which are a function of some system operators $\hat S_j$.

Expanding in the eigenstate basis of $ \hat {\mathcal H}_{\boldsymbol \varphi   }$, the master equation approximately becomes
\begin{eqnarray}
 \frac{d}{dt} \rho_{\mu_1,\mu_2} &\approx&  \left(i\epsilon_{\mu_2,\boldsymbol \varphi-\frac{\boldsymbol\chi}{2} } -i\epsilon_{\mu_1 ,\boldsymbol \varphi+\frac{\boldsymbol \chi}{2} } -\gamma_{\mu_1\mu_2, \mu_1\mu_2 }  \right) \rho_{\mu_1,\mu_2} \nonumber \\
 &+&   \sum_{\mu_3}   \gamma_{\mu_1\mu_2, \mu_3\mu_4 }  \rho_{\mu_3,\mu_4},
 \label{eq:masterEquation:eigenBasis}
\end{eqnarray}
where $\epsilon_{\mu ,\boldsymbol \varphi }$ denote the energies of the eigenstates $ \ket{u_{\mu ,\boldsymbol \varphi } } $. We note that the relation $ \braket{u _{\mu_1 ,\boldsymbol \varphi}  }{ u_{\mu ,\boldsymbol\varphi-\frac{\boldsymbol\chi}{2}}  } \approx \delta_{ \mu_1 ,  \mu_2} $ is only approximately fulfilled for small $\boldsymbol \chi$. Yet, in  the approximated master equation in Eq.~\eqref{eq:masterEquation:eigenBasis} we rely on the findings in Ref.~\cite{Engelhardt2024c} that the cumulants are dominated by the energies $ \epsilon_{\mu,\boldsymbol \varphi}$ and not by the counting-field dependence of the eigenstates.

We use the following ansatz to solve Eq.~\eqref{eq:masterEquation:eigenBasis},
\begin{equation}
\rho_{\mu_1,\mu_2} (t )  =\tilde \rho_{\mu_1,\mu_2} e^{\lambda_{\boldsymbol\chi} t},
\label{eq:densityMatrixAnsatz}
\end{equation}
where the time independent $\tilde \rho_{\mu_1,\mu_2}$ is a function of the counting field, and $ \lambda_{\boldsymbol\chi}$ is the cumulant-generating function in the long-time limit  as explained in Eq.~\eqref{eq:auxGenFkt-longTimeLimit}.  Inserting into Eq.~\eqref{eq:densityMatrixAnsatz} and resolving for $\rho_{\mu_1,\mu_2}^{(0)}$ we find
\begin{eqnarray}
\tilde \rho_{\mu_1,\mu_2}
&=& \frac{  \sum_{(\mu_3, \mu_4 ) \neq (\mu_1, \mu_2 )   }   \gamma_{\mu_1\mu_2, \mu_3\mu_4 }  \tilde \rho_{\mu_3,\mu_4}  }{  i\epsilon_{\mu_2,\boldsymbol \varphi-\frac{\boldsymbol \chi}{2} }-i\epsilon_{\mu_1,\boldsymbol \varphi-\frac{\boldsymbol \chi}{2} }- \gamma_{\mu_1\mu_2, \mu_1\mu_2 }  - \lambda_{\boldsymbol \chi} }.\nonumber \\
\end{eqnarray}
From this relation we learn that the coherence elements $\tilde \rho_{\mu_1,\mu_2}$ with $\mu_1 \neq \mu_2$ are strongly suppressed in comparison to the diagonal terms with  $\mu_1 = \mu_2$, as  we consider the weak dissipation regime characterized by $\epsilon_{\mu_2,\boldsymbol \varphi-\frac{\boldsymbol \chi}{2} }-\epsilon_{\mu_1,\boldsymbol \varphi-\frac{\boldsymbol \chi}{2} }\gg \gamma_{\mu_1\mu_2, \mu_1\mu_2 }$.

For this reason, we can restrict the differential equation in Eq.~\eqref{eq:masterEquation:eigenBasis} to the diagonal elements, which can be written in a vector form as 
\begin{equation}
\frac{d}{dt} \boldsymbol p_{\boldsymbol \chi} (t) =    \left[  \boldsymbol  K_{\boldsymbol \chi} +  \boldsymbol \Gamma  \right] \boldsymbol p_{\boldsymbol \chi} (t).
\label{eq:generalizedRateEquation}
\end{equation}
Thereby, $\boldsymbol p_{\boldsymbol \chi}  =  \left( p_{\boldsymbol \chi,1},\dots,  p_{\boldsymbol \chi,D}  \right)$ is the vector of diagonal elements, which are subject to the normalization condition $  \sum_\mu  { p}_{\boldsymbol \chi=0,\mu}=1$, where $ { p}_{\boldsymbol \chi=0,\mu}$ is just the probability to be in eigenstate $\mu$  in the stationary state. As we will see, the following  considerations are independent of the  normalization condition for $\boldsymbol\chi\neq 0$. Moreover, we have defined the matrices
\begin{eqnarray} 
\left[  \boldsymbol   K_{\boldsymbol \chi} \right]_{\mu,\mu} &=& i\left( \epsilon_{\mu,\boldsymbol \varphi-\frac{\boldsymbol \chi}{2} } - \epsilon_{\mu,\boldsymbol \varphi +\frac{\boldsymbol \chi}{2}}\right) \nonumber ,\\
\left[  \boldsymbol   \Gamma \right]_{\mu_1,\mu_2} &=& \gamma_{\mu_1\mu_1, \mu_2\mu_2 }.
\end{eqnarray}
Equation~\eqref{eq:generalizedRateEquation} describes quantum jumps between the eigenstates of the closed Hamiltonian system. Depending on each eigenstate, the measurement properties are determined by the terms $\left[  \boldsymbol   K_{\boldsymbol \chi} \right]_{\mu,\mu} $, which can be interpreted as the cumulant-generating function under the condition, that  the system is in eigenstate $\mu$. The first derivative of $\left[  \boldsymbol   K_{\boldsymbol \chi} \right]_{\mu,\mu}$ with respect to  the counting fields predicts the mean-photon number change (i.e., photon flux) of the photonic laser pulses, conditioned on the eigenstate $\mu$. It is easy to show that the second derivatives of  $\left[  \boldsymbol   K_{\boldsymbol \chi} \right]_{\mu,\mu} $ with respect to the counting field at $\boldsymbol \chi =0$ vanishes. Thus, the photon number variance remains constant as long as the system stays in a specific eigenstate. The increase of the variance can thus only occur when the system jumps between different eigenstates, each featuring a different mean photon flux. The rates of these quantum jumps  are given  by the matrix elements of $ \boldsymbol\Gamma $.

The ansatz in Eq.~\eqref{eq:densityMatrixAnsatz} can now be  written as
\begin{equation}
	\boldsymbol p_{\boldsymbol \chi} (t)   =    e^{\lambda_{\boldsymbol \chi} t }  \tilde {\boldsymbol p}_{\boldsymbol \chi},
\end{equation}
which leads to  the eigenvalue equation
\begin{equation}
\lambda_{\boldsymbol \chi}  \tilde {\boldsymbol p}_{\chi} =    \left[  \boldsymbol  K_{\boldsymbol \chi} +  \boldsymbol \Gamma  \right] \tilde{ \boldsymbol p}_{\boldsymbol \chi} ,
\label{eq:stationaryDiagonalEquation}
\end{equation}
containing the information about all cumulants in the long-time limit.

To simplify the notation in the following, we consider only a single counting field $\chi$, whose cumulants shall be determined.  The following considerations are a simplified version of the derivations in Refs.~\cite{Emary2007,Hussein2014}. Deriving Eq.~\eqref{eq:stationaryDiagonalEquation} with respect to $\chi$ and using the probability normalization condition, we find
\begin{equation}
\lambda_{0}^\prime \tilde {  \boldsymbol p}_{0} =      \boldsymbol  K_{ 0}^\prime \tilde{ \boldsymbol p}_{0}  +  \boldsymbol \Gamma  \tilde{ \boldsymbol p}_{0} ^\prime,
\label{eq:stationaryDiagonalEquationDerivative}
\end{equation}
where the prime denotes the derivative with respect to $\chi$ evaluated at $\chi=0$.
Summing over the states $\mu$, we find  an expression for the first cumulant aptitude
\begin{equation}
	\kappa^{(\delta)}_{1} = -i\lambda_{0}^\prime  = \sum_\mu  \epsilon_{\mu,\varphi}^\prime\tilde { p}_{0,\mu} .
\end{equation}
In doing so, we took advantage of $\sum_{\mu} \left[  \boldsymbol   \Gamma \right]_{\mu,\mu_1}=0$ for all $\mu_1$.

In a similar fashion, we obtain the second cumulant aptitude. To this end, we derive Eq.~\eqref{eq:stationaryDiagonalEquation} two times and obtain
\begin{equation}
\lambda_{0}^{\prime\prime }  \boldsymbol p_{0}  + \lambda_{0}^\prime   \boldsymbol p_{0}^\prime =      \boldsymbol  K_{0}^\prime \tilde{ \boldsymbol p}_{0}^\prime  +  \boldsymbol \Gamma  \tilde{ \boldsymbol p}_{0} ^{\prime\prime}.
\end{equation}
 We can neglect the second term on the right-hand side, as it will vanish once we sum over the eigenstates. Moreover, from Eq.~\eqref{eq:stationaryDiagonalEquationDerivative} we find that 
\begin{equation}
\tilde{ \boldsymbol p}_ {0}^\prime   = 	\boldsymbol\Gamma^+ \left[ \lambda_{0}^\prime-   \boldsymbol  K_{0}^\prime  \right] \tilde{ \boldsymbol p}_ {0} = -   \boldsymbol  K_{0}^\prime\tilde{ \boldsymbol p}_ {0} ,
\label{eq:stationaryDiagonalEquation2ndDerivative}
\end{equation}
in which $\boldsymbol \Gamma^+$ denotes the pseudoinverse of $\boldsymbol \Gamma$. Inserting this expression for $\tilde{ \boldsymbol p}_ {0}^\prime $ into Eq.~\eqref{eq:stationaryDiagonalEquation2ndDerivative} and summing over the eigenstates, we finally obtain an expression for the second cumulant aptitude
\begin{equation}
\kappa^{(\delta)}_{2} = -\lambda_{0}^{\prime\prime } = \sum_{\mu_1 \mu_2 } \left[   \Gamma^+    \right]_{\mu_1 \mu_2}  \epsilon_{\mu_1,\varphi}^\prime\epsilon_{\mu_2,\varphi}^\prime\tilde { p}_{0,\mu} ,
\end{equation}
which is a function of the  energy derivatives $\epsilon_{\mu ,\varphi}^\prime$, i.e., the fluxes conditioned on the eigenstate. As this expression involves the pseudoinverse of the rate matrix $\boldsymbol \Gamma$, it will diverge  for a vanishing dissipation.

 \bibliography{projectLibrary}

\end{document}